%% file: manuscript.tex
\documentclass[manuscript,screen,nonacm]{acmart}
\setcopyright{none}
\AtBeginDocument{%
  }

\usepackage{algpseudocode}
\usepackage{algorithm}
\usepackage{quantikz}
\usepackage{tikz}
\usepackage{subcaption}
\usepackage[table]{xcolor}% http://ctan.org/pkg/xcolor
\usepackage{pifont}% http://ctan.org/pkg/pifont
\usepackage{multirow}
\usepackage{hyperref}
\usepackage{booktabs}
\usepackage{longtable}
\newcommand{\cmark}{\ding{51}}%
\newcommand{\xmark}{\ding{55}}%
\begin{document}

\title{MOSAIQC: Mixed-topology-aware Optimization for Scalable Approximate noise-Informed Quantum circuit Cutting}

\author{Koen Mesman}
\email{k.j.mesman@tudelft.nl}
\orcid{0000}

\affiliation{%
  \institution{Delft University of Technology}
  \city{Delft}
  \country{Netherlands}
}

\author{Yinglu Tang}
\orcid{https://orcid.org/0000-0003-3075-6610}
\email{y.tang-5@tudelft.nl}
\affiliation{%
  \institution{Delft University of Technology}
  \city{Delft}
  \country{Netherlands}
}
\author{Matthias M\"{o}ller}
\orcid{https://orcid.org/0000-0003-0802-945X}
\affiliation{%
  \institution{Delft University of Technology}
  \city{Delft}
  \country{Netherlands}
}
\author{Boyang Chen}
\orcid{https://orcid.org/0000-0001-7393-4363}
\affiliation{%
  \institution{Delft University of Technology}
  \city{Delft}
  \country{Netherlands}
}
\author{Sebastian Feld}
\orcid{https://orcid.org/0000-0003-2782-1469}
\affiliation{%
  \institution{Delft University of Technology}
  \city{Delft}
  \country{Netherlands}
}

\renewcommand{\shortauthors}{Mesman et al.}

\begin{abstract}
  Current quantum computers do not yet have the required qubit resources to meet the demands of most practical quantum algorithms. To circumvent this constraint, the practice of dividing these algorithms into parts through quantum circuit cutting has been explored. Many of these works either show exponential scaling or are far from optimal solutions. In this paper, MosaiQC is presented as a novel framework to improve upon existing circuit cutting frameworks. A hybrid warmstart with refinement optimization is used to find cutting solutions, allowing the combination of both wire and gate cuts. Additionally, MosaiQC enables hardware partitions of mixed sizes. Furthermore, the refinement stage incorporates a fast approximate quadratic assignment solver to better place hardware partitions, demonstrating a mean local fidelity improvement of $+19.56 \% \pm 6.17\%$ over the baseline algorithm. In runtime and sampling overhead costs, improvements of $2.88 \times$ and an average of $16.84\%$ cut reduction (resulting in an average $5.83 \cdot 10^{11} \times$ overhead reduction) are observed. MosaiQC demonstrates a superior trade-off for run speed and solution quality, while adding fundamental features excluded by most competitors. With this, MosaiQC demonstrates that scalable heuristic optimization can substantially reduce the computational overhead of circuit-cut placement for increasingly large quantum circuits.
\end{abstract}





\maketitle

\include{Sections/Intro}
\include{Sections/Method}
\include{Sections/Results_ext}

\include{Sections/Discussion}
\include{Sections/Conclusion}
\bibliographystyle{ACM-Reference-Format}
\bibliography{bib.bib}

\include{Sections/Appendix}

\end{document}

%% file: Sections/Intro.tex
\section{Introduction}
Although quantum computers continue to increase in size, practical quantum algorithms remain well beyond the qubit capacity and fidelity of current NISQ hardware. A major challenge is the projected hardware requirements for practical usage in terms of qubits and (gate) fidelity. Current hardware is restrained by a limited number of qubits and subject to noise, limiting the scope of realizable computations.

To enable larger computations with small to moderate quantum computers, methods such as distributed computing \cite{tomesh2023divide, CALEFFI2024110672, escofet2023hungarian, fujii2022deep} and circuit knitting \cite{mitarai2021constructing, peng2020simulating, bechtold2023patterns} have been developed.

%Circuit cutting

\begin{table}[h]
    \centering
    \begin{tabular}{|c|c|c|c|c|}
        \hline
         & k wire cuts & \multicolumn{3}{c|}{k gate cuts} \\
        & & CNOT & swap & $CR(\theta)$ \\
       \hline
       
        Sampling overhead & $16^k$ & $9^k$ & $49^k$ & $(1+2|sin(\theta)|)^{2k}$\\
        
        \hline
    \end{tabular}
    \caption{Sampling costs circuit cutting for exact reconstruction, based on \cite{Brandhofer2024OptimalPartitioning}.}
    \label{tab:sampling}
\end{table}

\begin{figure}[h]
    \centering
    \includegraphics[trim={0 0 0 1cm}, clip, width=0.8\linewidth]{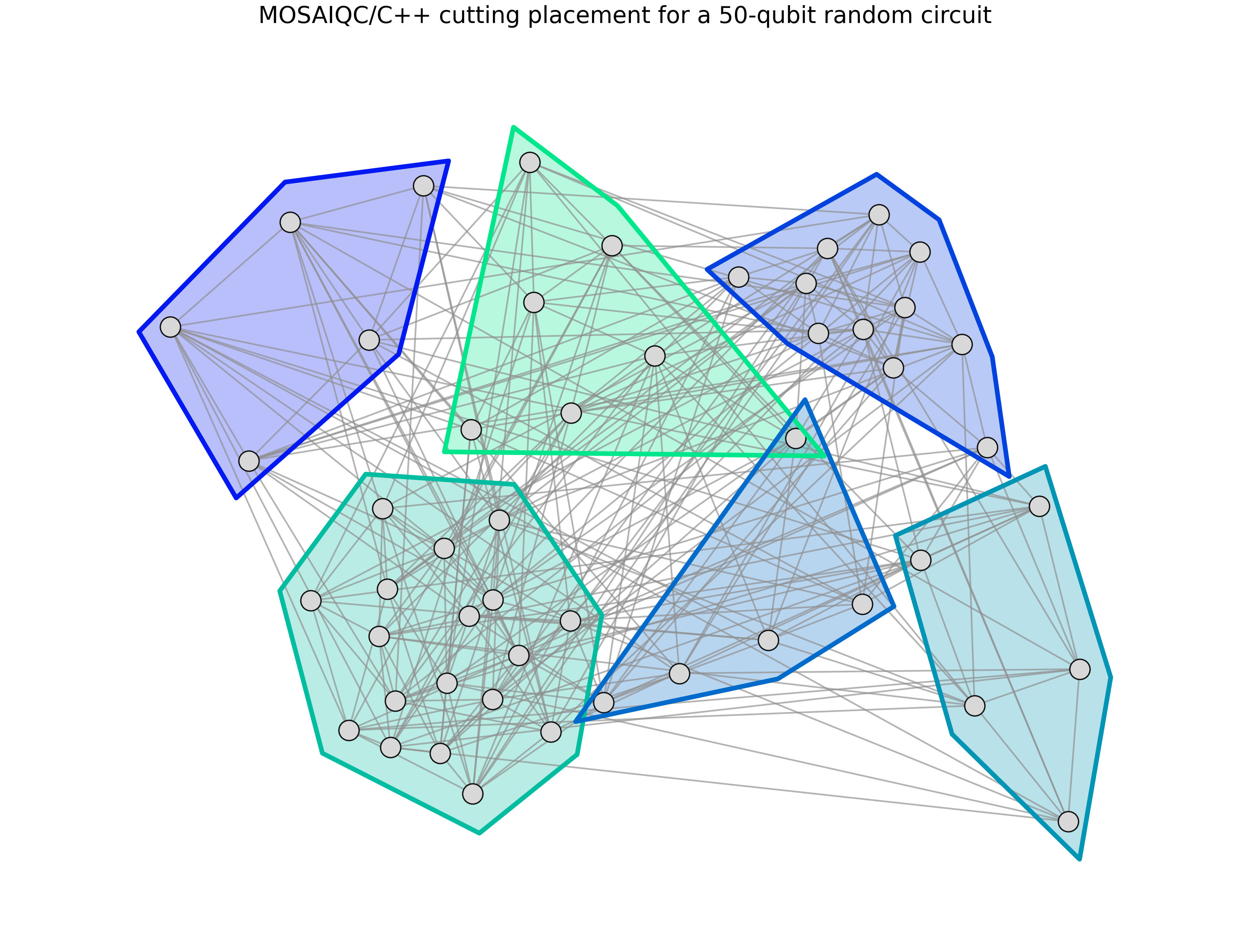}
    \caption{Example of partition mapping by MosaiQC of smaller-capacity hardware backends to a large quantum circuit. The vertices represent qubits, and the edges represent 2-qubit gates. Each polygon represents a different quantum hardware backend.}
    \label{fig:placement}
\end{figure}

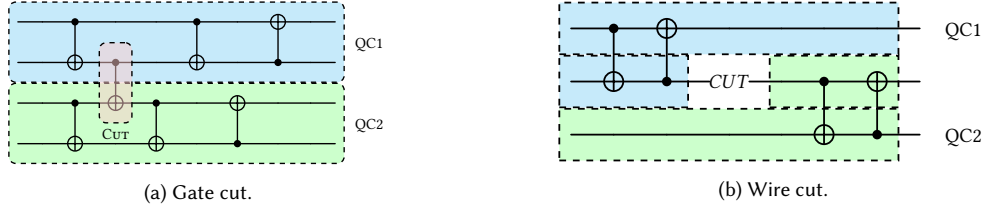
\begin{figure*}[h]
    \centering
    \begin{subfigure}[t]{0.5\textwidth}
    \centering
    \scalebox{0.65}{
    \begin{quantikz}
    \gategroup[2,steps=10,style={dashed,rounded
corners,fill=cyan!20, inner
xsep=2pt},background,label style={label
position=right,anchor=north,xshift=0.5cm}]{{\sc
QC1}}&&\ctrl{1}&&& \ctrl{1}& &\targ{}&&\\
        & &\targ{}&\ctrl{1}\gategroup[2,steps=1,style={dashed,rounded corners, fill=red!20,  fill opacity=0.50, draw opacity=1, text opacity=1, inner xsep=2pt},label style={label position=below,anchor=north,yshift=-0.2cm}]{{\sc Cut}} & &\targ{} &&\ctrl{-1}&&\\
       \gategroup[2,steps=10,style={dashed,rounded corners,fill=green!20, inner xsep=2pt},background,label style={label position=right,anchor=north,xshift=0.5cm}]{{\sc QC2}} & &\ctrl{1}&\targ{}&\ctrl{1}&&\targ{}&&&\\
        & &\targ{}&&\targ{}& & \ctrl{-1} && &
    \end{quantikz}
    }
    \caption{Gate cut.}
    \label{fig:gatecuts}
    \end{subfigure}
    ~ 
    \begin{subfigure}[t]{0.5\textwidth}
    \centering
    \scalebox{0.85}{
    \begin{quantikz}
    \gategroup[1,steps=7,
    style={dashed,fill=cyan!20,inner xsep=2pt},
    background, label style={label
position=right,anchor=north,xshift=1cm}]{{\sc QC1}} & \ctrl{1} & \targ{} & & & & &\\
     \gategroup[1,steps=3,
    style={dashed,fill=cyan!20,inner xsep=2pt},
    background]{}& \targ{} & \ctrl{-1} & \push{CUT}  & \gategroup[1,steps=3,
    style={dashed, fill=green!20,inner xsep=2pt},
    background]{}& \ctrl{1} & \targ{}&\\
    %\gategroup[1,steps=5, style={dashed,rounded corners,fill=green!20,inner xsep=2pt}, background label style={label position=right,anchor=north,xshift=0.5cm}]{{\sc QC2}}
    \gategroup[1,steps=7,
    style={dashed, fill=green!20,inner xsep=2pt},
    background,label style={label position=right,anchor=north,xshift=1cm}]{{\sc QC2}} & & & & &\targ{} & \ctrl{-1}&
    \end{quantikz}
    }
    \caption{Wire cut.}
    \label{fig:wirecut_a}
    \end{subfigure}
    \caption{Circuit cutting examples.}
    \label{fig:cuts}

\end{figure*}

In circuit knitting, a large quantum circuit is mapped to multiple quantum computers of smaller qubit capacity (Figure \ref{fig:placement}). This could enable the execution of quantum circuits normally not executable with available resources. To split the quantum algorithm into partitions, two cutting techniques are available: Gate cutting \cite{mitarai2021constructing} and Wire cutting \cite{peng2020simulating}, visualized in Figure \ref{fig:cuts}. Both techniques require state reconstruction in post-processing, for which an exponential sampling overhead is required, as presented in Table \ref{tab:sampling}.  The cost of quantum circuit knitting is the exponential sampling overhead for each required cut and is therefore critical to minimize for practical execution. In recent works, these measurement overheads are being reduced \cite{lowe2023fast, nakamura2025improved}, even showing polynomial scaling when some approximation error is allowed \cite{harada2025exponentialtopolynomialscalingmeasurementoverhead}. In this work, we will exclusively focus on circuit knitting as a more general framework compared to distributing calculations on an algorithmic level.

Several works have developed optimization approaches to find the optimal locations for placing circuit cuts.  Different approaches consider either gate-cutting or wire-cutting. The combination of these is rarely demonstrated, as it is computationally significantly more costly to optimize. For example, Tang et al. demonstrate wire cut placement using Mixed Integer Programming (MIP), which scales as $O(n!)$ \cite{Tang2021CutQC, Tang2022ScaleQC} with $n$ being the number of cuts. In comparison, Brandhofer et al. demonstrated cutting both wires and gates using an SMT solver, scaling exponentially \cite{Brandhofer2024OptimalPartitioning}. Both computational complexities severely limit the size of circuits to which exact approaches can be applied.

While these cuts are optimal in terms of sampling overhead, the effect of noise is often neglected, as it vastly increases the complexity of the problem.
Some approaches either combine the noise cost globally, locally, or exclusively focus on the optimization for noise reduction.
Combining both noise and sampling overhead optimization is a more challenging multi-objective approach, resulting in a more computationally costly optimization routine.

As these optimizations are part of a compilation subroutine for running the quantum circuit, it is crucial that these optimizations are both fast and scalable. Hardware mapping is known to be an NP-hard optimization problem \cite{ito2023algorithmic, zhu2022complexity}, and cut placement is shown to be an NP-complete problem \cite{idan2026quantumcircuitcuttingcomplexity}. Therefore, scalable circuit cutting requires heuristic optimization strategies that balance solution quality with runtime.% The problem of both noise mapping (initial placement and routing) and optimal cut placement are both NP-hard problems. Finding an exact solution in polynomial time is therefore unfeasible. Instead, we must rely on heuristic approaches for close-to-optimal but scalable methods.

Existing heuristic partitioning methods generally assume homogeneous hardware partitions. In practice, users increasingly execute quantum workloads across multiple cloud-accessible quantum processors with different qubit capacities and calibration characteristics. Supporting heterogeneous hardware therefore substantially broadens the applicability of circuit cutting.

We propose MosaiQC, a heuristic optimization framework for scalable circuit cutting. The framework jointly optimizes sampling overhead, hardware-aware placement, and mixed gate/wire cuts while maintaining polynomial runtime complexity. We publish our optimization framework as a Python package, with the optimization implemented in C++ through PyBind11 for ease of integration and fast runtimes. Current optimization frameworks become a significant compilation bottleneck for circuits approaching the thousand-qubit regime. Our goal is therefore not only to reduce sampling overhead, but also to ensure that cut placement itself remains computationally practical for large-scale circuits.

The contributions of this paper are:
%\begin{itemize}
%\item A heuristic approach with low computational overhead for assessing the wire cut, in addition to gate cuts, allowing for optimizing the combination of both cutting methods
%\item A local wire-cut evaluation strategy that reduces optimization complexity from DAG-based representations to qubit-level graphs.
%\item Support for heterogeneous hardware backends with varying qubit capacities and noise characteristics.
%\item An accurate hardware-aware placement method based on doubly weighted graph matching using Frank–Wolfe optimization for quadratic assignment, improving circuit execution fidelity.
%\item A hybrid global/local search algorithm structure is used to find global approximate solutions in polynomial computational scaling
%\end{itemize}

\begin{itemize}
\item A heuristic framework for jointly optimizing mixed wire- and gate cuts within a single optimization procedure.
\item A local wire cut evaluation strategy based on qubit-level graph representations, reducing the computational complexity of repeated DAG-based optimization.
\item A hardware abstraction supporting heterogeneous quantum backends with varying qubit capacities and hardware characteristics.
\item A hardware-aware placement method based on doubly weighted graph matching, solved using Frank-Wolfe optimization for quadratic assignment.
\item A hybrid optimization strategy that combines global METIS partitioning for warm-start initialization with local Tabu Search refinement, enabling scalable approximate optimization with polynomial runtime.
\end{itemize}

This paper is organized as follows: Section \ref{sec:related} summarizes existing work on quantum circuit cutting and positions our work in comparison. Section \ref{sec:method} details the method of the MosaiQC framework. In Section \ref{sec:res}, the results are presented and further discussed in Section \ref{sec:discussion}. Finally, in Section \ref{sec:conclusion}, the conclusion is delivered.

\section{Related work}
\label{sec:related}
\subsection{Exact methods}
As the sampling overhead for quantum circuit cutting scales exponentially with the number of cuts, it makes sense to invest additional computational effort into finding the exact minimal cut solution. Several publications have demonstrated exact approaches to this end. 

\textbf{MIP} solvers (CutQC \cite{Tang2021CutQC}, ScaleQC \cite{Tang2022ScaleQC})  have been proposed for exact wire-cut optimization. ScaleQC further introduces a state merging framework to reduce classical memory requirements. Complementary approaches have addressed exact gate-cut optimization using \textbf{Clifford cutting}, as implemented in SuperSim \cite{Smith2023SuperSim}. Finally, \textbf{SMT solver} \cite{Brandhofer2024OptimalPartitioning} and \textbf{tensor contractions} \cite{ayral2021quantum} can be used to solve for both gate and wire cuts.
In \cite{ayral2021quantum}, noise sources are investigated as well, giving insight into how circuit cutting can affect the fidelity.

Although these methods provide exact cut placements, their exponential or factorial computational complexity limits their applicability to large-scale circuits.

\subsection{Heuristics}
Heuristic methods sacrifice exact solutions in exchange for substantially improved scalability, making them suitable for larger circuit instances.

Several heuristic approaches formulate circuit cutting as a graph-based partitioning problem. FitCut \cite{Kan2024FitCut} optimizes gate cut placement through weighted graph clustering, but does not use exact hardware qubit capacities and therefore generally leaves hardware resources underutilized. Similarly, Cambiucci et al. \cite{Cambiucci2025SpatialTemporalCutting} and QDisLib \cite{Tejedor2025Qdislib} employ (hyper)graph partitioning heuristics such as Kernighan-Lin and METIS. While these methods scale efficiently, they optimize either wire cuts or gate cuts rather than both simultaneously.

Alternative heuristic strategies have also been explored. QuantCut \cite{soloviev2025quantcut} applies the evolutionary optimizer EDAspy to optimize gate cut placement, whereas LarQucut \cite{Dou2025LarQucut} uses a pruned min-heap search for distributed quantum computing (DQC). Although worst-case exponential runtime, pruning substantially improves its practical performance.

Structure-aware heuristics have also been proposed. ACK \cite{johnson2026distributedquantumcomputingadaptive, mohseni2026buildquantumsupercomputerscaling} partitions circuits using entropy measures to balance computational load across homogeneous resources. CiFold \cite{kan2025circuit} instead exploits modular circuit structure and is one of the few heuristic frameworks supporting both gate and wire cuts. Similar to FitCut, however, it does not explicitly account for the exact capacities of the target hardware.

Collectively, these methods demonstrate that heuristic optimization enables scalable circuit cutting, but existing approaches generally sacrifice support for mixed cut types, heterogeneous hardware, or hardware-aware optimization.

\subsection{Noise reduction}
Rather than minimizing sampling overhead, several works instead optimize circuit cutting for hardware fidelity by exploiting device-specific noise characteristics. Here, less focus is placed on reducing sampling overhead but rather on finding the partition placement such that the overall circuit execution noise on the mapped hardware is minimized. One such implementation is FragQC \cite{Basu2024FragQC}, which constructs weighted hardware topology graphs from average device error rates to guide noise-aware partitioning, demonstrating the importance of hardware-aware optimization.
Similarly, DevQCC \cite{Sahu2025DevQCC} assumes partitions that fit within the available hardware and evaluates the best placement of partitions on hardware after cut placement. The best placements are determined based on the transpiled circuit dimension, device coupling map, and device noise model.
Another approach to this is finding partition placements to minimize SWAP insertions. Ren et al. \cite{Ren2024HardwareAwareGateCutting} compare circuit and hardware topology graphs using Graph Edit Distance (GED) as a heuristic proxy for routing overhead, demonstrating a strong correlation between GED and the number of inserted SWAP gates.

Beyond partition placement, several works consider how circuit fragments are scheduled across distributed hardware. In other words, given a circuit cutting fragmentation, when and on which hardware a fragment is executed. The work by Bhoumik et al. \cite{Bhoumik2023DistributedScheduling} targets this specific problem, incorporating device-specific noise parameters, queueing delays, and communication patterns to find optimal scheduling.
In configurations where sparse connections between quantum hardware is available through EPR pairs, other constraints need to be considered. This specific configuration is targeted by DisMap \cite{Du2024DisMap}, and also takes into account noise of the hardware to optimize placement.

These approaches demonstrate that hardware-aware optimization can substantially improve execution fidelity, but generally treat noise optimization independently from sampling-overhead minimization. In contrast to these approaches, our work jointly optimizes sampling overhead and hardware-aware placement within a single optimization framework.

\subsection{Application specific}
If circuit cutting is applied to a specific problem, the knowledge of the structure can be used to find more efficient methods to cut. Publications have been presented for Quantum Neural Networks \cite{Marchisio2025CuttingAllYouNeed, periyasamy2025cutreglossregularizerenhancing} and QAOA (QDCA) \cite{tomesh2023divide}. In this paper, we focus on a problem-agnostic framework. Therefore, we do not compare ourselves to these approaches. Regardless, we believe such works give valuable insights.

\subsection{Comparison}
%To position our work with related circuit cutting frameworks, we present a comparison in Table \ref{tab:comparison}. Here, we present the main elements of each implementation. We compare the capabilities of optimizing wire and gate cuts, whether the optimization considers hardware noise, allows for hardware of different qubit capacities, solution accuracy, and scalability. In some cases, optimization can be done for either wire or gate cuts, but not both at the same time (marked blue). Notably, for exact solvers, the scaling is always exponential or factorial, whereas for approximate solvers, the solution time is generally polynomial. For LarQucut, we mark the scaling as blue, as formally the worst-case scaling is exponential, but in practice it is closer to polynomial. Similarly, the Qiskit CKT can run exponentially if it is allowed to run an exhaustive search, but by default it will stop early, giving it an approximate result. For FitCut, no explicit computational complexity is stated, but as the solver is a constrained Louvain approach, it can be assumed that it is $O(k\cdot n_e)$ for $n_e$ edges and $k$ iterations. In the work of ACK, no computational complexity is explicitly defined, but it can be assumed from the method that it scales polynomially. In our work, the main solver requires $k$ iterations at a cost of $n$, with a subroutine of $m$ evaluations, for which $m << n$ of which each evaluation costs $n^3/\epsilon$ for the Frank-Wolfe algorithm applied to the Hungarian assignment with error $\epsilon$.
To position our work within the current literature, Table~\ref{tab:comparison} compares representative quantum circuit cutting frameworks in terms of supported cut types, hardware awareness, heterogeneous hardware support, solution accuracy, and computational complexity.

The comparison highlights that existing frameworks typically optimize only a subset of these capabilities. In particular, support for mixed wire and gate cuts, heterogeneous hardware, and hardware-aware placement is rarely combined within a single scalable optimization framework. MosaiQC addresses this gap while maintaining polynomial computational complexity of $O(k \cdot n_q +m\cdot n_q^3)$, where $n_q$ denotes the number of qubits, $k$ the number of graph partitioning iterations, and $m$ the number of candidate solutions refined using quadratic assignment.

\begin{table}[]
    \centering
    \begin{tabular}{|c|c|c|c|c|c|c|c|}
     \hline
     \multirow{2}{*}{publication} & optimization & wire & gate  & noise  & mixed  & computational & cut placement  \\
      & method &  cuts &  cuts &  aware & hw & scaling & accuracy \\
     
     \hline

     MosaiQC % verified
     & Metis + Frank-Wolfe
     & \cellcolor{green!25}\cmark & \cellcolor{green!25}\cmark 
     & \cellcolor{green!25}\cmark & \cellcolor{green!25}\cmark
     & \cellcolor{green!25}$O(k \cdot n_q +m\cdot n_q^3)$
     & \cellcolor{cyan!25}approximate \\

     \hline
     
     CutQC \cite{Tang2021CutQC, Tang2022ScaleQC} % verified
     & MIP
     & \cellcolor{green!25}\cmark & \cellcolor{gray!25}\xmark & \cellcolor{gray!25}\xmark &  \cellcolor{gray!25}\xmark 
     & \cellcolor{gray!25} factorial
     & \cellcolor{green!25}exact \\

    Ayral \cite{ayral2021quantum}
    & Tensor contraction
    & \cellcolor{green!25}\cmark & \cellcolor{green!25}\cmark & \cellcolor{green!25}\cmark & \cellcolor{gray!25}\xmark 
    & \cellcolor{gray!25}exponential
    & \cellcolor{green!25}exact \\

     SuperSim \cite{Smith2023SuperSim}
     & Clifford cutting
     & \cellcolor{gray!25}\xmark & \cellcolor{green!25}\cmark & \cellcolor{gray!25}\xmark  & \cellcolor{gray!25}\xmark 
     & \cellcolor{gray!25}exponential
     & \cellcolor{green!25}exact \\

     Brandhofer \cite{Brandhofer2024OptimalPartitioning} % verified
     & SMT
     & \cellcolor{green!25}\cmark & \cellcolor{green!25}\cmark & \cellcolor{gray!25}\xmark & \cellcolor{gray!25}\xmark
     & \cellcolor{gray!25}exponential 
     & \cellcolor{green!25}exact \\

     FragQC \cite{Basu2024FragQC} % verified
     & Quantum Annealing 
     & \cellcolor{green!25}\cmark & \cellcolor{gray!25}\xmark 
     & \cellcolor{green!25}\cmark & \cellcolor{gray!25}\xmark 
     & \cellcolor{gray!25} $O(e^{\sqrt{n}})$
     & \cellcolor{cyan!25}approximate \\

     Ren \cite{Ren2024HardwareAwareGateCutting} % verified
     & Karger-Stein
     & \cellcolor{gray!25}\xmark & \cellcolor{green!25}\cmark 
     & \cellcolor{green!25}\cmark &  \cellcolor{gray!25}\xmark
     & \cellcolor{green!25} polynomial
     & \cellcolor{cyan!25}approximate \\

    FitCut \cite{Kan2024FitCut}
    & Community detection
    & \cellcolor{green!25}\cmark & \cellcolor{green!25}\cmark 
    & \cellcolor{gray!25}\xmark & \cellcolor{green!25}\cmark
    & \cellcolor{green!25}polynomial
    & \cellcolor{cyan!25}approximate \\

     DevQCC \cite{Sahu2025DevQCC} % verified
     & MIP 
     & \cellcolor{green!25}\cmark & \cellcolor{gray!25}\xmark 
     & \cellcolor{green!25}\cmark & \cellcolor{green!25}\cmark
     & \cellcolor{gray!25} factorial
     & \cellcolor{green!25}exact \\

    Cambiucci \cite{Cambiucci2025SpatialTemporalCutting}
    & Kernighan-Lin
    & \cellcolor{cyan!25}\cmark / \xmark & \cellcolor{cyan!25}\cmark / \xmark & \cellcolor{gray!25}\xmark  & \cellcolor{gray!25}\xmark 
    & \cellcolor{green!25} polynomial
    & \cellcolor{cyan!25}approximate \\

    QDisLib \cite{Tejedor2025Qdislib}
    & Kernighan-Lin, METIS
    & \cellcolor{cyan!25}\cmark / \xmark & \cellcolor{cyan!25}\cmark / \xmark & \cellcolor{gray!25}\xmark &  \cellcolor{green!25}\cmark
    & \cellcolor{green!25} polynomial
    & \cellcolor{cyan!25}approximate \\

    LarQucut \cite{Dou2025LarQucut}
    & Pruning
    & \cellcolor{green!25}\cmark & \cellcolor{gray!25}\xmark 
    & \cellcolor{gray!25}\xmark & \cellcolor{green!25}\cmark
    & \cellcolor{cyan!25}exp to polynomial
    & \cellcolor{cyan!25}approximate \\

    ACK \cite{johnson2026distributedquantumcomputingadaptive, mohseni2026buildquantumsupercomputerscaling}
    & entropy based
    & \cellcolor{green!25}\cmark & \cellcolor{gray!25}\xmark 
    & \cellcolor{gray!25}\xmark & \cellcolor{gray!25}\xmark 
    & \cellcolor{green!25}polynomial
    & \cellcolor{cyan!25}approximate \\

    QuantCut \cite{soloviev2025quantcut}
    & EDAspy
    & \cellcolor{gray!25}\xmark & \cellcolor{green!25}\cmark 
    & \cellcolor{gray!25}\xmark & \cellcolor{gray!25}\xmark 
    & \cellcolor{green!25} polynomial
    & \cellcolor{cyan!25}approximate \\

    CiFold \cite{kan2025circuit}
    & Folding
    & \cellcolor{green!25}\cmark & \cellcolor{green!25}\cmark 
    & \cellcolor{gray!25}\xmark & \cellcolor{gray!25}\xmark
    & \cellcolor{green!25}polynomial
    & \cellcolor{cyan!25}approximate \\

    Qiskit CKT \cite{qiskit-addon-cutting}
    & Dijkstra's
    & \cellcolor{green!25}\cmark & \cellcolor{green!25}\cmark 
    & \cellcolor{gray!25}\xmark & \cellcolor{gray!25}\xmark
    & \cellcolor{cyan!25} exp to polynomial
    & \cellcolor{cyan!25}approximate \\
    
    \hline

    \end{tabular}
    \caption{Comparison of representative quantum circuit cutting frameworks with respect to supported optimization objectives, hardware capabilities, solution accuracy, and computational complexity. A \cmark ~indicates native support, and \xmark ~no support. Mixed (\cmark/\xmark) indicates both wire and gate cuts are supported but not the combination.}
    \label{tab:comparison}

\end{table}

% Note: DisMap not included as it is focused on EPR pair linked QC
% DevQCC not considered as it is a scheduling framework (not cutting)

%% file: Sections/Method.tex
\section{Methodology}
\label{sec:method}

%\subsection{Structure Overview}

%MosaiQC is designed as a hierarchical optimization framework that progressively refines circuit cut placement through increasingly accurate optimization stages. Rather than directly optimizing the complete objective, the framework first constructs an efficient initial solution using inexpensive global heuristics before applying local refinement techniques where they provide the greatest benefit. This balances computational efficiency with solution quality while maintaining polynomial runtime complexity.

%An overview of the optimization pipeline is shown in Figure~\ref{fig:workflow}. The framework consists of four sequential stages:

\subsection{Structure Overview}

MosaiQC is designed as a hierarchical optimization framework that progressively refines circuit cut placement through increasingly accurate optimization stages. Rather than directly optimizing the complete objective, the framework first constructs an efficient initial solution using inexpensive global heuristics before applying more computationally intensive optimization steps only where they provide the greatest benefit. This balances computational efficiency with solution quality while maintaining polynomial runtime complexity.

An overview of the optimization pipeline is shown in Figure~\ref{fig:workflow}. The framework consists of four optimization stages:

\begin{enumerate}
    \item \textbf{Graph representation.} The quantum circuit is transformed into a compact graph representation that captures the qubit interaction required for cut placement optimization while reducing the computational overhead associated with repeated graph evaluation.

    \item \textbf{Initial partitioning.} The graph is partitioned into candidate subcircuits using a global graph partitioning heuristic, producing an initial circuit partition that satisfies the hardware capacity constraints.

    \item \textbf{Local refinement.} Starting from the initial partition, local modifications are iteratively evaluated and accepted when they improve the overall optimization objective. During this process, the hardware-aware placement cost is periodically updated to balance sampling overhead and execution fidelity.

    \item \textbf{Hardware-aware partition assignment.} During local refinement, circuit partitions are periodically assigned to available quantum hardware through a quadratic assignment formulation that accounts for hardware connectivity and noise characteristics. The resulting placement score is incorporated into the optimization objective while limiting the computational overhead of repeated hardware assignment.
\end{enumerate}

Each stage produces the input required for subsequent optimization, progressively improving the solution while avoiding the computational cost of directly optimizing the complete problem. The following subsections describe each stage of the framework in detail.

%The following subsections describe each stage of the framework in detail. The stages are ordered from inexpensive global optimization toward increasingly expensive local refinement to maximize scalability.

\subsection{Graph representation}
%Preprocessing
%- sparse quant circ to weighted connectivity graph
%- parse hardware objects to connectivity graphs based on noise values
%- 2-q gates are stores as LUTs for wire cut calculations
%- Stored as c++ objects through PyBind11

To enable efficient cut placement optimization, the quantum circuit and hardware topology are first transformed into compact graph representations. While many approaches use the Directed Acyclic Graph (DAG) of the quantum circuit in order to calculate both wire and gate cuts \cite{Kan2024FitCut, Tang2022ScaleQC, Tejedor2025Qdislib, Brandhofer2024OptimalPartitioning}, we instead opt to split this into two separate elements. First is the weighted connectivity graph. Here, qubits are represented by nodes and edges by 2-qubit gates. The weight of both are the number of 1-q gates and 2-q gates on nodes and edges, respectively.

The connectivity graph is needed to optimize gate cuts. However, to include wire cuts, the temporal connectivity needs to be stored as well. The evaluation of wire cuts occurs in our framework when two hardware topologies share the same qubit in placement. For each qubit, we store the outgoing 2-qubit gates. When evaluating the wire cuts, we can index these outgoing gates by their respective partition (outgoing gates not included in either partition are not included here). The number of wire cuts that can be implemented here is the count of outgoing edge pairs that both belong to different partitions.
        
Existing DAG-based representations accurately capture circuit dependencies but become computationally expensive for repeated cut-placement evaluations. We therefore construct a qubit-level graph that preserves the information required for optimization while substantially reducing graph size.

The hardware topology graph is a straightforward connectivity graph based on the hardware topology. The weights are assigned by the 1-q and 2-q gate noise to nodes and edges, respectively. 
For a more accurate noise representation, non-existent edges are filled with an alternative error path through a Floyd-Warshall shortest path algorithm. With this, the costs of non-existent connections are still evaluated instead of being excluded from the calculation. Here, the edge weights are set as:
\begin{equation}
    w_{i, j} = 3 \cdot \text{shortest\_path\_error}+\alpha \cdot \text{shortest path hops} \cdot \text{mean backend cx error}
\end{equation}

The $$3* \text{shortest\_path\_error}$$ approximates the SWAPs inserted to realize the qubit routing, and $$\alpha * \text{shortest path hops} * \text{mean backend cx error} $$ serves as a small noise-aware discouraging factor to avoid unnecessary SWAPs. Factor $\alpha$ is chosen empirically, where $\alpha_{path} = 0.25$ demonstrates a reasonable balance where reroutes are not favored over existing connections but also not overly discouraged. The overhead of finding the additional routing costs is approximately $O(n^3)$ for $n$ hardware qubits. As this is only executed once in preprocessing, this overhead can be considered negligible. 

In our framework, we allow for multiple hardware backends of varying sizes (or simulators) with different noise topologies in parallel. %This allows users who do not have access to a fully heterogeneous quantum multicore system to still fully optimize for their system. This situation is expected for many users in the current NISQ era, where generally third-party quantum computers are accessed in parallel, each with different qubit capacities and noise topologies. 
In order to make optimal use of the hardware, we implement a mapping of quantum capacities where the minimal sum of the available hardware is assigned, favoring larger hardware to reduce overall cuts. If not enough qubits are available to fully run the algorithm in parallel, the hardware capacities are re-used for the sequential running of circuit partitions.
Specifically, we assign hardware from set $$S = \{s_1, s_2, ..., s_k \}$$ for $k$ different hardware instances such that:
\begin{equation}
    S_{\text{qubits}} = \sum_{i=1}^k s_i
\end{equation}
\begin{equation}
        R = \lfloor C_{\text{qubits}}/S_{\text{qubits}} \rfloor
\end{equation}
\begin{equation}
        \text{remainder} = C_{\text{qubits}} \text{mod} (R)
\end{equation}
For the remaining assignments, the hardware backends are selected such that:
\begin{equation}
    S^* = \arg\min_{S}\left( |S|,\ \sum_{s_x \in S} s_x \right) \quad
    \text{s.t.} \quad \sum_{s_x \in S} s_x \ge \text{remainder} 
\end{equation}

For the implementation of these graphs, we initialize both the circuit and the hardware as Qiskit objects \cite{javadiabhari2024quantumcomputingqiskit}, which are then converted to lightweight C++ graph objects through PyBind11 \cite{pybind11}. For the hardware noise information, we make use of Qiskit's (Fake)Backend objects.

% Figure examples hardware and circuit graphs
%-------------------------------------------------------------------------------

%Here, the workflow of the optimization is elaborated. The optimization starts with a warm-start in order to get a fast good initial guess, to be later refined using a more thorough optimization process. The optimization is fully implemented in C++ to ensure high performance.

\subsection{Initial partitioning}
The initial warm-start uses the circuit connectivity graph and the available hardware capacities, i.e., the sizes of the allowed partitions. Using the METIS library \cite{karypis1997metis}, we partition the connectivity graph. This gives a good initial solution for only the gate cuts or only wire cuts, as the partitioning method does not allow for overlaps.

\subsection{Refinement}
The warm-start is passed to a Tabu-search \cite{glover1989tabu}, which optimizes through local partition assignment moves:
\begin{itemize}
    \item Add circuit node to partition assignment (if qubit capacity allows)
    \item Move circuit node to a different partition
    \item Swap circuit nodes between partitions at partition edges
\end{itemize}
For each move that results in an overlap between partitions, the wire-cut lookup is evaluated.

Local evaluation intentionally replaces repeated global graph analysis, substantially reducing the computational cost of evaluating neighboring solutions.

\begin{figure}
    \centering
\scalebox{0.7}{
\begin{quantikz}
\lstick{} &        & \ctrl{1} &        &               & \ctrl{1} & \ctrl{1} &        & \ctrl{1} & \ctrl{1} &  
\\
\lstick{} & \ctrl{1} & \targ{}  & \ctrl{1} & \ctrl{1} &        \targ{}  & \targ{}  &     \ctrl{1}   & \targ{}  & \targ{}  &  
\\
\lstick{} & \targ{}  &        & \targ{}  & \targ{}  &               &        & \targ{}  &        &        &  
\end{quantikz}
}
\quad
$\Longrightarrow$
\quad
\scalebox{0.7}{
\begin{quantikz}
\lstick{} & \qw & \ctrl{1}
\gategroup[2,steps=1,
style={dashed,rounded corners,fill=red!20,
fill opacity=0.50,draw opacity=1,text opacity=1,
inner xsep=2pt}]{} 
& \qw & \qw& \qw &\ctrl{1} & \ctrl{1} & \qw & \ctrl{1} & \ctrl{1} & \qw
\\
\lstick{} & \ctrl{1} & \targ{} & \ctrl{1} & \ctrl{1} & \push{CUT} &\targ{} & \targ{} & \ctrl{1} \gategroup[2,steps=1,
style={dashed,rounded corners,fill=red!20,
fill opacity=0.50,draw opacity=1,text opacity=1,
inner xsep=2pt}]{} & \targ{} & \targ{} & \qw
\\
\lstick{} & \targ{}    & \qw   & \targ{}  & \targ{}  & \qw   & \qw   & \qw    & \targ{}  & \qw     & \qw     & \qw
\end{quantikz}
}

    \caption{Example of mixed gate and wire cut optimization.}
    \label{fig:mixed_cut}

\end{figure}
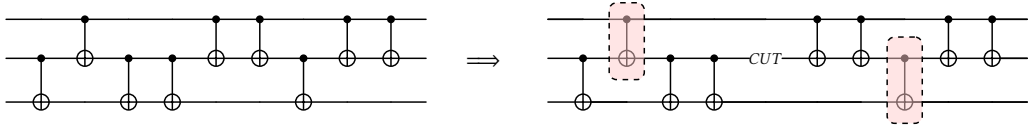

Computing the wire cut overhead from gate lists can be done efficiently in $O(m)$ for per-qubit gate depth $m$. An example of such an optimization for mixed wire and gate cuts is shown in Figure \ref{fig:mixed_cut}. Here, the combination of cuts reduces, e.g., four wire cuts to two wire cuts and one gate cut.

The pseudocode for retrieving the measurement overhead cost of performing wire cuts is defined in Appendix \ref{app:wirecut}.
The computational overhead for these evaluations is $O((a+1)p \text{log}(p))$, where $p$ is the longest gate chain on any qubit and $a$ is the number of unused ancilla qubits. This $O((a+1)p \text{log}(p))$ computational cost is performed for $n$ qubits where partition mappings overlap. We can safely assume that $n\ll N$, and $a << N$; therefore, few evaluations will be needed in practice. The at most $n$ overlaps can easily be computed in parallel when needed.

As both wire and gate cuts are included in the optimization, the cost function is adjusted to reflect the measurement overhead. Using the overhead costs presented in Table \ref{tab:sampling}, we can write the cost function as:
\begin{equation}
    \text{overhead} = (\text{gate cost})^g \cdot 16^w
\end{equation}
for $g$ gate cuts and $w$ wire cuts. To make this evaluation computationally cheaper, we can rewrite this as:
\begin{equation}
    \ln(\text{overhead}) = g \cdot \ln(\text{gate cost}) + w \cdot \ln(16)
\end{equation}
Here, $ln(\text{gate cost})$ and $ln(16)$ can be mostly pre-processed as constants (except for parameterized gates), and the actual overhead can be retrieved in post-processing.

\subsection{Hardware noise mapping}
To evaluate the noise mapping, a comparison must be made between the topology graph of the hardware and the assigned circuit partition. Ren et al. suggest using Graph Edit Distance (GED) to assess this as a heuristic ($O(n^2)$ - $O(n^3)$ for $n$ qubits) proxy \cite{Ren2024HardwareAwareGateCutting}. GED calculates the minimal number of edges and nodes to be added and removed from a graph in order to achieve the same graph as it is compared to. As we are using weights for both the circuit and hardware topology graphs, we find (weighted) GED to be inaccurate as the weights are of different scales. Instead, we intend to optimize for matrices $A$ and $B$ (circuit partition and topology graph) to find a permutation matrix $P$:
\begin{equation}
    \min \mkern9mu \text(trace)(APB^TP^T)
\end{equation}
 To this end, a Quadratic Assignment Problem (QAP) solver, the Frank-Wolfe algorithm \cite{Frank1956AnAF}, is used for a more accurate placement. This algorithm can better handle the different weights, while still being an efficient heuristic, and is generally considered to give close to optimal results \cite{jaggi2013revisiting}.

For the implementation, the placement score is constructed from the hardware noise topology. For each weight on the topology graph, the hardware noise  $\epsilon_{\text{noise}}$ is transformed as:
\begin{equation}
    w = \ln(1-\epsilon_{noise})
\end{equation}

 denoted as $w_{i}$ for qubit $i$ and $w_{i,j}$ for connections between qubits $i$ and $j$. For the quantum circuit, we denote vertices as $v_{i}$ and edges as $e_{i,j}$. The QAP algorithm then finds the maximum value placement score $s$ :
\begin{equation}
    s = \sum_{i,j} w_{i, j} \cdot e_{i,j} + \sum_{i} w_{i} \cdot v_i
\end{equation}

Where the maximum value corresponds to the highest expected fidelity, i.e., the lowest combined noise.

The combined fidelity could be reconstructed by taking the exponent such that:
\begin{equation}
  f_{\text{combined}} = e^s \in [0, 1]  
\end{equation}

However, as this compresses the scores to a very small range in practice, we instead express the local mapping score as:
\begin{equation}
    w_{\text{local}} = -\frac{s}{\sqrt{\sum_i v_i+\sum_j e_j }}  \in [0, 1] \label{eq:placement}
\end{equation}

Then, for the combined placement score of all placements, the placement scores are aggregated by the mean squared error:
\begin{equation}
   S_{\text{QAP}} = 1-\text{mean}((1-w_{\text{local}})^2)
\end{equation}

This gets a balanced mapping of circuit partitions to hardware.

The Frank-Wolfe algorithm still requires a computational complexity of $O(n^3)$ and is therefore the most costly of the routines. To alleviate this bottleneck, an adaptive cost function evaluation is employed. The local refinement will search for better cuts through the Tabu-search optimization. Every $I_{interval}$ iterations, it will evaluate a set number of top candidate solutions for their QAP scores. This limitation on the number of evaluations is to reduce the costly QAP calls.

As the optimization now becomes a multi-objective optimization, the cost function for the adaptive cost is adjusted to:
\begin{equation}
    S_{\text{adapt}} = S_{\text{cuts}}/S_{\text{upper limit}} + \beta \cdot S_{\text{QAP}}
\end{equation}
Here, $S_{\text{upper limit}}$ is a worst-case cost (overhead for all gates cut) to normalize the cut solution. As the QAP solution $S_{\text{QAP}}$ is effectively a product sum of fidelities, this value will be bounded by $[0, 1]$. By normalizing the cut solution, a balanced multi-objective cost can be evaluated. The $\beta$ factor ($\geq 0$) is included as a user-defined parameter to increase or reduce the weight on the noise placement score.

%\begin{algorithmic
%\State $\text{topology graph} \gets \Call{PyBind11}{\text{Qiskit %hardware backend}}$
%\State $\text{connectivity graph, temporal graph} \gets \Call{PyBind11}{\text{Qiskit quantum circuit}}$
%\end{algorithmic}

\input{Sections/tikz_flowchart}

%% file: Sections/tikz_flowchart.tex
\usetikzlibrary{arrows.meta, positioning, shapes.geometric, calc}

\definecolor{inputblue}{RGB}{175,179,255}
\definecolor{parserorange}{RGB}{255,216,176}
\definecolor{processred}{RGB}{255,179,178}
\definecolor{decisiongreen}{RGB}{179,254,174}
\definecolor{arrowgray}{RGB}{162,177,195}
\definecolor{textgray}{RGB}{51,51,51}

\tikzset{
    >={Stealth},
    line/.style={thick, draw=arrowgray, -{Stealth}},
    input/.style={
        trapezium,
        trapezium left angle=70,
        trapezium right angle=110,
        minimum height=1.2cm,
        text width=2.4cm,
        align=center,
        draw=black,
        fill=inputblue,
        text=textgray,
        inner sep=2pt
    },
    parser/.style={
        rectangle,
        minimum height=1.1cm,
        text width=8.8cm,
        align=center,
        draw=black,
        fill=parserorange,
        text=textgray,
        inner sep=3pt
    },
    process/.style={
        rectangle,
        rounded corners,
        minimum height=1.1cm,
        text width=3.8cm,
        align=center,
        draw=black,
        fill=processred,
        text=textgray,
        inner sep=3pt
    },
    smallprocess/.style={
        rectangle,
        minimum height=1.1cm,
        text width=3.4cm,
        align=center,
        draw=black,
        fill=parserorange,
        text=textgray,
        inner sep=3pt
    },
    decision/.style={
        diamond,
        aspect=2,
        minimum height=1.5cm,
        text width=3.8cm,
        align=center,
        draw=black,
        fill=decisiongreen,
        text=textgray,
        inner sep=1pt
    },
    output/.style={
        rectangle,
        rounded corners,
        minimum height=1.1cm,
        text width=4.8cm,
        align=center,
        draw=black,
        fill=processred,
        text=textgray,
        inner sep=3pt
    }
}

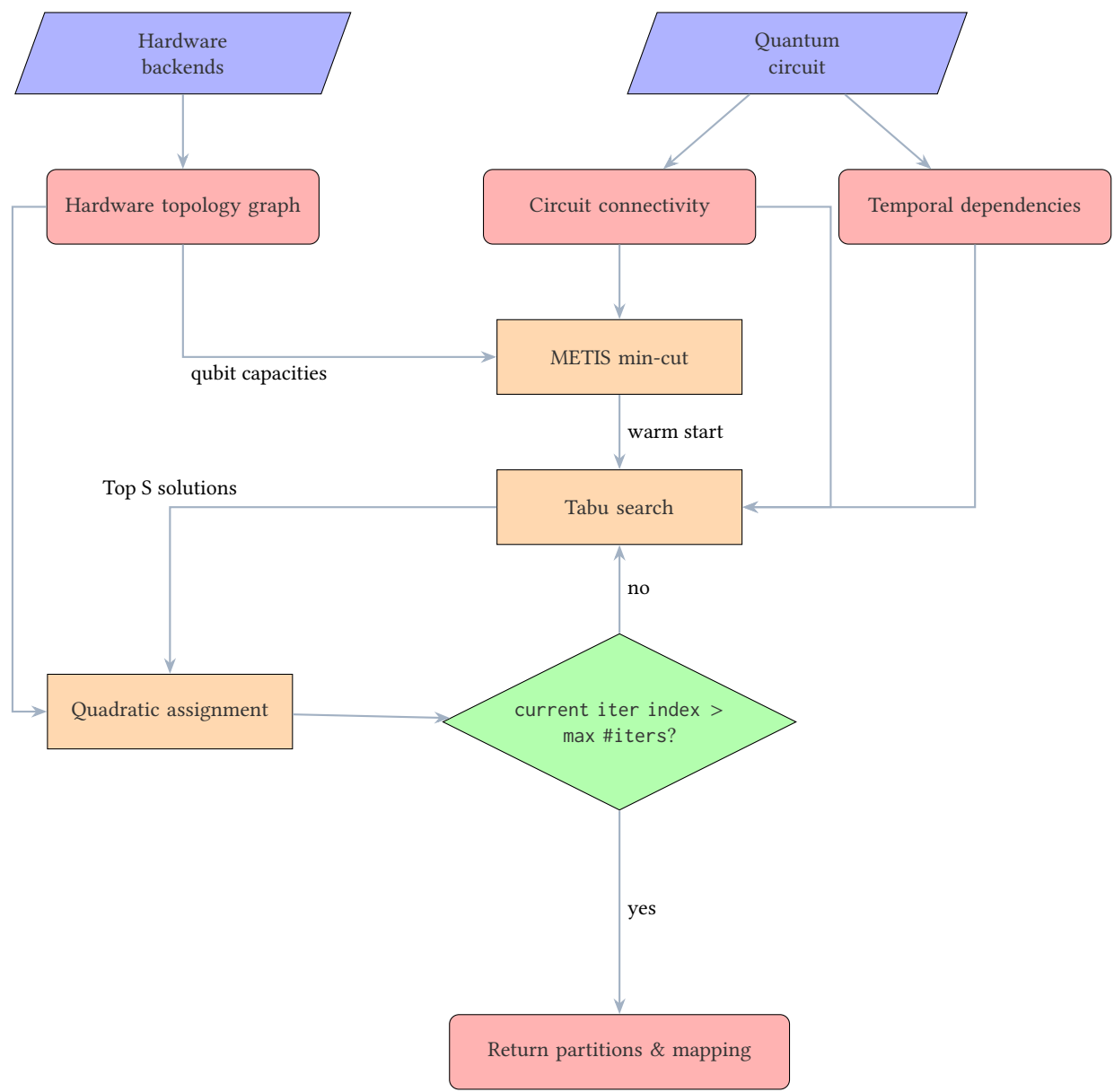
\begin{figure}[t]
\centering
\begin{tikzpicture}[node distance=1.1cm and 0.2cm]

% Top row
\node[input] (backend) {Hardware backends};
\node[input, right=4.5cm of backend] (circuit) {Quantum \\ circuit};

% Parser
%\node[parser, below=1.0cm of $(backend)!0.5!(circuit)$] (parser) {Pybind parser};

% Parsed objects
\node[process, below =1.1cm of backend] (topology) {Hardware topology graph};
\node[process, below left =1.1cm and 0cm of circuit] (connectivity) {Circuit connectivity};
\node[process, below right=1.1cm and 0cm of circuit] (temporal) {Temporal dependencies};

% Main flow
\node[smallprocess, below=1.1cm of connectivity] (metis) {METIS min-cut};current iter
\node[smallprocess, below=1.1cm of metis] (tabu) {Tabu search};
%\node[decision, below left=1.75cm and 3cm of tabu] (check)
%{Top S solutions};

\node[decision, below=1.3cm of tabu] (iter)
{$\texttt{current iter index} > \texttt{max \#iters}$?};

\node[smallprocess, below left=1.9cm and 3cm of tabu] (assign) {Quadratic assignment};

\node[output, below=3cm of iter] (result) {Return partitions \& mapping};

% Arrows
%\draw[line] (backend) -- (parser);
%\draw[line] (circuit) -- (parser);

\draw[line] (backend) -- (topology);
\draw[line] (circuit) -- (connectivity);
\draw[line] (circuit) -- (temporal);

\draw[line] (topology) |- node[below right] {qubit capacities} (metis);

\draw[line] (connectivity) -- (metis);
\draw[line] (metis) -- node[right] {warm start} (tabu);
%\draw[line] (check) --  node[right] {yes} (assign);
\draw[line] (assign) -- (iter);
\draw[line] (tabu)  -| node[above] {Top S solutions}  (assign);
\draw[line] (iter)  -- node[right] {no} (tabu);
%\draw[line] (check)  -- node[above] {no}  (iter);

\draw[line] (temporal.south) |- (tabu.east);
\draw[line] (connectivity.east) -| ++(1.1,0) |- (tabu.east);
\draw[line] (topology.west) -- ++(-0.5,0) |- (assign.west);

\draw[line] (iter.south) -- node[right] {yes} (result.north);

\end{tikzpicture}
\caption{Partitioning and mapping workflow of MosaiQC.}
\label{fig:workflow}

\end{figure}

%% file: Sections/Results_ext.tex
%New Layout
% OVERALL
% - setup/parameter settings
% - hardware layout
%   - 5q
%   - 27q
%   - 127q
%   - mixed?
% - comparison algorithms
%   - Exact
%   - Dijkstras
%   - MosaiQC init
%   - MosaiQC refinement (random)
%   - MosaiQC full

% Runtime comparisons

% Solution quality

% Mapping fidelity

% Refinement convergence

\section{Results}
\label{sec:res}
%Algorithms
To evaluate the performance of MosaiQC, sampling overhead, runtime performance, and reconstruction fidelity are benchmarked. To investigate the contributions of all elements of MosaiQC, we test the initial METIS warmstart, the refinement from random placement, and the combined MosaiQC workflow. As a baseline, we compare against the Qiskit circuit cutting add-on \cite{qiskit-addon-cutting}, which uses Dijkstra’s best-first search algorithm. The Qiskit circuit-cutting add-on was selected as the baseline, as it represents the most widely adopted open-source circuit-cutting framework.
We compare against the exhaustive search, which results in optimal cutting placements. We also test against the default parameters of 1000 backjumps and 1024 max gamma, effectively resulting in an early-stopping heuristic search. All samples for all methods are run with a time-out of 500 seconds. 
The QAP calculations at the large-scale (150-1000 qubits), the 127 qubit mapping optimizations have a large performance impact with current precision settings. While the QAP evaluations are only done at a set interval, as QAP evaluations take on a larger portion of the overall runtime, QAP evaluations were turned off at this scale to facilitate a fair runtime and overhead cost comparison.
All computations are done using a 12th Gen Intel Core i7-1265U × 12 with 6GB memory.

We evaluate the MosaiQC algorithm for several benchmark algorithms, which are common in circuit cutting literature \cite{Tang2022ScaleQC, Sahu2025DevQCC, Ren2024HardwareAwareGateCutting, Yang2024ScalabilityCircuitCutting, Cambiucci2025SpatialTemporalCutting, Kan2024FitCut}. The algorithms tested are:
\begin{itemize}
    \item Adder: the quantum Cuccaro, Draper, Kutin, and Moulton (CDKM) full adder
    \item BV: Bernstein-Vazirani
    \item AQFT: Approximate Quantum Fourier Transform
    \item HWAE: Hardware-efficient Ansatz (efficient su2)
    \item Supremacy: an instance of random circuits used in the Google supremacy experiment \cite{arute2019quantum}
    \item Quantum Approximate Optimization Algorithm (QAOA) for the maxcut problem
    \item Variational Quantum Eigensolver (VQE) for the Ising problem using the HWEA
    \item VQE for the Ising problem using the Unitary Coupled Cluster with Single and Double excitations (UCCSD) ansatz
\end{itemize}

For each benchmark, we determine 3 cutting scenarios:
\begin{itemize}
    \item small: 6-30 qubit algorithms, partitioned to 5-qubit hardware
    \item medium: 30-150 qubit algorithms, partitioned to 27-qubit hardware
    \item large: 150-1000 qubit algorithms, partitioned to 127-qubit hardware
\end{itemize}
To facilitate a fair comparison, MosaiQC is evaluated using equal-sized hardware partitions in all benchmark scenarios, matching the limitations of the Qiskit circuit-cutting add-on, even though MosaiQC also supports heterogeneous hardware capacities.

For most benchmarks, we used gate-cuts for the warm-start approach, as it generally gives better results. For the BV benchmark, however, the solutions are generally wire-cut heavy, and as such, a wire-cut initialization gives better results. As such, we introduce wire-cut initialization for use cases where substantially more wire cuts than gate cuts are expected.

For the Tabu search, we use the number of algorithm qubits $N$ as tabu tenure and set the maximum number of tabu iterations at $N^2$. For the hardware, we use the noise profiles of the IBM Lima (5q), Algiers (27q), and Brisbane (127q) quantum backends. These partition sizes should be representative of real-life use cases.

\subsection{Runtime comparisons}
The runtime of the partitioning algorithm is crucial, as it determines the size for which it can still find a good partition. The trade-off generally lies between speed and solution quality, where exact methods have the worst scaling but provide exact solutions. In contrast, heuristics provide better scaling at the cost of finding imperfect solutions. As we have implemented a heuristic solver, our aim is to find an overall better trade-off regarding the runtime to solution quality balance.

For the largest tasks, not all partitions will be possible to calculate, either due to the cost value of the sampling overhead overflowing the floating-point representation, due to one of the solvers spending too much time, or due to memory overflow. The latter occurs frequently for the Qiskit exhaustive search and severely limits the maximum circuit sizes it can partition.

%We present the aforementioned benchmarks for all three scales in detail in Appendix \ref{app:runtime}. The results have been grouped by average $\text{runtime}/\text{qubit}$, scaled as a percentage of the longest $\text{runtime}/\text{qubit}$ in Figure \ref{fig:runtime_average} for a more comprehensive comparison. As the summarized figure hides most of the details, the representation has its limits. The detailed figures in Appendix \ref{app:runtime}, show that the MosaiQC results continue beyond the results of the baseline, due to the baseline timing out. From these detailed results, we compare the geometric mean improvement for all sizes completed by both MosaiQC and the baseline and find a runtime improvement of $2.88 \times$ over all benchmarks. This in itself allows us to partition larger circuits compared to the baselines.

\begin{figure}[h]
    \centering
    \includegraphics[width=\linewidth]{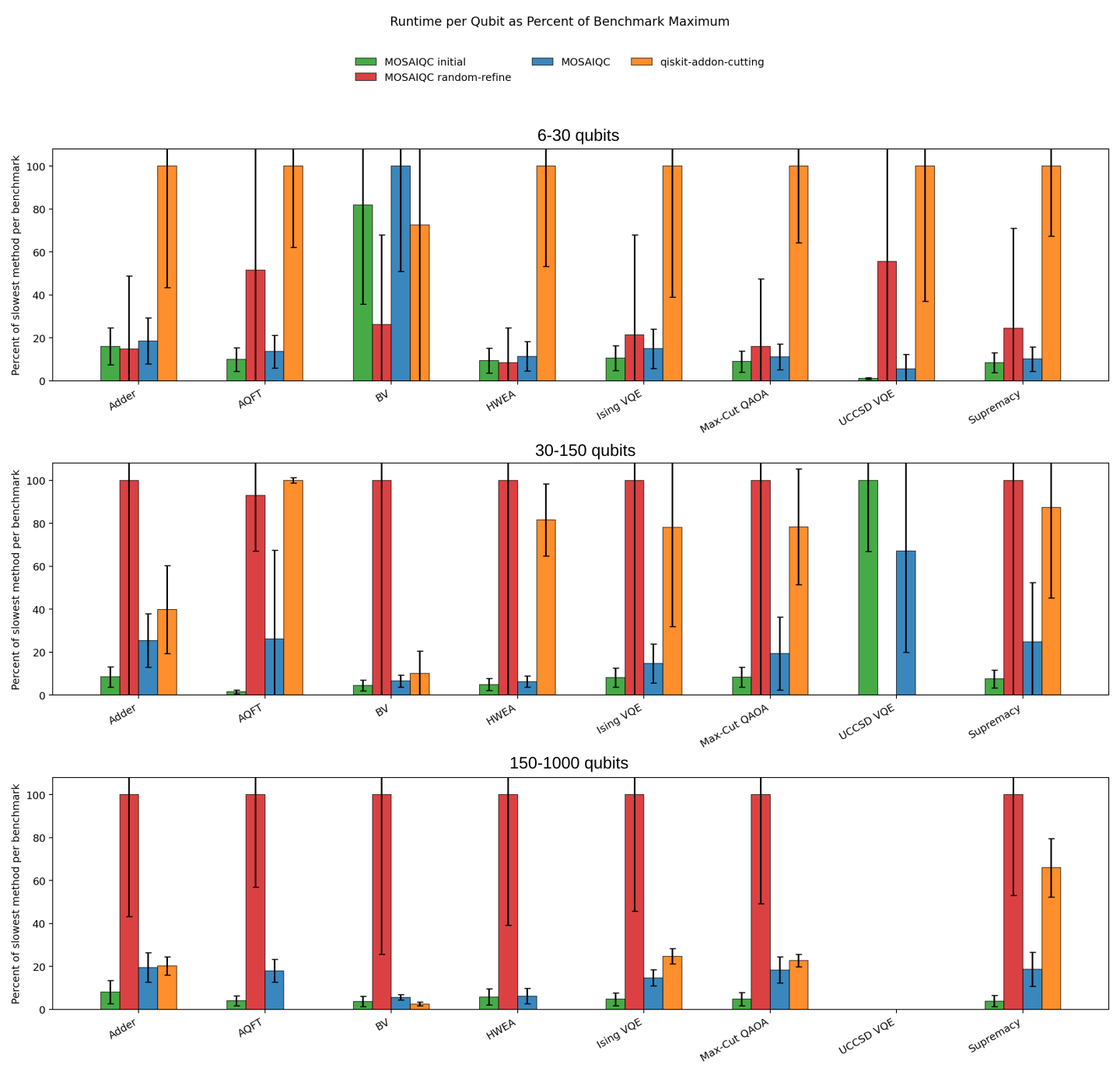}
    \caption{Runtime comparison of MosaiQC and its subcomponents against the baseline Dijkstra algorithm (early stopping and exact), scaled as $\text{runtime}/\text{num\_qubits}$ and shown as a percentage of the highest runtime per benchmark.}
    \label{fig:runtime_average}
\end{figure}

Figure \ref{fig:runtime_average} summarizes the runtime per qubit across all benchmark families, normalized to the longest runtime per qubit for comparison. While this representation highlights the overall runtime trends, it necessarily omits the full extent of the scalability differences between the methods. The detailed results presented in Appendix \ref{app:runtime} show that MosaiQC continues to successfully optimize circuit partitions well beyond the largest instances completed by the Qiskit circuit-cutting add-on, which frequently reaches the imposed timeout. Across all benchmark instances completed by both methods, MosaiQC achieves a geometric mean runtime improvement of 2.88×. More importantly, this computational advantage enables MosaiQC to efficiently optimize substantially larger circuits, extending the practical applicability of circuit cutting to problem sizes that are infeasible for the baseline.

Existing optimization frameworks become a compilation bottleneck for circuits approaching the thousand-qubit regime, often requiring hours of optimization or failing to terminate. In contrast, MosaiQC maintains optimization times on the order of seconds to minutes across all evaluated benchmarks.

Overall, we observe that the refinement stage does not induce a large overhead over the initialization stage. Especially for the benchmarks under 150 qubits, this is to be expected, as most of the optimization is already solved by the initializer, and QAP evaluations are limited. Without initialization, the refinement-only results often display the highest runtime, highlighting the importance of the initialization stage. We expand on these findings in section \ref{sec:convergence}.

The results also highlight some of the challenges of the individual benchmarks. For example, the UCCSD VQE circuits are much deeper compared to other benchmarks, resulting in high cut costs. As a result, the Dijkstra approach either times out or encounters a memory overflow. For the 30–150 qubit regime, these results demonstrate that MosaiQC continues to find high-quality cut placements even after the baseline becomes computationally infeasible, demonstrating the practical scalability of the proposed optimization framework.

\subsection{Sampling overhead}
As determined in the previous section, the overall runtime is significantly faster than the Dijkstra search. This result is only favorable if the solution quality is mostly retained, such that a good trade-off is preserved.

\begin{figure}[h]
    \centering
    \includegraphics[width=\linewidth]{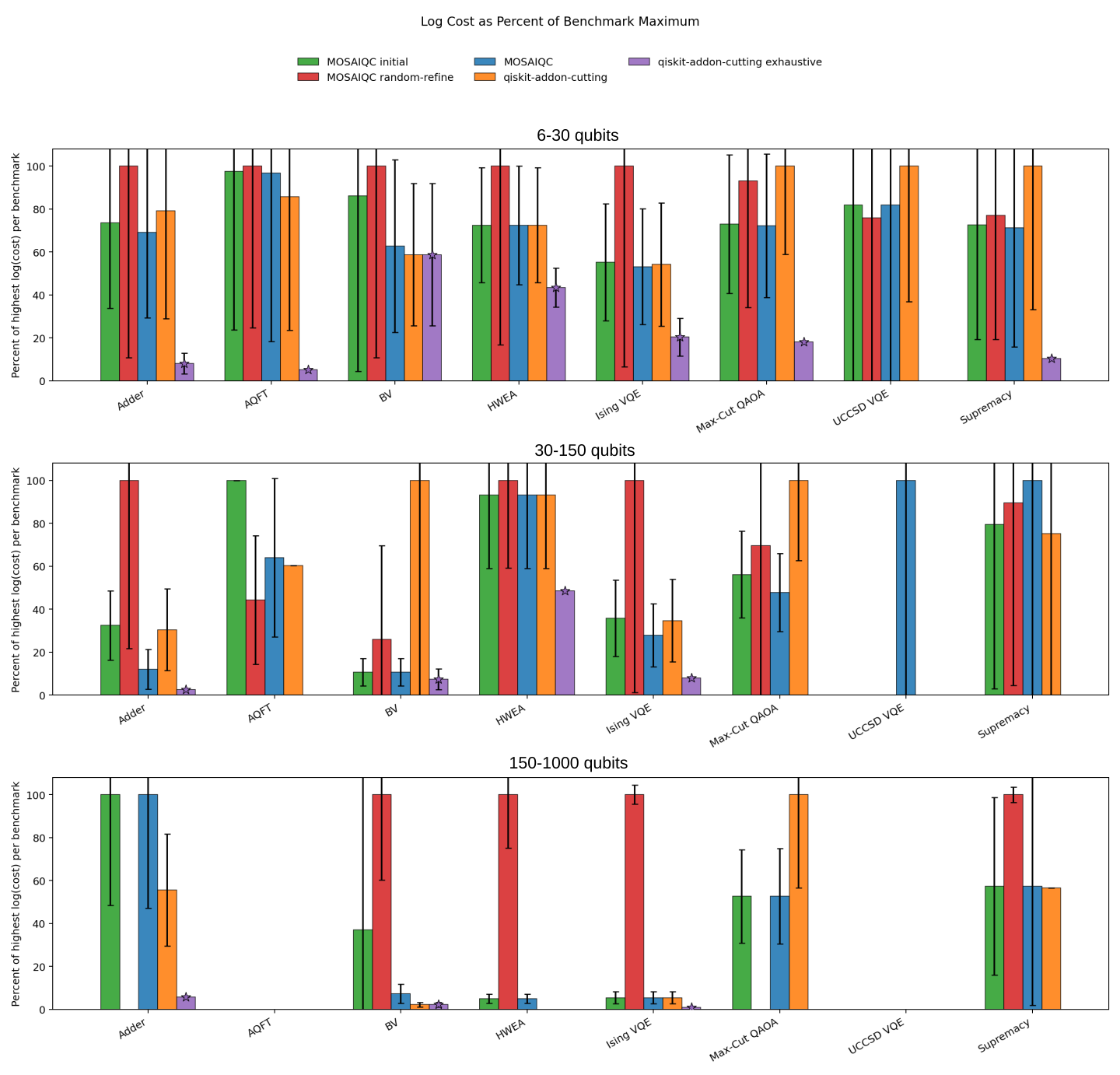}
    \caption{Sampling overhead cost comparison of MosaiQC and its subcomponents against the baseline Dijkstra algorithm (early stopping and exact), scaled as $\text{sampling\_overhead}/\text{num\_qubits}$ and shown as a percentage of the highest runtime per benchmark.}
    \label{fig:cost_average}
\end{figure}

In Figure \ref{fig:cost_average}, the cost comparison is shown for the benchmark algorithms. In general, we see MosaiQC outperform the baseline algorithm, finding lower sampling overheads for all partition sizes. For the HWEA benchmark, the baseline results are not presented, as the baseline solver timed out. For the AQFT and supremacy benchmark, the solutions for 27 and 127-qubit partitions overflowed the floating-point number representation.

Figure \ref{fig:cost_average} shows the percentage difference against the most costly method to give a compact comparison. Notably, this results in large error bars due to the scaling of the linear progression ($\text{csost}/\text{qubits}$), whereas the actual progressions are either polynomial for the Qiskit add-on case or almost constant for the MosaiQC case. For the comparison of the actual per-instance sampling overhead, we refer to the data presented in Appendix \ref{app:overhead}.
For all benchmarks, we find an average per-instance cut reduction of $16.84\% \pm 5.09\%$, resulting in a geometric mean improvement of the sampling overhead of $5.38 \cdot 10^{11} \times$. The detailed results presented in Appendix \ref{app:overhead}, which give a clearer picture, as Qiskit results are not always available for larger circuits due to either run timeouts or memory usage exceeding limits. With these results, we show that for most benchmarks, the solution quality matches or improves on the Dijkstra search with default parameters. 

As Figure \ref{fig:cost_average} gives a summarized comparison, the solution quality difference between MosaiQC and the exhaustive search is not clearly visible. The Dijkstra exhaustive search reaches the time limit quickly, and only a few small instances can be compared. For these smaller instances, MosaiQC can often match the solution quality as the search space is not too large. However, this gap is expected to quickly grow for larger instances.

This, however, highlights the importance of heuristic approaches. While exact methods can find substantially better solutions for cutting placement, if the solution cannot be found in a reasonable time, there is limited use for such methods.

%For comparison, the averages $log(cost)/qubit$ for all benchmarks and sizes are plotted in Figure \ref{fig:cost_average}. The large error bars are due to the expectation of linear progression ($val/qubits$), whereas the actual progressions are either polynomial for the Qiskit add-on case or almost constant for the MosaiQC case.

\subsection{Placement score subroutine}
It is essential for achieving higher fidelity to have a subroutine that finds an optimal placement of the hardware noise maps on the quantum circuit graph. While alternative methods such as GED can serve as a good proxy to minimize swap overhead \cite{Ren2024HardwareAwareGateCutting}, the available noise topologies can be used to make an even more informed proxy. In MosaiQC, quadratic assignment is used to match these noise topologies to the circuit graph weighted by the number of 1- and 2-qubit gates.

\begin{figure}[h]
\centering

\begin{subfigure}{0.49\textwidth}
\includegraphics[width=\linewidth]{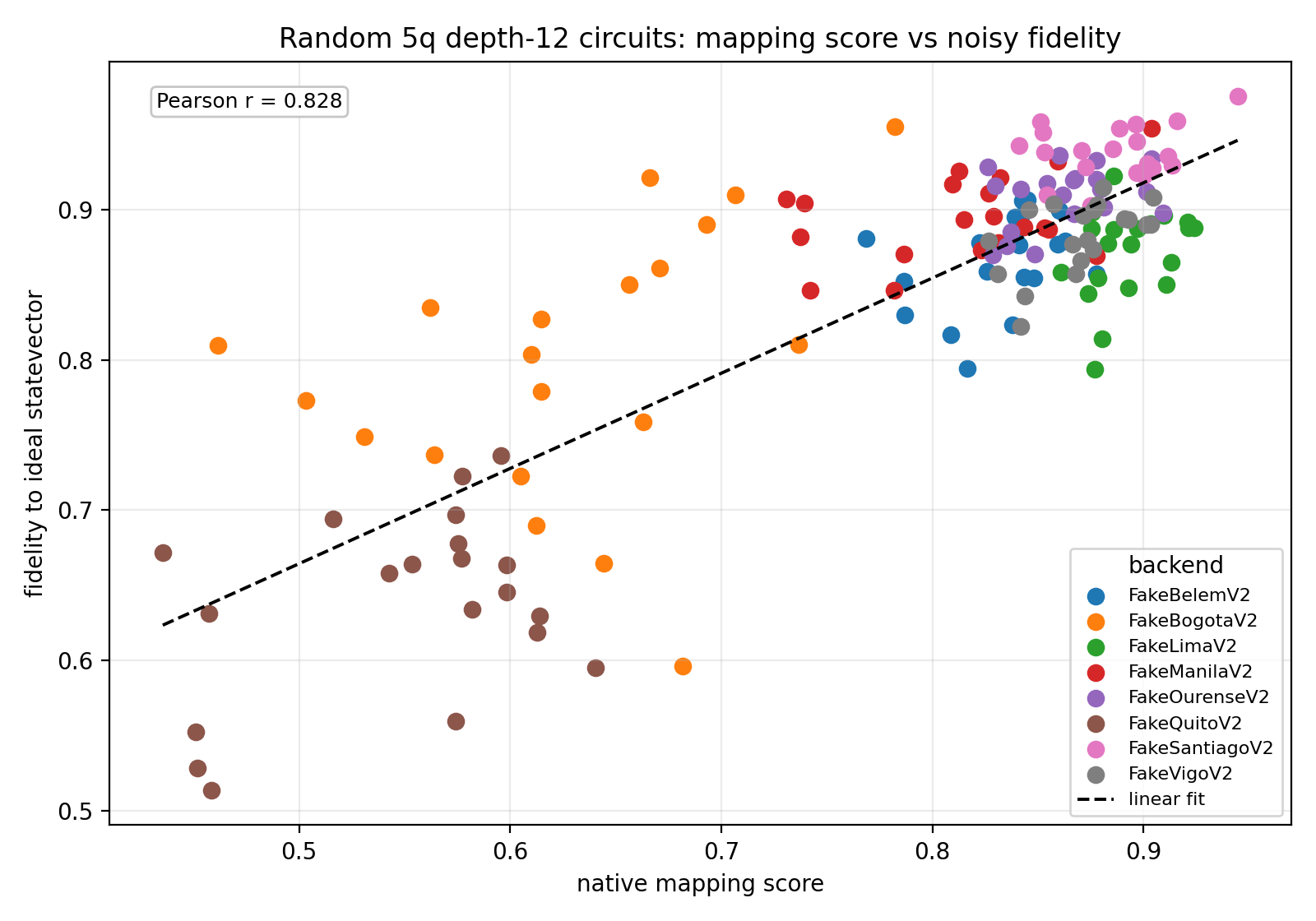}
\caption{Depth 12}
\label{fig:QA_a}
\end{subfigure}
\begin{subfigure}{0.49\textwidth}
\includegraphics[width=\linewidth]{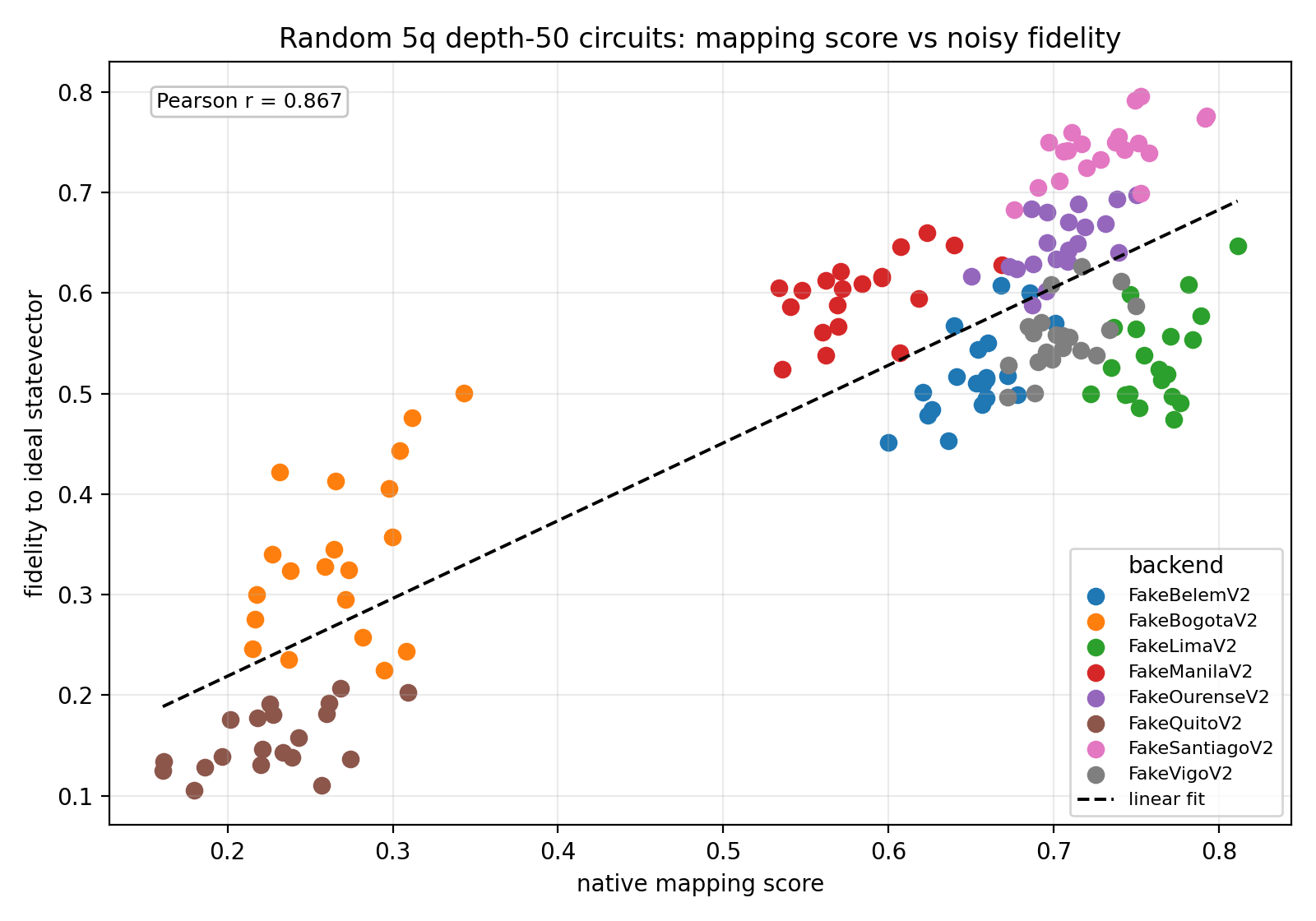}
\caption{Depth 50}
\label{fig:QA_b}
\end{subfigure}

\caption{Placement score correlation with sampled fidelity for different quantum circuit depths.}
\label{fig:QA}

\end{figure}

This proxy is validated by generating random quantum circuits of identical width and depth and comparing the quadratic assignment score (QAS) to the sampled fidelity on the noisy system. The random circuits are sampled on a variety of quantum noise profiles of existing IBM hardware.
The samples are compared to the perfect vector state to obtain the fidelity. In Figure \ref{fig:QA}, the calculated QAS is plotted against the measured fidelity in a scatter plot for circuit depths 12 and 50. From these measurements, we observe a strong Pearson correlation of $r=0.828$ for depth 12 and $r=0.867$ for depth 50. By varying the depth, we observe a slightly stronger correlation as the circuit depth increases. This corresponds with the expected result that as the number of gates increases, the effect of noise will be more pronounced, which in turn is well captured by our model.

\subsection{Mapping Fidelity}
To validate the impact of the QAP-fidelity correlation on a full mapping, the average fidelity was measured for a full circuit spanning multiple partitions. To do so, random circuits were generated for 20 qubits and depth 20, to be compared in total QAP score and average measured fidelity.

\begin{figure}
\centering
\begin{minipage}{.5\textwidth}
  \centering
    \includegraphics[width=\linewidth]{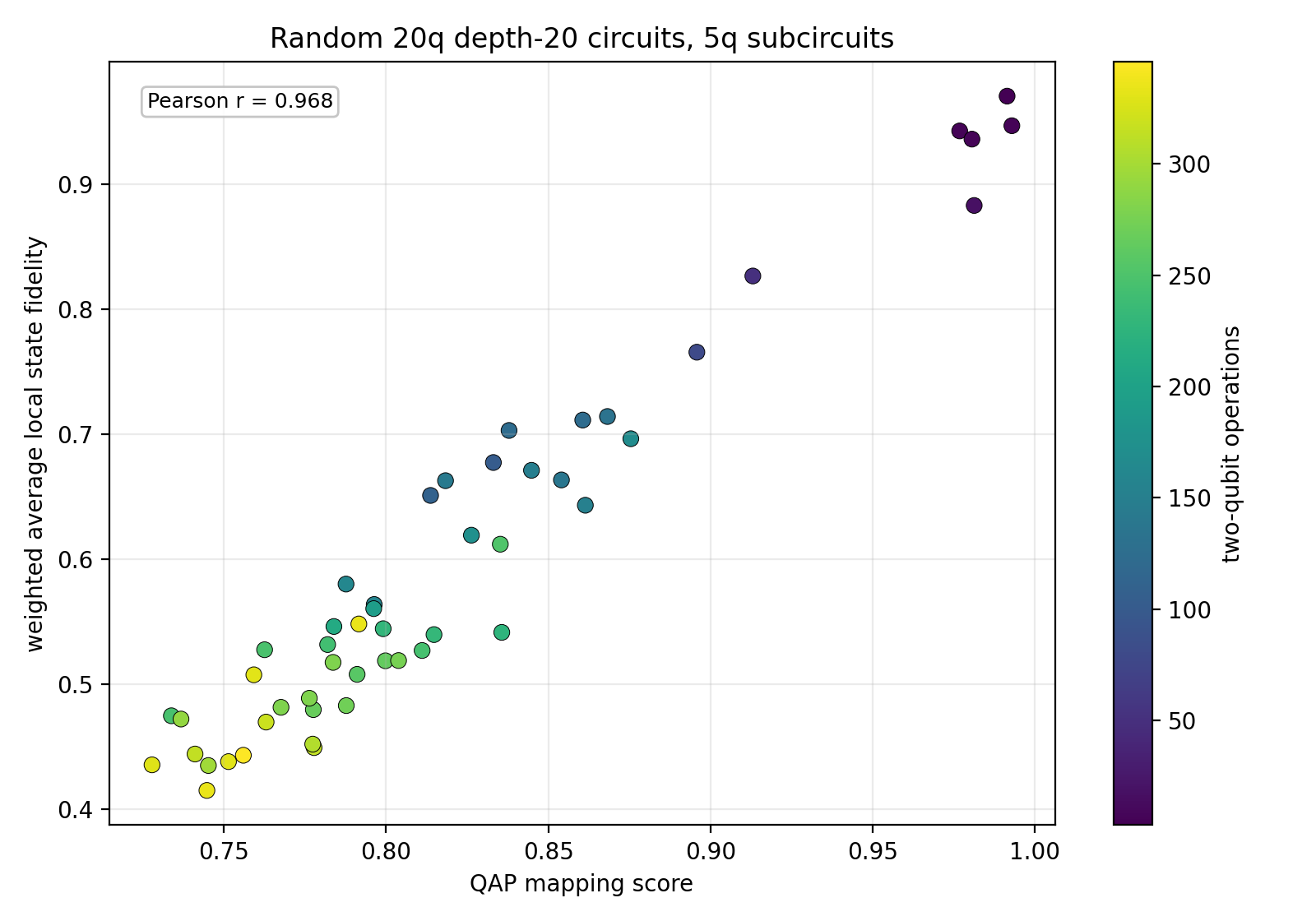}
    \caption{Scatter plot of QAP score against measured fidelity for circuits each having 20 qubits and a depth of 20.}
    \label{fig:QAP_scatter}
\end{minipage}%
\begin{minipage}{.5\textwidth}
  \centering
    \includegraphics[trim={15cm 0 0 0.7cm}, clip, width=\linewidth]{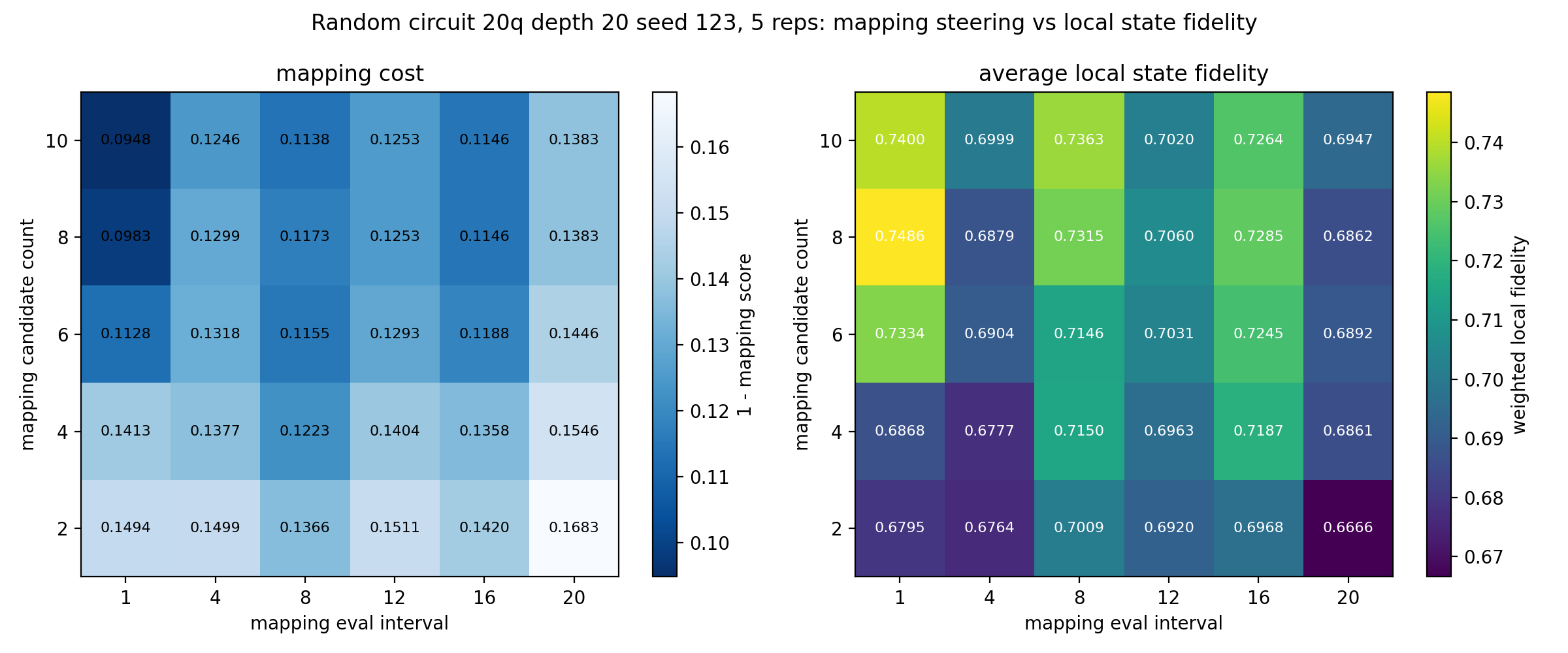}
    \caption{Heatmap of average local fidelity for the number of candidates and iteration interval for $\beta=50$. Fidelity placement score and number of QAP evaluations are shown for the 20-qubit random circuits partitioned to various 5-qubit hardware topologies. Each point is averaged over 5 samples.}
    \label{fig:heatmap}
\end{minipage}
\end{figure}

In Figure \ref{fig:QAP_scatter}, the correlation between the total QAP score (Equation \ref{eq:placement}) and the average fidelity can be observed. Overall, a positive Pearson correlation of $r=0.986$ can be found, saturating at placement scores of $1$. 

As the compute time of the fidelity evaluations can be greatly reduced by limiting the QAP calls, we investigate the effects of the evaluation interval and the number of solution candidates. 

For Figure \ref{fig:heatmap} the adder algorithm is tested for 20 qubits on 5-qubit hardware partitions. The results show that by increasing the number of candidates and decreasing the interval between evaluations, the placement score improves accordingly. This allows trade-offs to significantly reduce the memory and runtime cost of optimizing for fidelity.
%For reconstruction fidelity, we test the cutting algorithms on small-scale instances of the benchmarks. These instances are cut to 5q partitions, to be sampled on a noise model of IBM's Lima 5-qubit quantum computer. The measured samples are then compared for fidelity to a noiseless (non-partitioned) simulation.

%We note that the Qiskit add-on has no integrated subroutine for finding the best mapping in terms of fidelity. This is, however, common to most solvers who do not directly target fidelity as the cost function. After cutting, partitions assigned to hardware are transpiled to this hardware for the best placement, but are not optimized during the cut-finding algorithm itself (with exceptions to some \cite{Ren2024HardwareAwareGateCutting}). 

%\begin{figure}[h]
%    \centering
%    \includegraphics[trim = {0 10.1cm 0 0}, clip, width=0.8\linewidth]{Fidelity_results/%local_fidelity_bar_charts.png}
%    \caption{State fidelity compared to exact simulation for 12-qubit benchmarks. For each benchmark, the algorithm for 12 qubits was cut up to 5-qubit partitions to be sampled on 5-qubit noisy backend simulators. The AQFT algorithm uses an approximation degree of 6.}
%    \label{fig:fidelity}
%\end{figure}

\begin{figure}
    \centering
    \includegraphics[width=\linewidth]{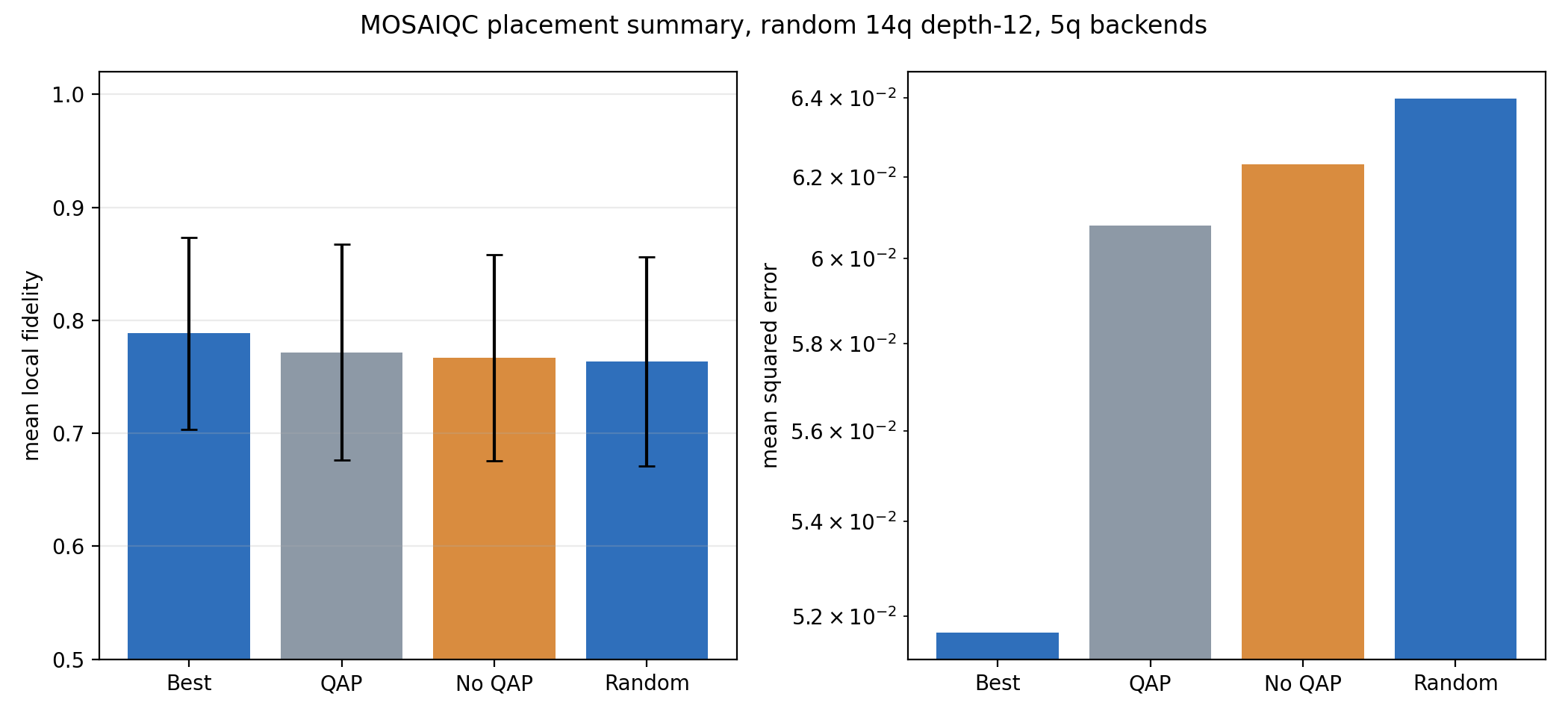}
    \caption{Comparison of local state fidelity of QAP subroutine enabled ($\beta=50$) against disabled, with brute force best placement as baseline.}
    \label{fig:fidelity}
\end{figure}

To compare the accuracy of both MosaiQC and the baseline, the output states of the partitioned circuits are sampled to calculate the fidelity. By comparing the sampled states directly, as opposed to comparing the reconstruction fidelity, larger circuits can be sampled. Reconstruction evaluation is strongly limited by the sampling overhead, making small circuits quickly intractable. By allowing larger circuits, the influence of noise becomes more pronounced, allowing for a more effective comparison. 
%The comparison in fidelity is shown in Figure \ref{fig:fidelity}. A significant improvement in fidelity can be seen for the adder and HWEA benchmarks for the MosaiQC algorithm. On average over all benchmarks, the mean fidelity improvement is 0.014428. The small difference in fidelity is expected as the tested circuits have a very low circuit depth. As such, mapping is not a significant factor in increasing fidelity. For deeper circuits, this improvement is expected to be more evident.

The comparison in fidelity is shown in Figure \ref{fig:fidelity}. Here, random quantum circuits are generated, and mappings are compared for 5-qubit mixed noisy hardware on a 14-qubit circuit. Samples are taken from 20 different random circuits, all of depth 12. An improvement in fidelity can be seen compared to not using the QAP subroutine or simple random assignment. The improvement in fidelity is small, but nevertheless meaningful.
\begin{figure}
\begin{subfigure}[t][][t]{0.49\textwidth}
    \centering
    \includegraphics[width=\linewidth]{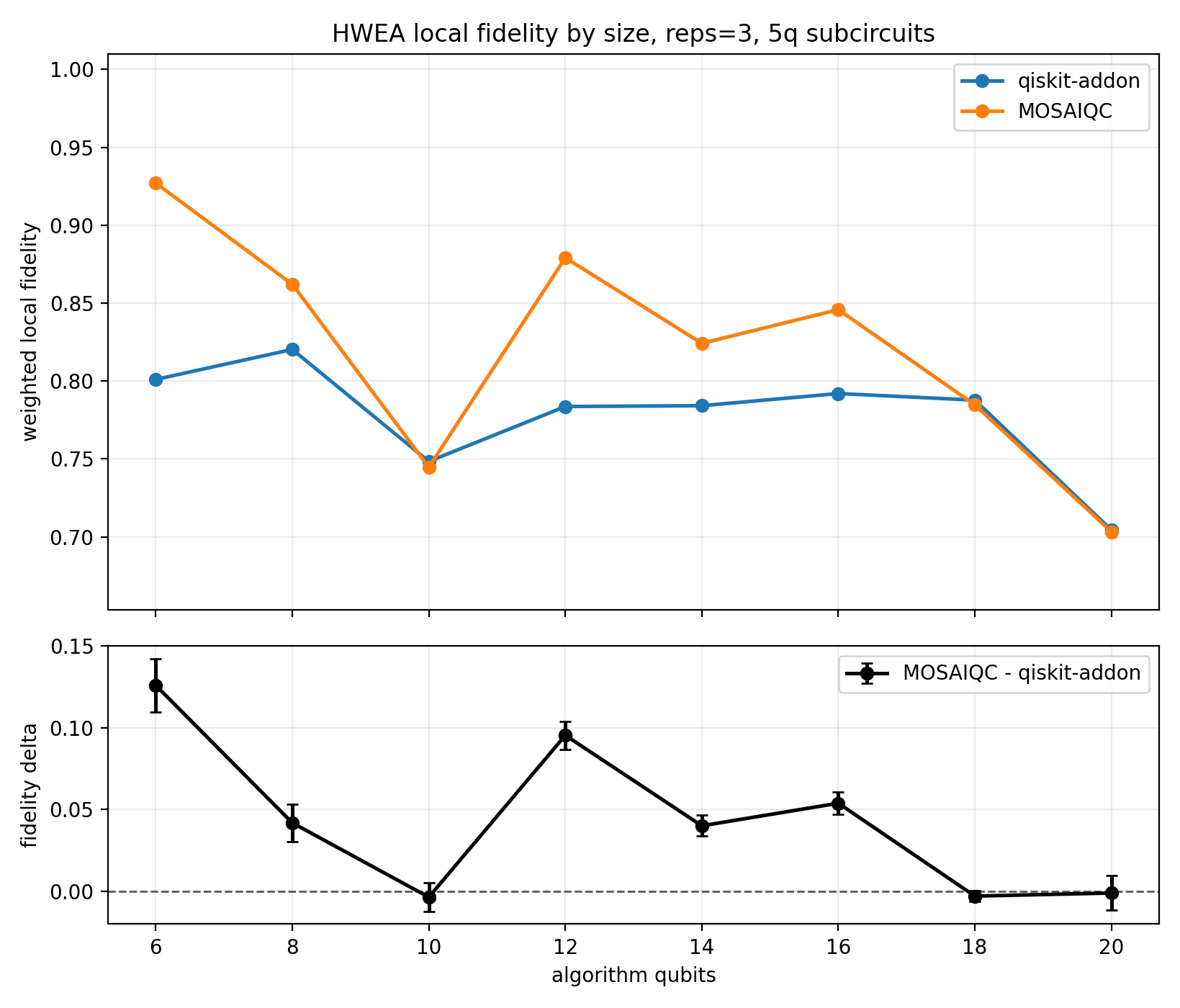}
    \caption{Mapping weight 1}
\end{subfigure}
\begin{subfigure}[t][][t]{0.49\textwidth}
    \centering
    \includegraphics[width=\linewidth]{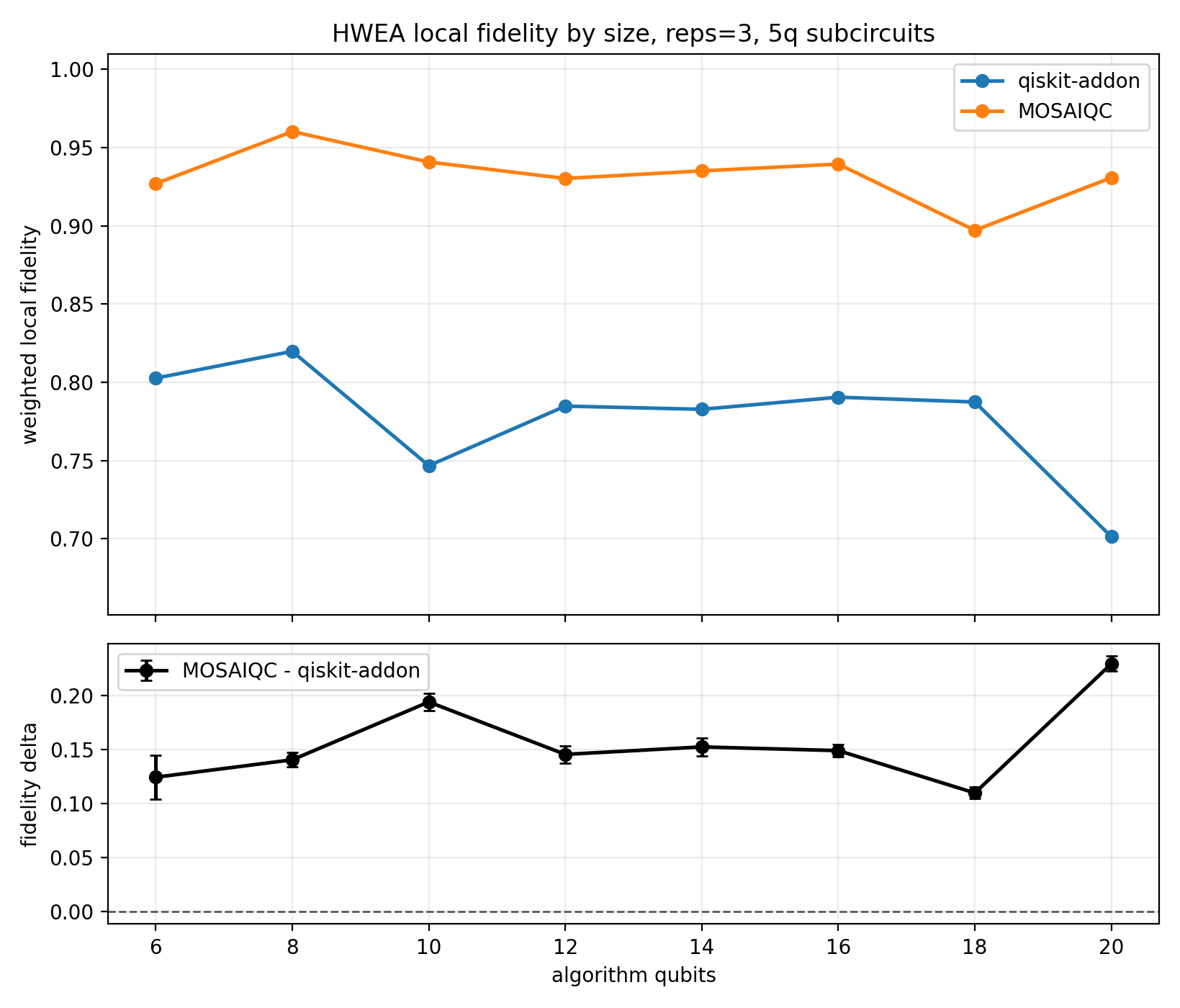}
    \caption{Mapping weight 50}
\end{subfigure}

    \caption{Fidelity comparison between MosaiQC and Qiskit addon for HWEA circuits (depth 3) ranging from 6 to 20 qubits for 5-qubit partitions. For each algorithm size, 10 samples were taken.}
    \label{fig:fideity_hwea}
    
\end{figure}

%Comparing fidelity of placements found by MosaiQC compared to the baseline is challenging for most benchmarks. As some benchmarks generate very different placements for each algorithm, the actual optimization for fidelity is hard to compare. In MosaiQC, the initial warm-start is based on minimizing the sampling oveerhead, meaning in any further refinement it is generally confounded by the local search space around this solution. If there is no nearby local solutions that match the already found cost, the optimizer is unlikely to find a better placement score placement unless this value is set extremely high. However, for some benchmarks such as the HWEA, the search space facilitates more movement as near solutions have costs in the same order. As such
We benchmark against the baseline for fidelities for the HWEA benchmark, as this benchmark permits sufficient placement diversity. The results are presented in Figure \ref{fig:fideity_hwea}. Here, we compare for a range of circuit sizes, all partitioned to 5-qubit hardware backends of varying noise topologies. We find that as the mean value, MosaiQC scores $0.0436 \pm 0.0460
$ ($+5.61 \% \pm 6.02\%$) higher in fidelity for mapping weight 1, and $ 0.1556 \pm 0.0375$ ($+19.56 \% \pm 6.17\%$) for mapping weight 50. Indicating that while for low weight the difference is statistically quite low, for a large weight it makes a significant improvement.

%However, we must note that the cutting solutions for these benchmarks also varied greatly between solvers. As the baseline required much more sampling overhead, some error may have also been introduced. Overall, the MosaiQC performs better for all benchmarks compared to the baseline, indicating that the quadratic assignment does reduce the hardware noise.

%For most of the benchmarks, the difference in fidelity is very minimal, which would be expected to be larger for larger problem sizes. For larger and longer algorithms, the effect of noise would be more pronounced, giving a clearer indication of the effectiveness. Unfortunately, the sampling overhead would grow to such high values that the computing power is not available to measure these results in a reasonable time.

\subsection{Optimizer convergence}
\label{sec:convergence}
\begin{figure}
    \centering
    \includegraphics[width=0.8\linewidth]{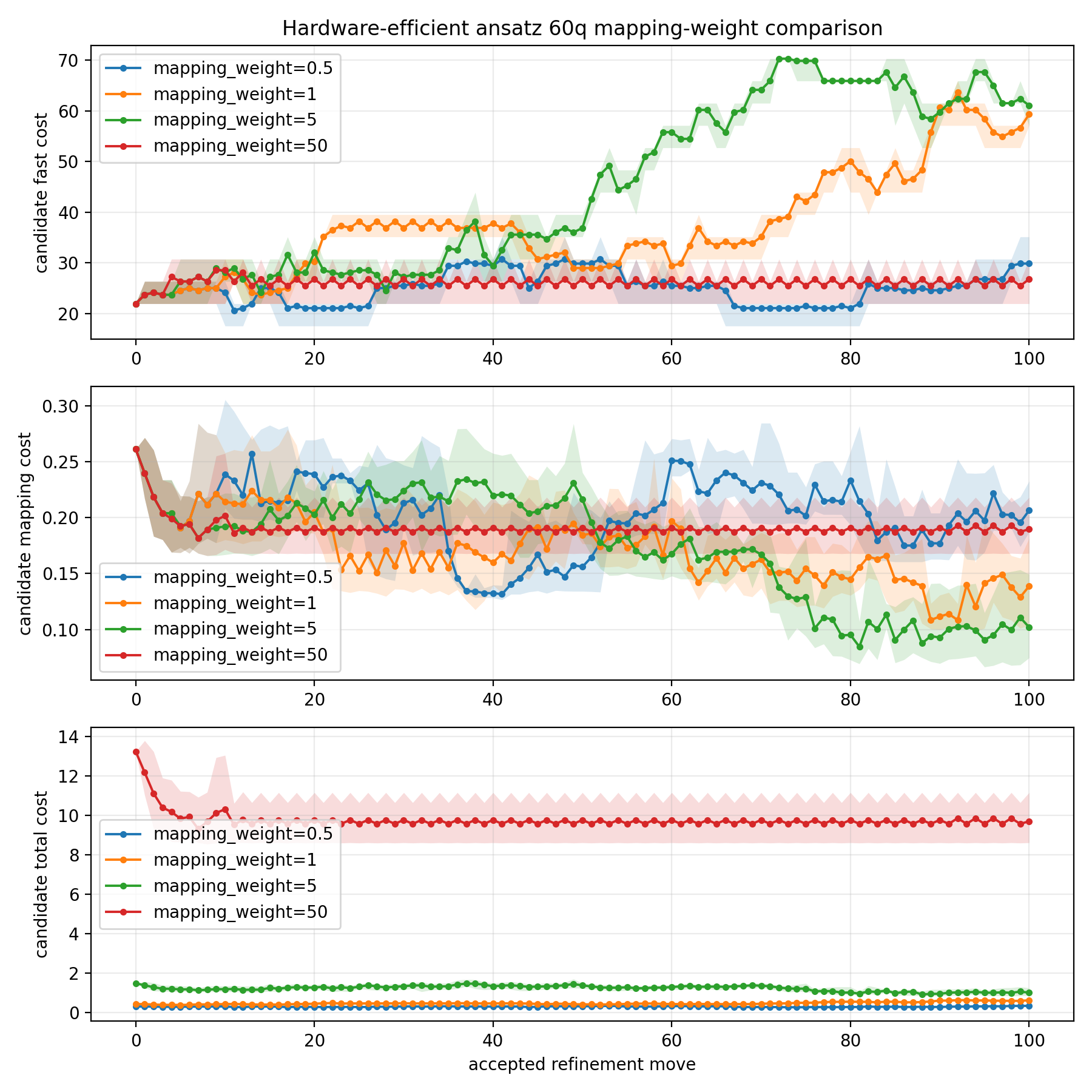}
    \caption{Convergence history for different placement scores on a 60-qubit HWEA, using 27q partitions. Results are shown for mapping weights of 0.5, 1, 2, and 5. The top image shows the log sampling overhead from cut placement. The middle image shows the progression of the mapping placement cost, which approximates fidelity (1-expected noise). The combined (weighted) cost is shown in the bottom figure. The transparent band indicates the tested QAP evaluations. The plotted line represents the chosen refinement move during optimization.}
    \label{fig:convergence}
\end{figure}

Monitoring the convergence of the optimizer gives us further insights into not only the refinement but also the impact of the warm start. For this, we compare the convergence for several benchmarks with and without a warm-start. We specifically test on the HWEA benchmark, as this is structured in such a way that there is more freedom to change cut placement. Figure \ref{fig:convergence} illustrates the search process during local refinement. At each iteration, multiple candidate moves are evaluated (translucent bands), while only accepted improvements are reflected by the solid lines. Rather than returning the final accepted state, the optimizer records and returns the lowest-cost solution encountered throughout the search.

In general, the minimal cutting cost (top) will be found within a few iterations, though it will continue to evaluate placements based on the set maximum number of iterations. This means that for the overhead minimization, there is little computational overhead required. However, for improving fidelity, the refinement stage will demand more iterations to find the local optima. As can be observed in Figure \ref{fig:convergence}, while the overall cost quickly remains stable, the individual cut and QAP costs still fluctuate. Here, the effect of the mapping weight in the cost function highlights a trade-off. If the mapping weight $\beta$ is high enough, the optimizer will accept a higher cutting cost if it results in a better mapping fidelity. Interestingly, a very high $\beta=50$  steers quickly to a local minimum, whereas a more balanced weight, such as $\beta=5$, will allow much more mobility.
%The mapping placement cost, which approximates fidelity (1-expected noise).
\begin{figure}
    \centering
    \includegraphics[width=0.8\linewidth]{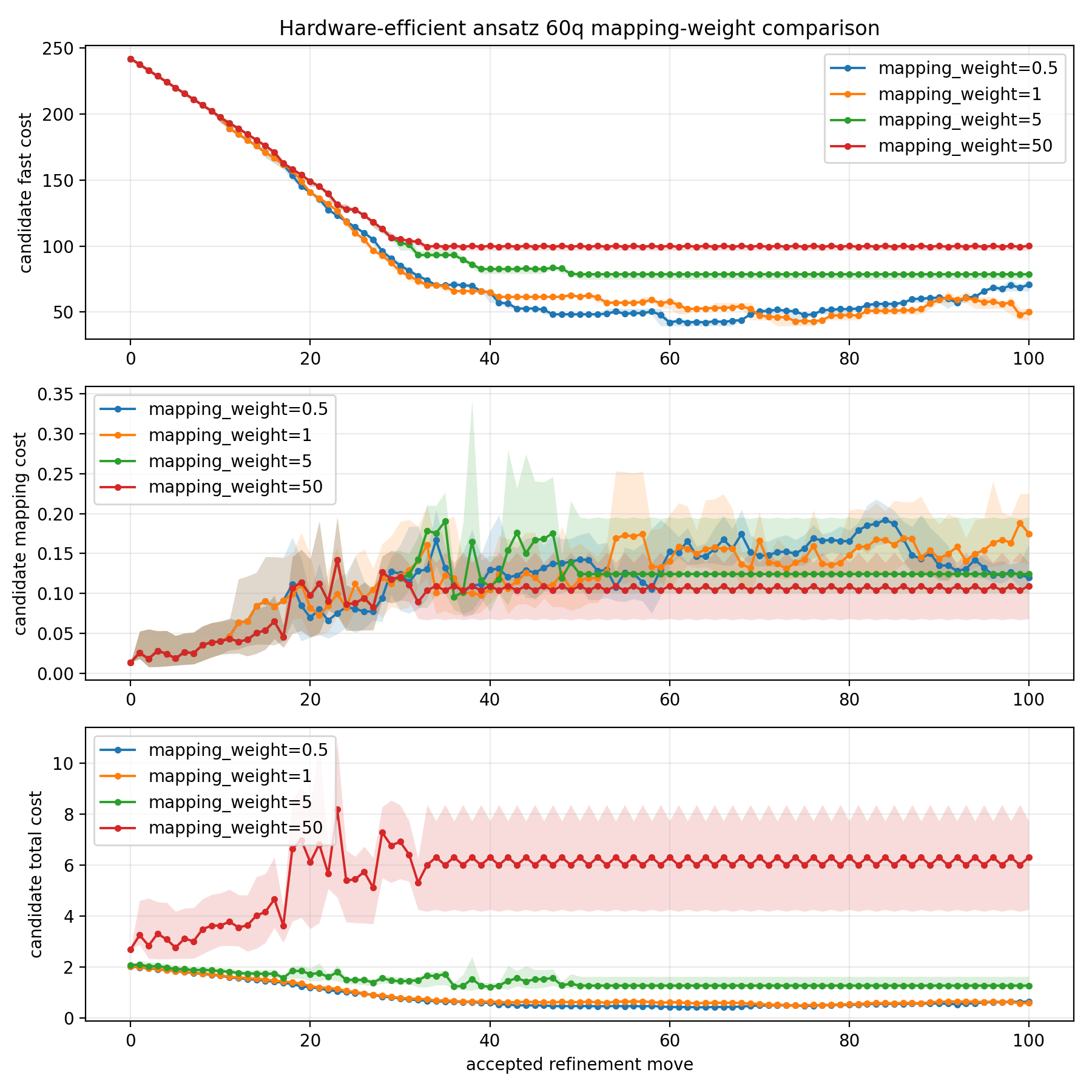}
    \caption{Cutting cost convergence without warm-start. Applied to 60-qubit HWEA for 27-qubit partitions. The transparent band indicates the tested QAP evaluations. The plotted line represents the chosen refinement move during optimization.}
    \label{fig:no-warm}
\end{figure}

The importance of the warm-start is visualized in Figure \ref{fig:no-warm}. While with warm-start the overall cost quickly converges in \~ 20 iterations, without it takes significantly more iterations (40-60). Furthermore, without warm-start the refinement stage stays local and generally finds higher cutting costs (50-150 compared to 20-30 ). This can be explained by the locality of the refinement optimizer. As moves keep the solution very local to the starting placement, it is hard to find good placements in a reasonable number of iterations. More so, the optimization is prone to being trapped in local minima. For this reason, it is crucial that the MosaiQC pipeline is executed in full to find usable cutting solutions.

%% file: Sections/Discussion.tex
%- Understanding the Scalability of Circuit Cutting Techniques for Practical %Quantum Applications \cite{Yang2024ScalabilityCircuitCutting}
%- Fast Quantum Circuit Cutting with Randomized Measurements %\cite{Lowe2023FastCuttingRandomizedMeasurements}

%The work by Nguyen \cite{Nguyen2025QCOFCC} considers both simulated annealing and a genetic algorithm to reorder the circuit based on ZX calculus, generating an overall simpler quantum circuit structure. This can enable better cut placements.

\section{Discussion}
\label{sec:discussion}
% Results
The results demonstrate that MosaiQC provides a favorable trade-off between optimization runtime and cut-placement quality. Across the evaluated benchmarks, MosaiQC consistently reduces optimization time while also lowering sampling overhead compared to the baseline Qiskit circuit-cutting add-on. This indicates that heuristic optimization does not merely improve scalability at the cost of solution quality, but can also find better cut placements when guided by an appropriate graph representation and local refinement strategy. As circuit sizes continue to grow, optimization itself should not become the next computational bottleneck. The runtime results therefore suggest that MosaiQC remains computationally practical for the evaluated circuits approaching 1000 qubits, where exact or exhaustive optimization approaches become increasingly unsuitable as compilation subroutines.

The hardware-aware placement experiments demonstrate two important findings. First, the strong correlation between the proposed QAP score and measured execution fidelity validates the QAP formulation as an effective surrogate objective for hardware quality during optimization. This enables MosaiQC to steer partition placement towards hardware assignments expected to improve execution fidelity without requiring costly quantum circuit simulations during optimization. Second, the comparison against existing placement strategies should be interpreted in the context of the overall optimization objective. While the QAP formulation effectively identifies high-quality hardware assignments, its individual contribution to end-to-end fidelity cannot be completely isolated, since partition selection simultaneously optimizes sampling overhead and hardware quality.

The current implementation of the QAP subroutine has not been included in the large-scale benchmark comparison due to its computational overhead. However, the Frank--Wolfe solver supports configurable approximation tolerances, providing a direct trade-off between runtime and assignment accuracy. Determining how this approximation affects the predictive power of the fidelity proxy remains an interesting direction for future work. More generally, an open question is how improvements in local hardware fidelity translate to the fidelity of the reconstructed quantum state, where sampling overhead and hardware noise jointly determine the final execution quality.

Several opportunities remain to further extend the framework. The current hardware-aware objective relies on static calibration data and does not account for dynamic hardware characteristics such as calibration drift. Incorporating time-dependent calibration information and additional hardware effects, such as crosstalk, could further improve the accuracy of the proposed fidelity proxy.

Although MosaiQC substantially improves the scalability of cut placement optimization, practical circuit cutting at very large scales remains fundamentally constrained by the exponential sampling overhead required for circuit reconstruction \cite{Yang2024ScalabilityCircuitCutting}. Consequently, optimization frameworks such as MosaiQC should be viewed as complementary to ongoing developments in sampling reduction techniques, including randomized measurements \cite{lowe2023fast}, dynamic shot allocation \cite{chen2025enhancedquantumcircuitcutting}, and the incorporation of classical side information \cite{Piveteau_2025}.

Taken together, these results demonstrate that multi-objective graph optimization provides a practical and scalable paradigm for quantum circuit cutting, opening opportunities for increasingly hardware-aware compiler optimizations.

%% file: Sections/Conclusion.tex
\section{Conclusion}
\label{sec:conclusion}
%In this paper, we present MosaiQC, a scalable and accurate quantum circuit cutting optimizer framework. MosaiQC outperforms the baseline Qiskit circuit cutting add-on in runtime with a $2.88\times$ speedup and $5.83 \cdot 10^{11} \times$ sampling overhead reduction. Our novel QAP subroutine allows high-fidelity placements within the local search space, with low computational overhead. On the HWEA baseline, we demonstrate a $5.06\%$ increase in fidelity on a small scale. As quantum circuit cutting becomes a more commonly used technique, our optimizer will facilitate better scalability, enabling larger quantum computations.

In this work, we introduced MosaiQC, a scalable multi-objective optimization framework for quantum circuit cutting. The proposed framework jointly optimizes wire and gate cut placement while incorporating hardware-aware partition assignment through a quadratic assignment formulation. By combining graph partitioning, local refinement, and hardware-aware optimization, MosaiQC balances sampling overhead, execution fidelity, and computational efficiency within a single optimization framework.

Experimental evaluation demonstrates that MosaiQC consistently outperforms the Qiskit circuit-cutting add-on in both optimization runtime and sampling overhead across a broad range of benchmark circuits. Furthermore, the proposed hardware-aware placement strategy shows a strong correlation between the QAP objective and measured execution fidelity, validating the use of graph-based surrogate objectives for hardware-aware optimization. Together, these results demonstrate that scalable heuristic optimization can improve both optimization efficiency ($2.88 \times$ speedup) and cut-placement quality ($16.84\%$ cut reduction, $19.65\%$ improved fidelity) for a better trade-off.

%Although practical circuit cutting at very large scales remains fundamentally constrained by sampling overhead, the presented results demonstrate that cut placement optimization itself need not become the limiting factor as quantum circuits continue to increase in size. We believe that MosaiQC, as a scalable, hardware-aware, multi-objective optimization framework, provides a practical direction for enabling circuit cutting on increasingly large and heterogeneous quantum computing systems.

%Although prac

Although practical circuit cutting at very large scales remains fundamentally constrained by sampling overhead, the presented results demonstrate that cut placement optimization itself does not need to be the limiting factor as quantum circuits continue to increase in size. By showing that scalability, hardware awareness, and cut quality can be optimized simultaneously, MosaiQC establishes a practical foundation for scalable circuit cutting for future heterogeneous quantum computing systems and demonstrates the potential of multi-objective optimization as a design paradigm for quantum compiler optimization.

\section{Code Availability}
To facilitate reproducibility and future research, we make the complete implementation of MosaiQC publicly available at: \href{https://https://github.com/koenmesman/MosaiQC}{https://https://github.com/koenmesman/MosaiQC}.

%% file: Sections/Appendix.tex
\section{Appendix}

\subsection{Wire cut optimization}
\label{app:wirecut}
\begin{figure}[h]
    \centering
    \includegraphics[width=0.55\linewidth]{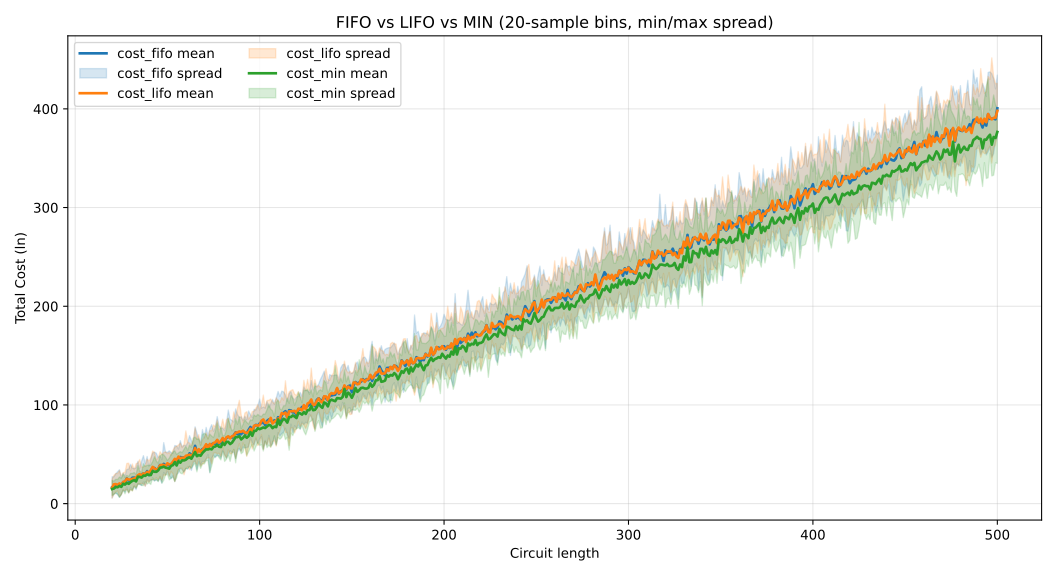}
    \caption{Comparison between wire cut placement methods.}
    \label{fig:app_wirecut}
\end{figure}

The wire cut calculation at partition placement overlaps uses a minimum cost first approach. Alternative methods are scanning from the first to the last gate (FIFO) or in reverse order (LIFO). These three methods have been tested and are shown in Figure \ref{fig:app_wirecut}. Comparisons are done for random gate target orders of random weights. In all cases, gate cuts can be inserted if this is cheaper compared to two wire cuts, or to one wire cut in boundaries (first/last gate). Overall, the minimum first approach is slightly better for a slightly higher computational cost. As this additional computational cost will be negligible, the minimal first approach is chosen for the better solution quality.

Furthermore, often there will not be enough ancillary qubits to allow all wire cuts. In this scenario, using min-cost first will generally improve the solution when only a few wire cuts are allowed.

\input{Sections/WireCut_alg}

\clearpage
\subsection{Detailed runtime results}
\label{app:runtime}

\begin{figure}[h]
\centering
\begin{minipage}{0.88\linewidth}  
\centering
%\setkeys{Gin}{width=\linewidth}
\begin{subfigure}{0.49\textwidth}
\includegraphics[width=\textwidth]{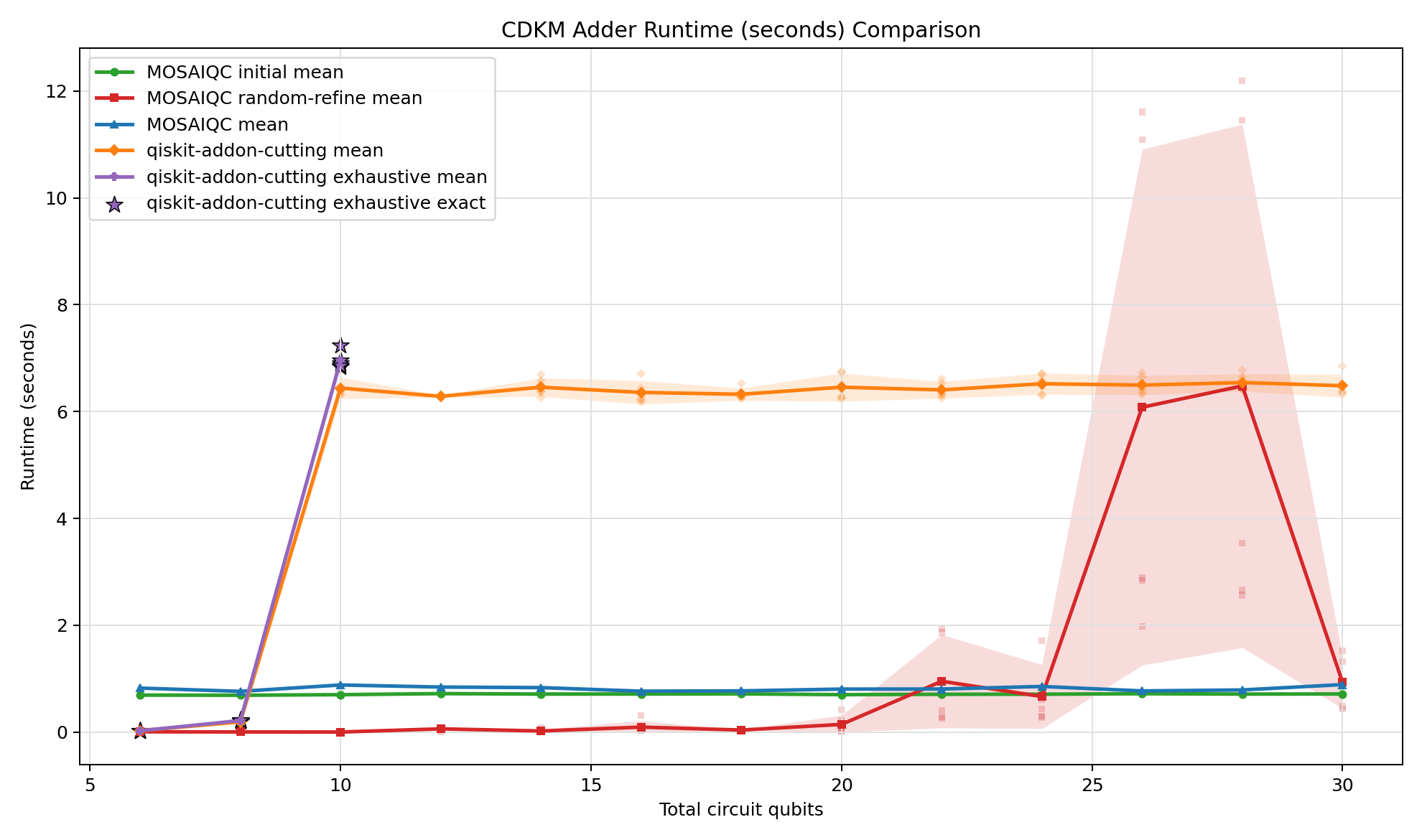}
\caption{Adder}
\label{fig:a_r_s}
\end{subfigure}
\begin{subfigure}{0.49\textwidth}
\includegraphics[width=\linewidth]{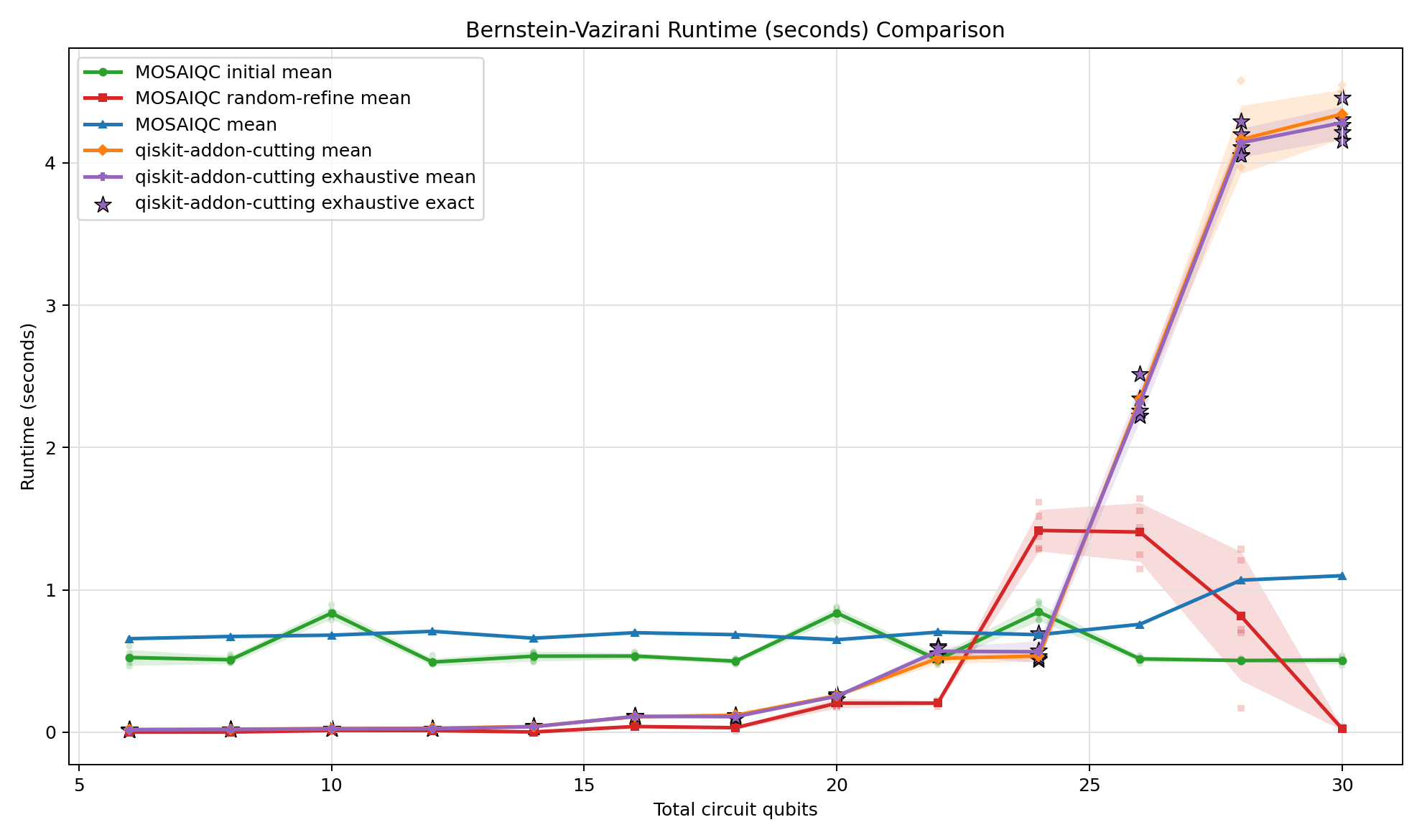}
\caption{BV}
\label{fig:b_r_s}
\end{subfigure}
\begin{subfigure}{0.49\textwidth}
\includegraphics[width=\linewidth]{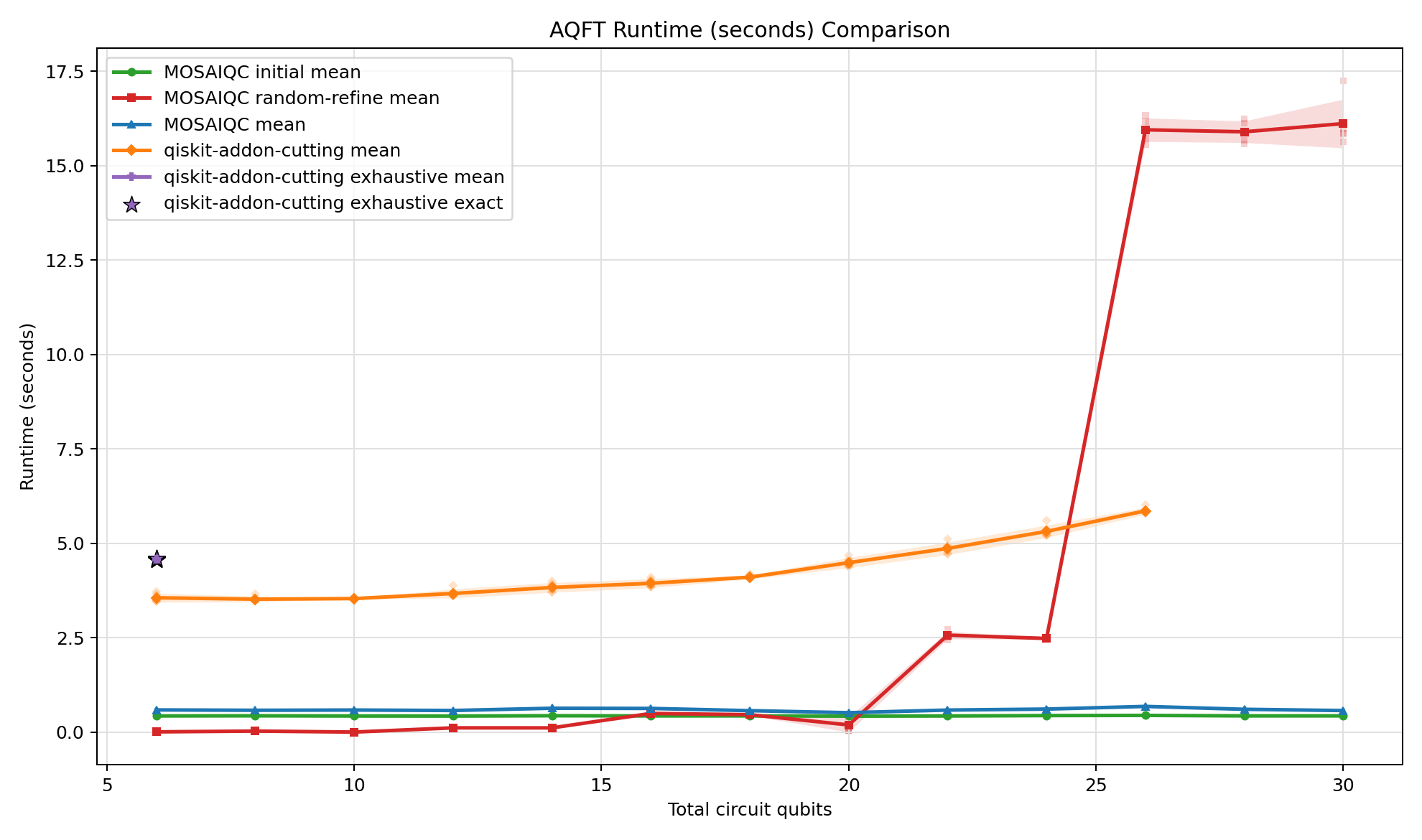}
\caption{AQFT}
\label{fig:c_r_s}
\end{subfigure}
\begin{subfigure}{0.49\textwidth}
\includegraphics[width=\linewidth]{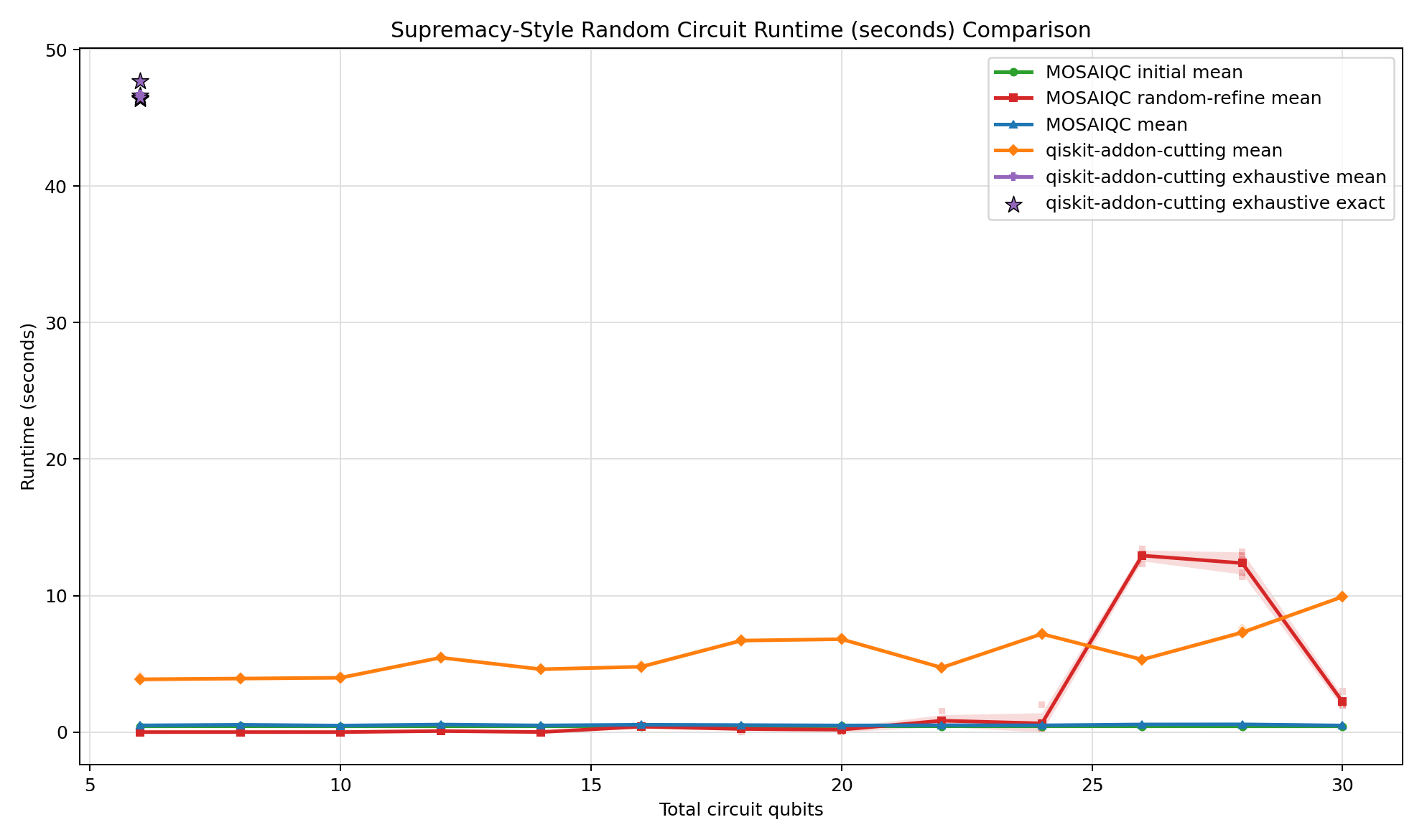}
\caption{Supremacy}
\label{fig:d_r_s}
\end{subfigure}
\begin{subfigure}{0.49\textwidth}
\includegraphics[width=\linewidth]{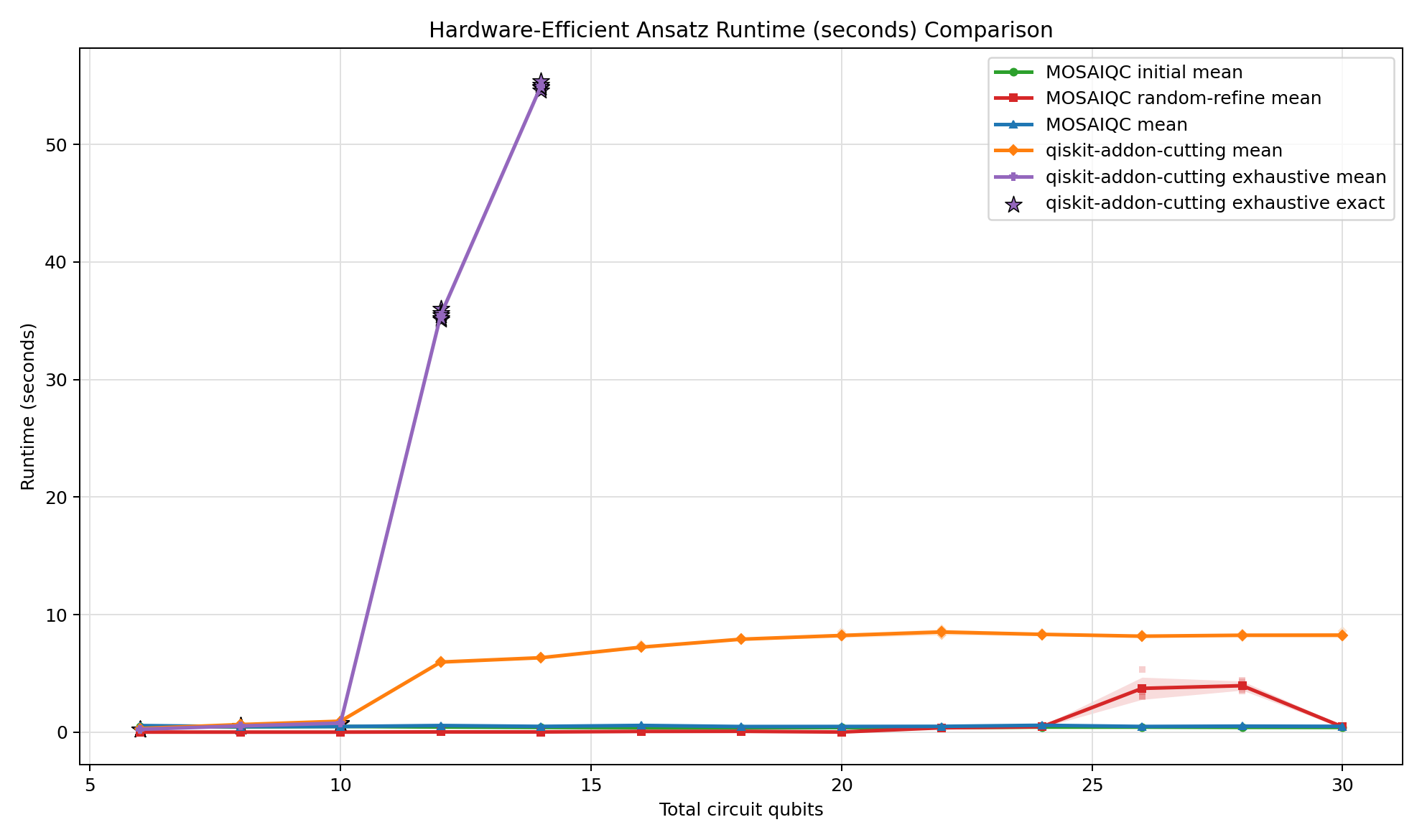}
\caption{HWEA}
\label{fig:e_r_s}
\end{subfigure}
\begin{subfigure}{0.49\textwidth}
\includegraphics[width=\linewidth]{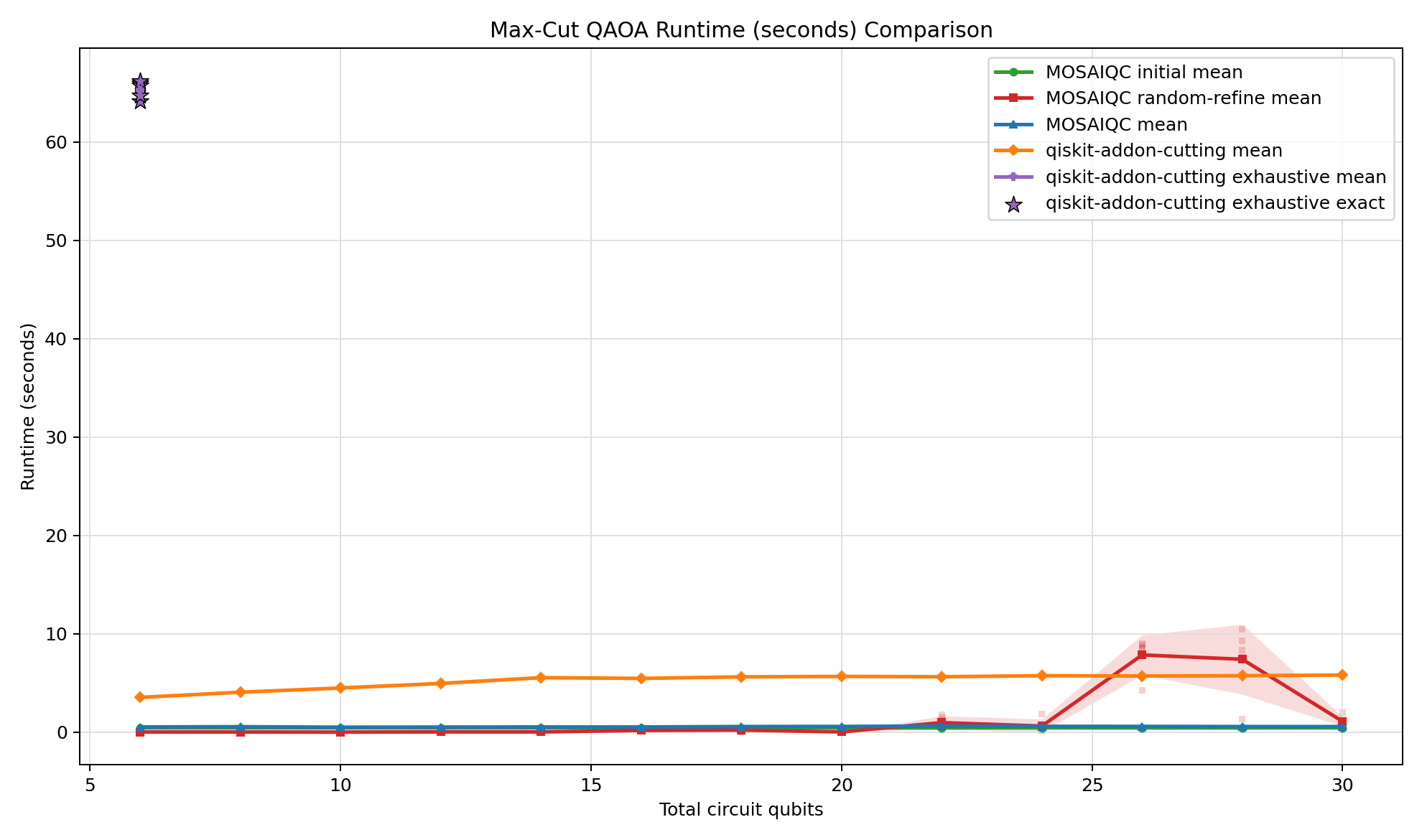}
\caption{QAOA MCP}
\label{fig:f_r_s}
\end{subfigure}
\\
\begin{subfigure}{0.49\textwidth}
\includegraphics[width=\linewidth]{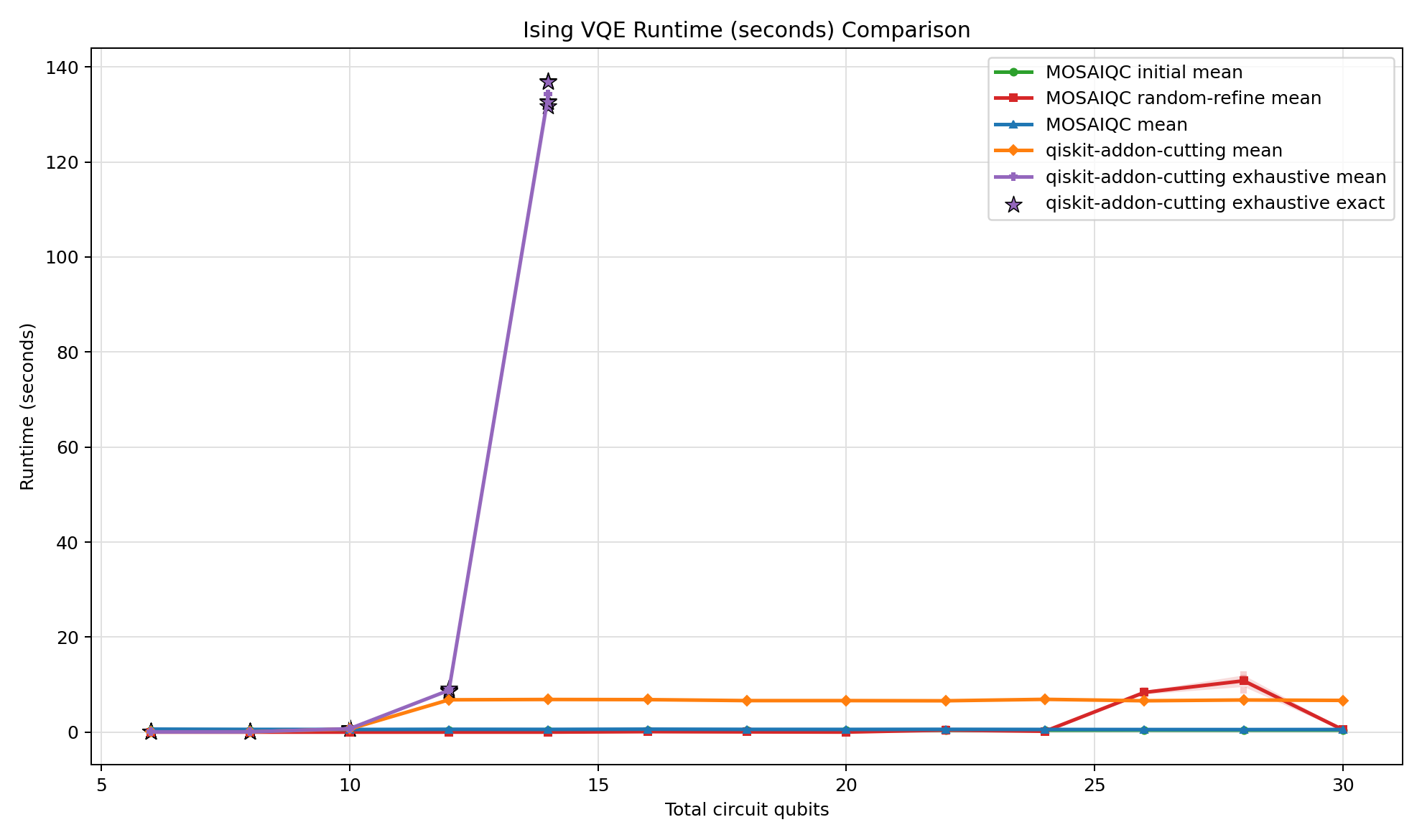}
\caption{Ising VQE HWAE}
\label{fig:g_r_s}
\end{subfigure}
\begin{subfigure}{0.49\textwidth}
\includegraphics[width=\linewidth]{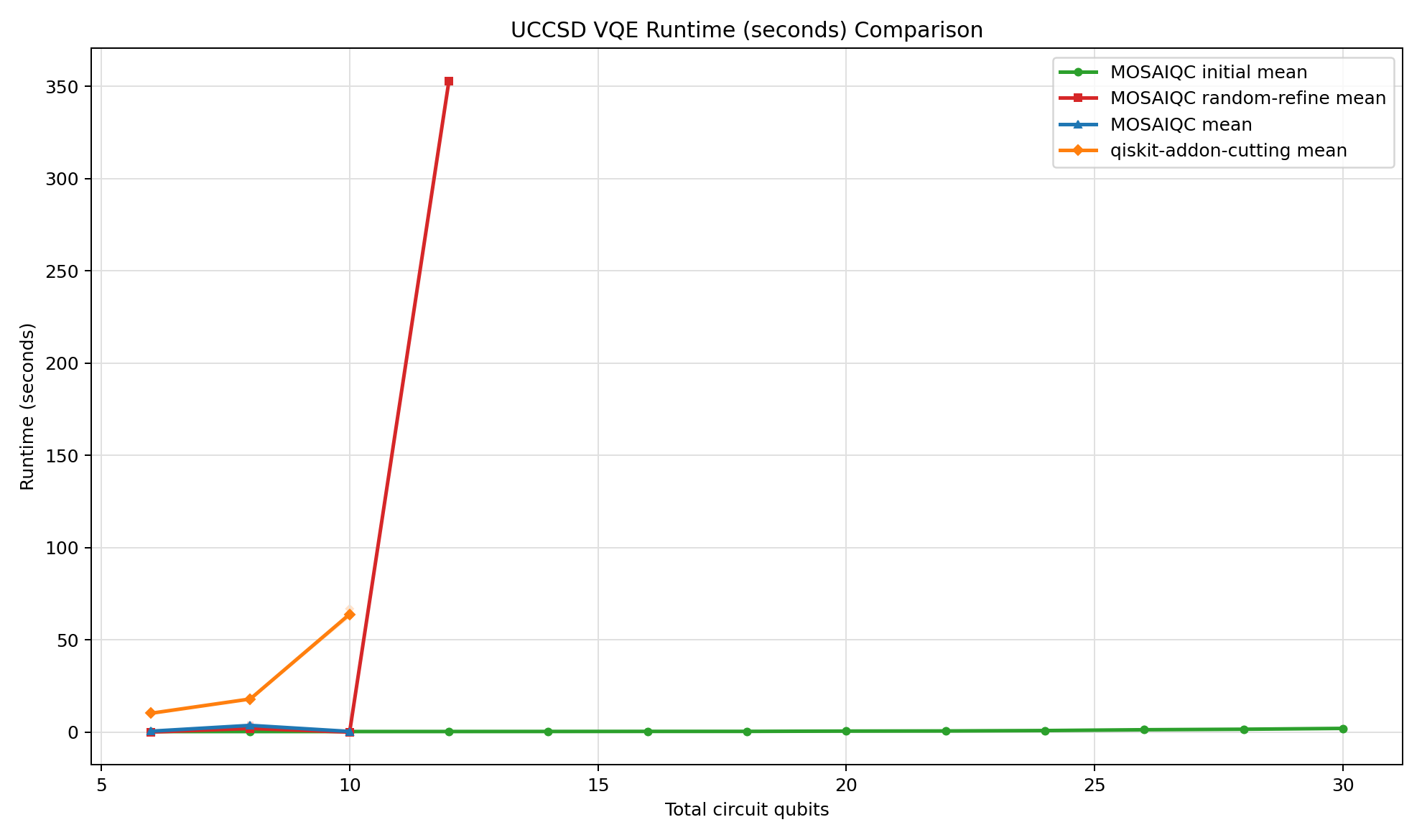}
\caption{Ising VQE UCCSD}
\label{fig:h_r_s}
\end{subfigure}

\end{minipage}
\caption{Runtime results for 6-30 qubit instances for 2-qubit increments.}
\label{fig:runtimes_s}

\end{figure}

\begin{figure}[h]
\centering

%\setkeys{Gin}{width=\linewidth}
\begin{subfigure}{0.49\textwidth}
\includegraphics[width=\textwidth]{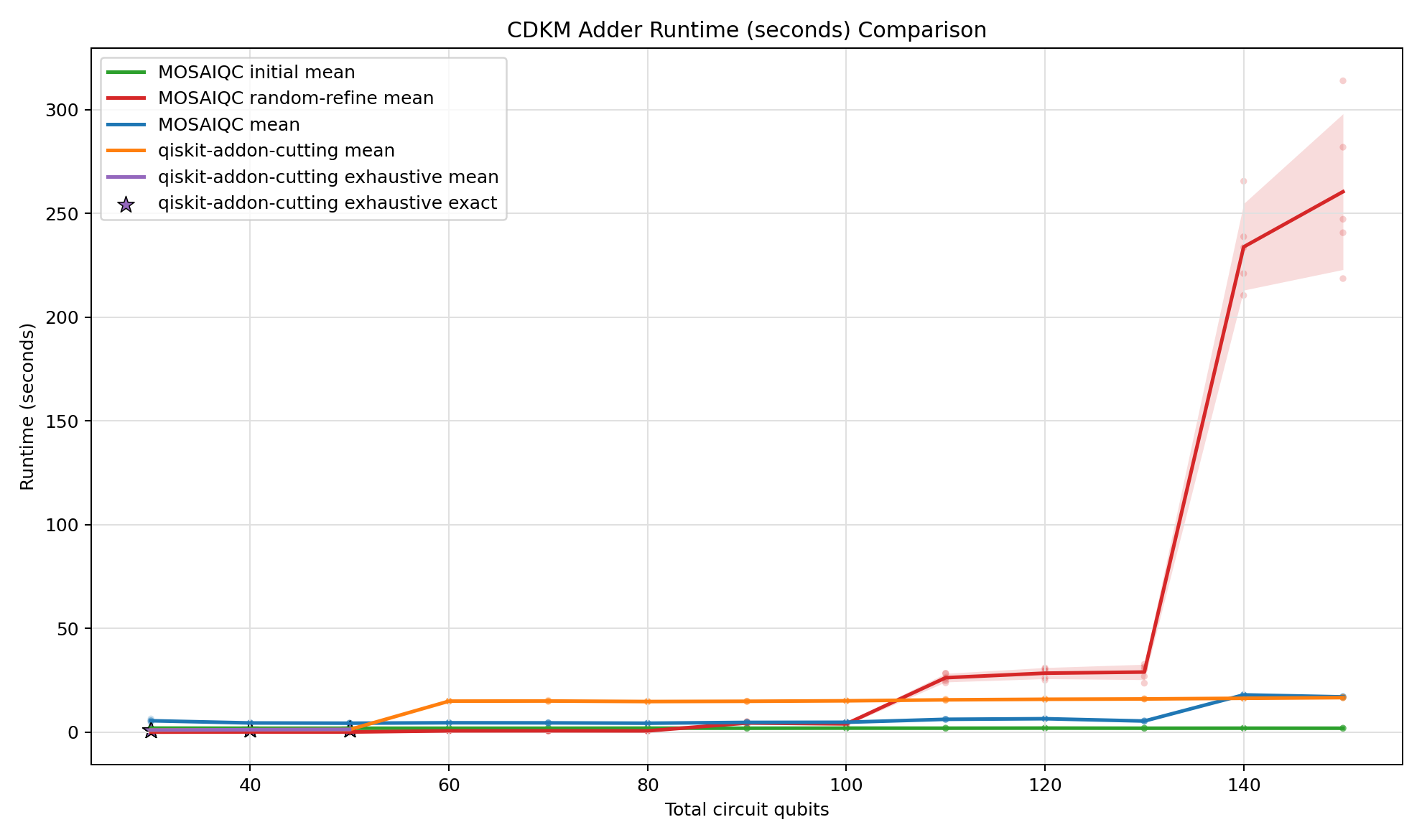}
\caption{Adder}
\label{fig:a_r_m}
\end{subfigure}
\begin{subfigure}{0.49\textwidth}
\includegraphics[width=\linewidth]{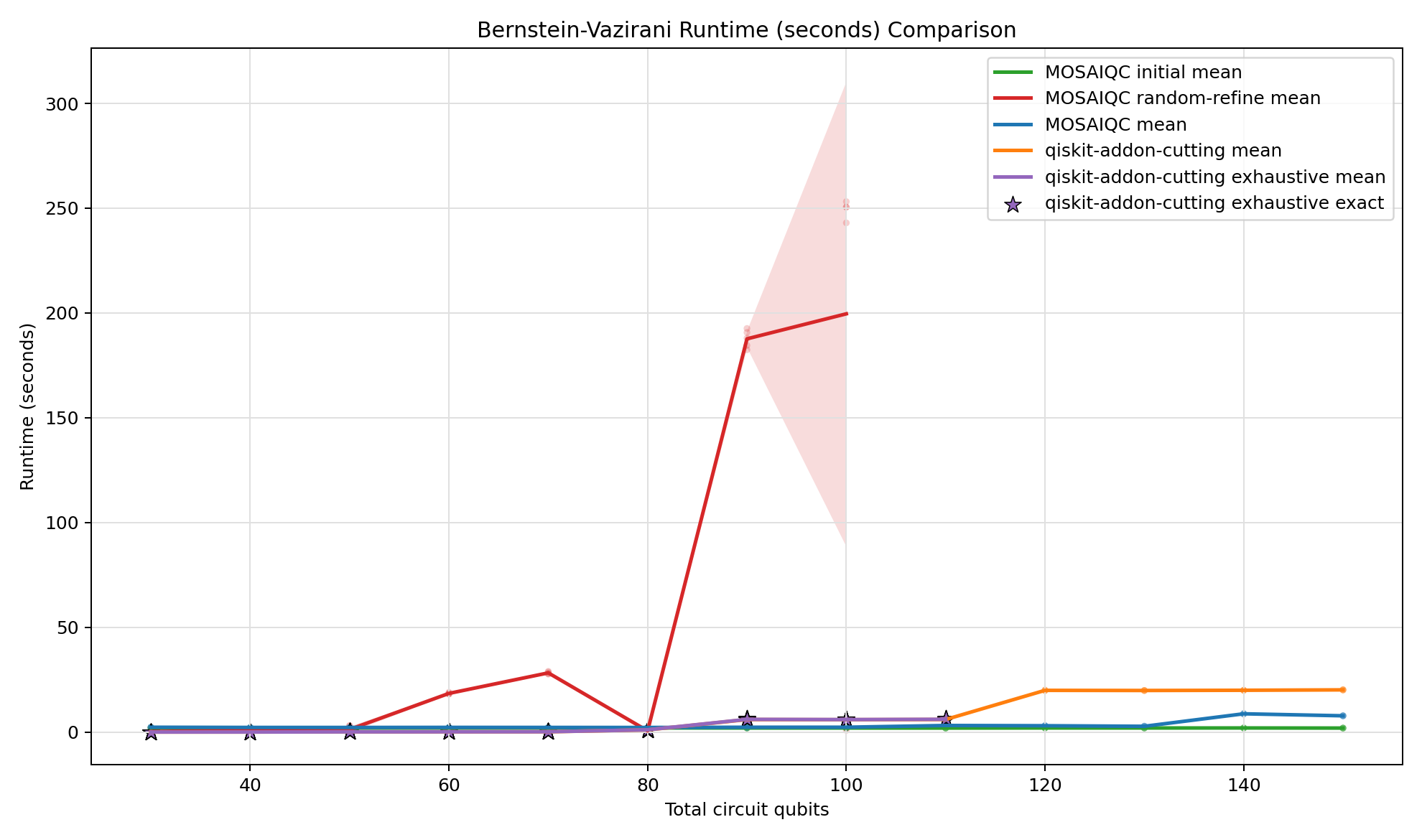}
\caption{BV}
\label{fig:b_r_m}
\end{subfigure}
\begin{subfigure}{0.49\textwidth}
\includegraphics[width=\linewidth]{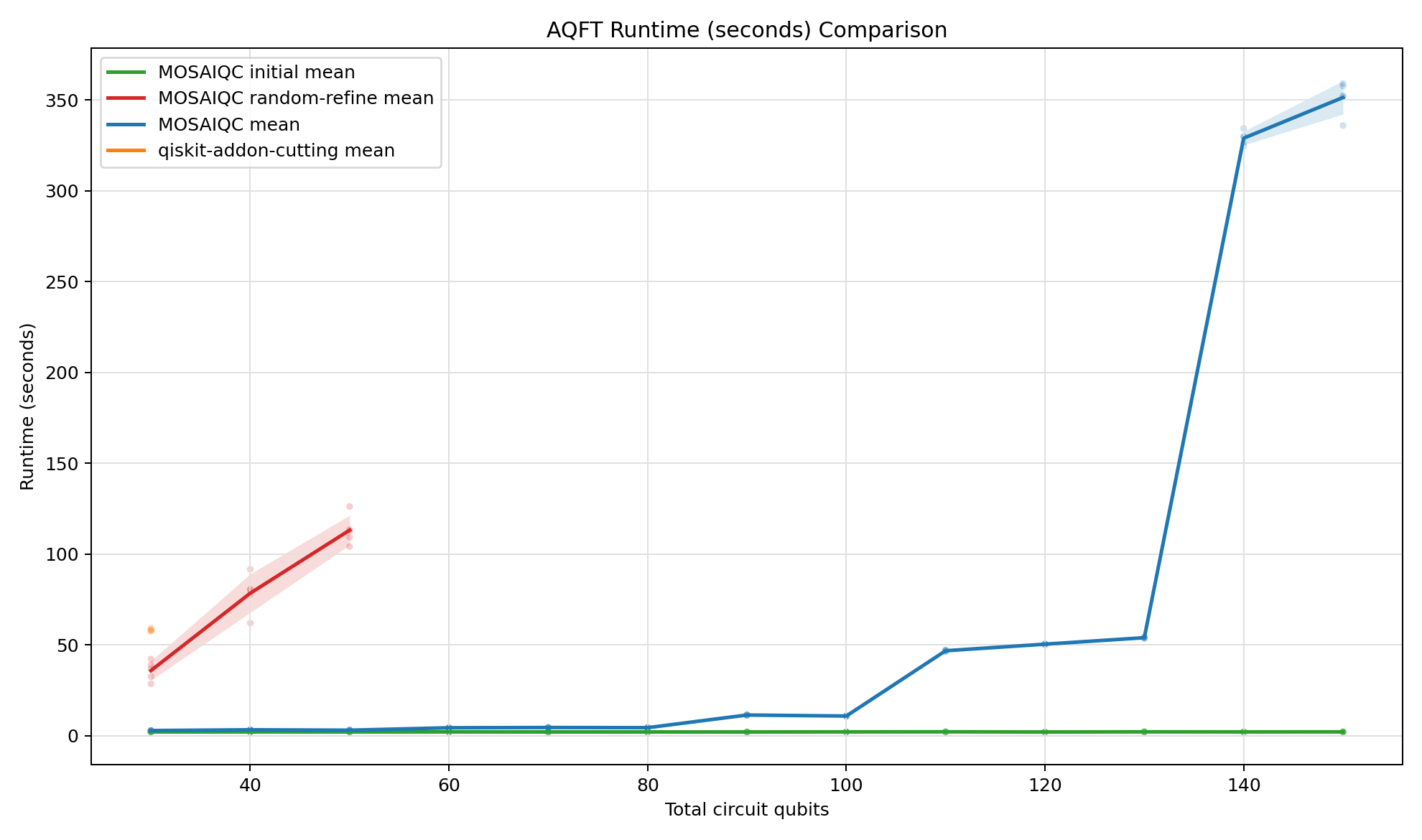}
\caption{AQFT}
\label{fig:c_r_m}
\end{subfigure}
\begin{subfigure}{0.49\textwidth}
\includegraphics[width=\linewidth]{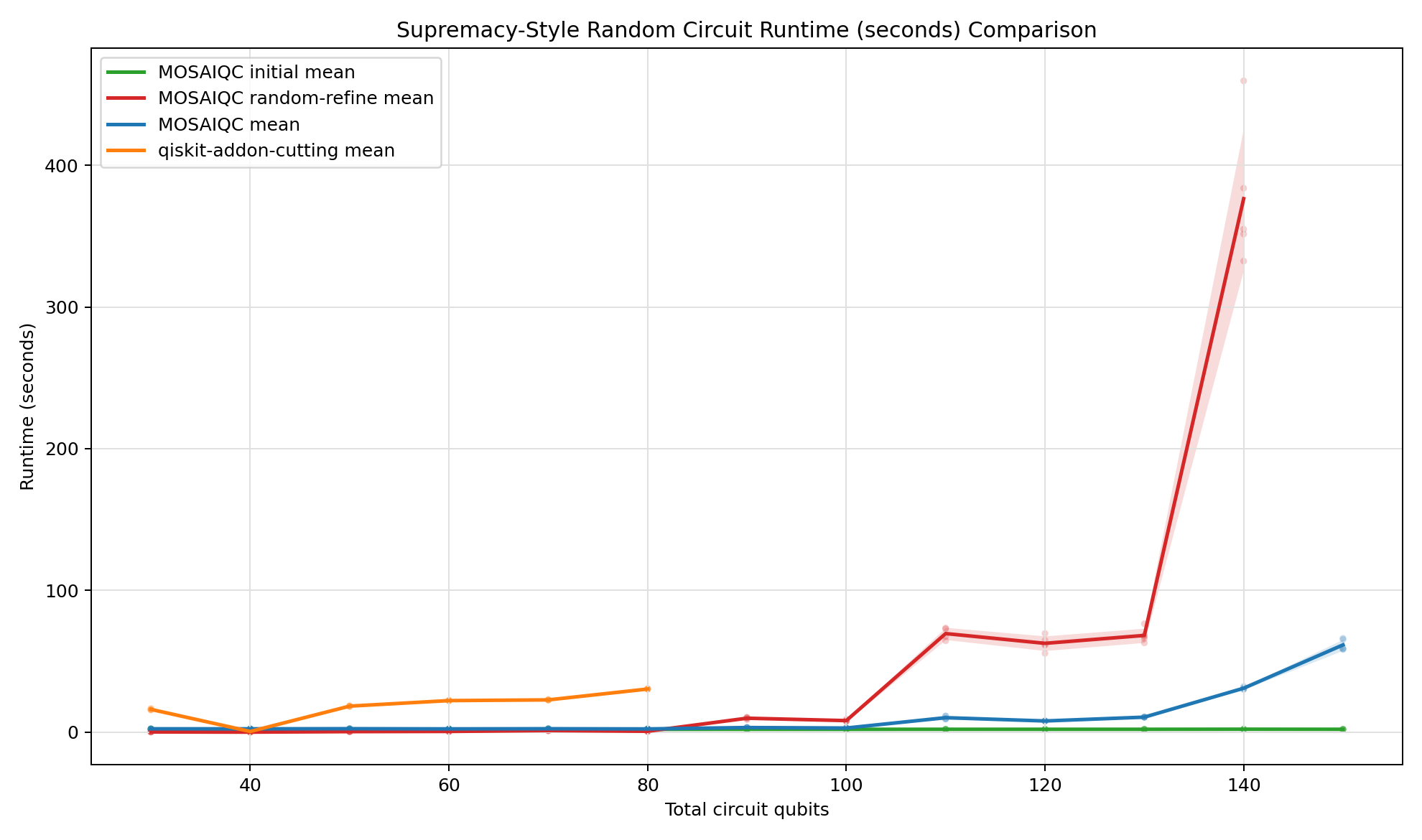}
\caption{Supremacy}
\label{fig:d_r_m}
\end{subfigure}
\begin{subfigure}{0.49\textwidth}
\includegraphics[width=\linewidth]{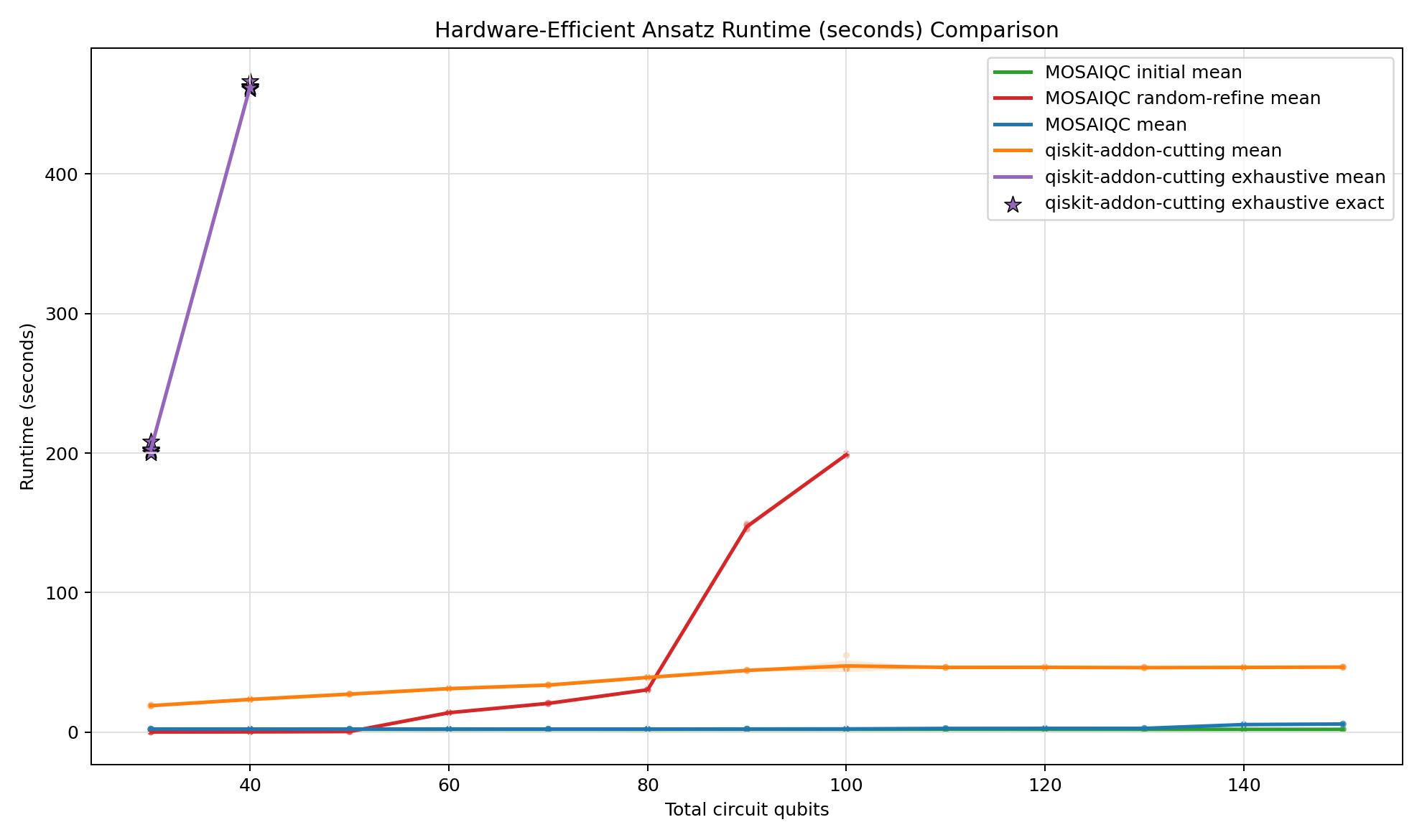}
\caption{HWEA}
\label{fig:e_r_m}
\end{subfigure}
\begin{subfigure}{0.49\textwidth}
\includegraphics[width=\linewidth]{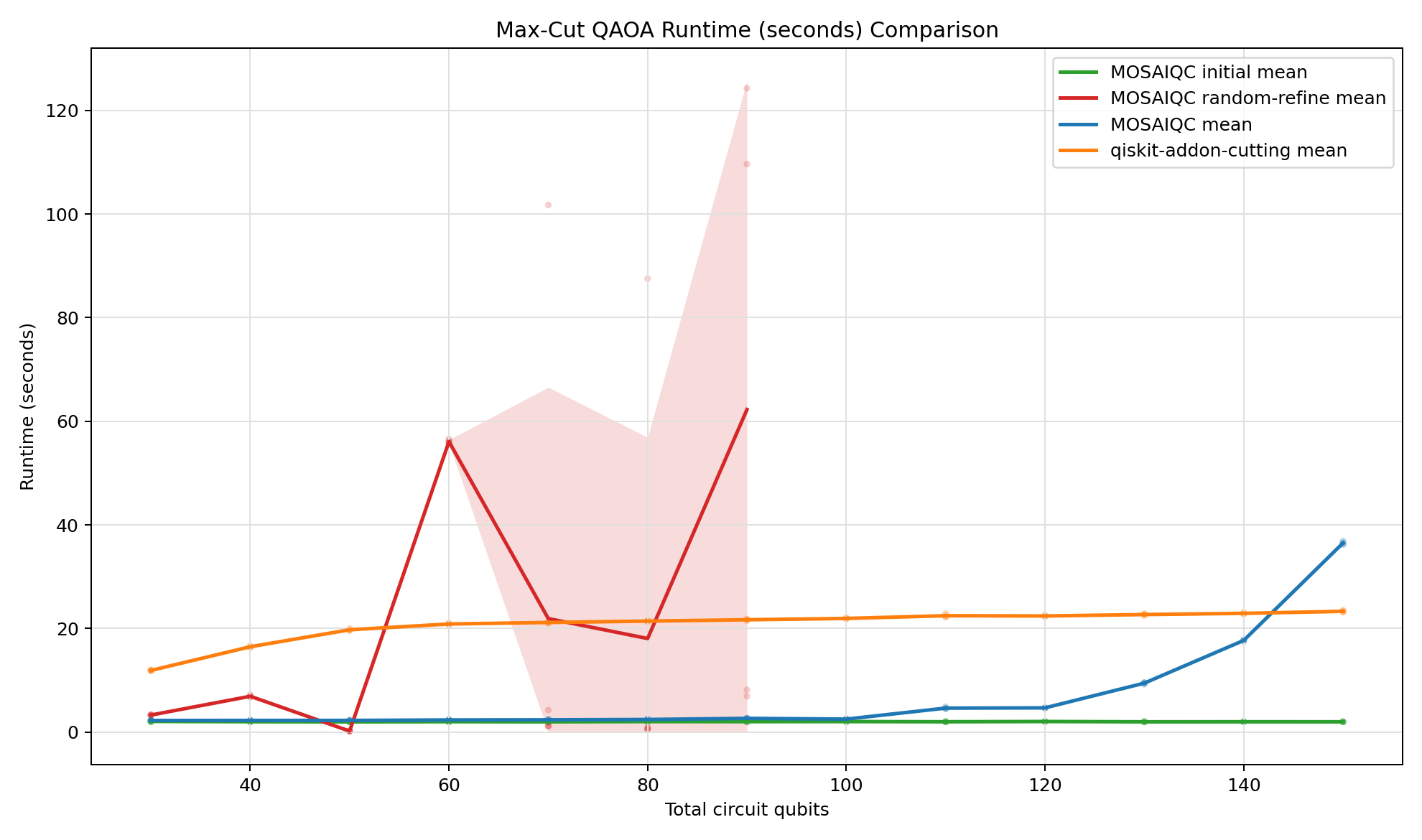}
\caption{QAOA MCP}
\label{fig:f_r_m}
\end{subfigure}

\begin{subfigure}{0.49\textwidth}
\includegraphics[width=\linewidth]{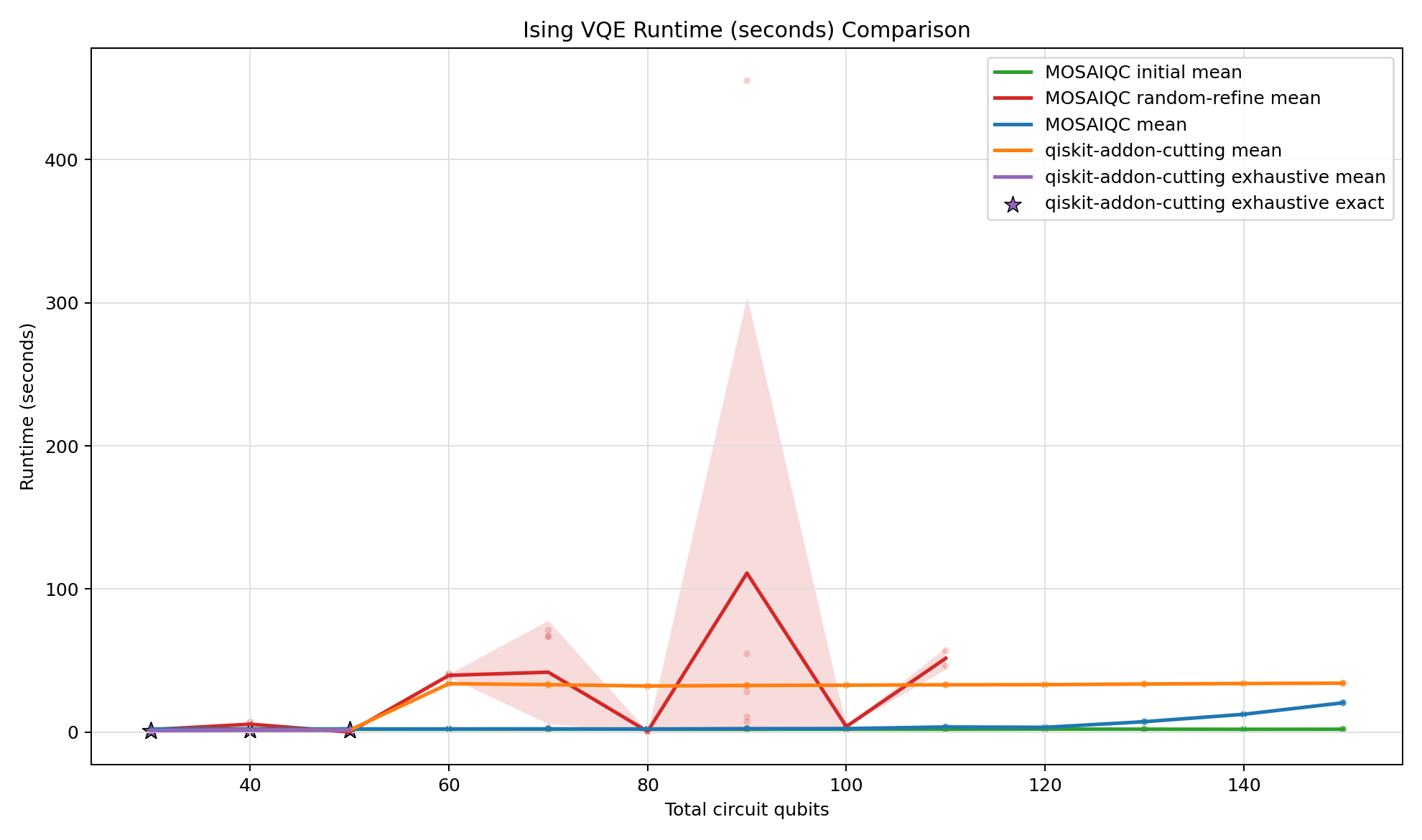}
\caption{Ising VQE HWAE}
\label{fig:g_r_m}
\end{subfigure}
\begin{subfigure}{0.49\textwidth}
\includegraphics[width=\linewidth]{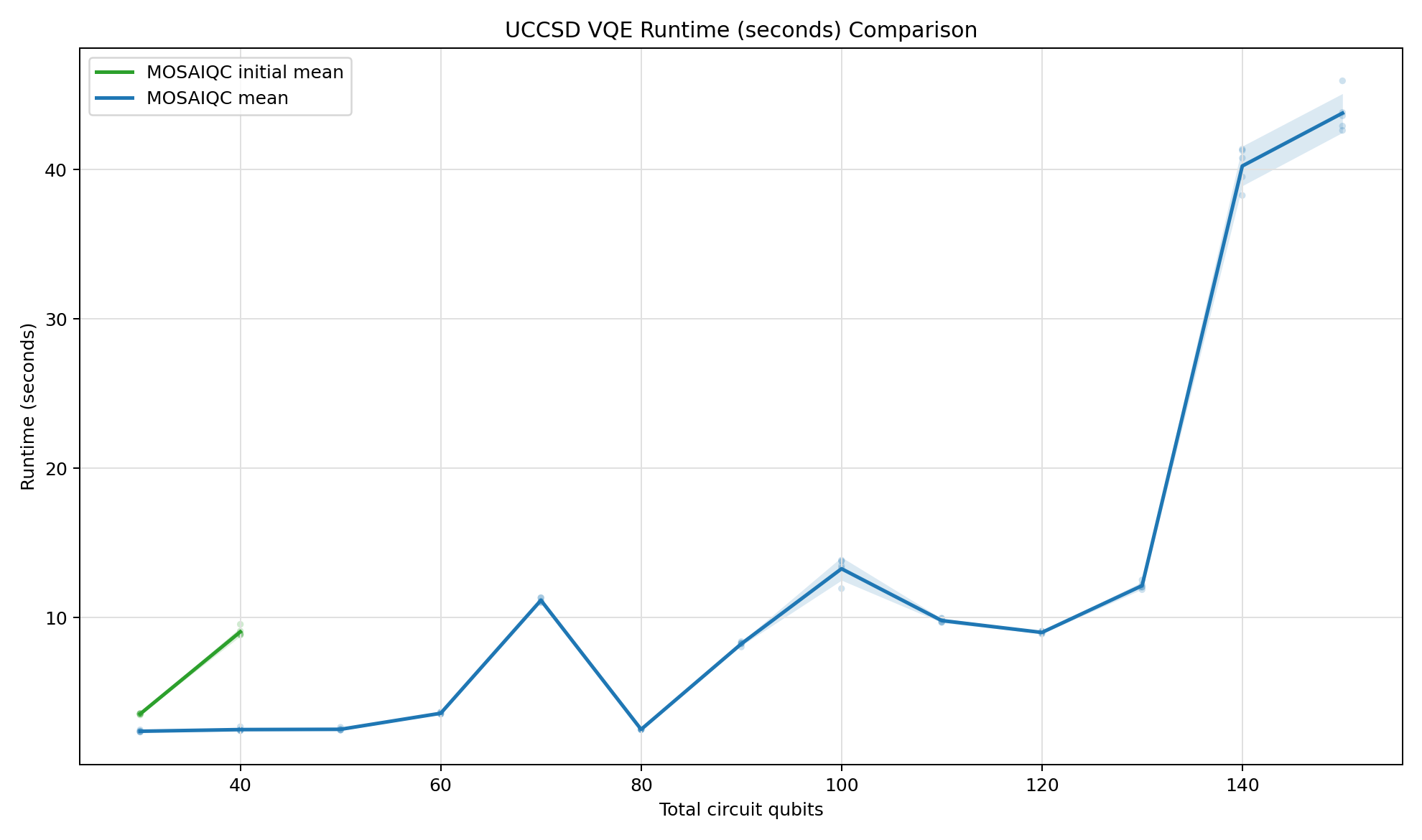}
\caption{Ising VQE UCCSD}
\label{fig:h_r_m}
\end{subfigure}
\caption{Runtime results for 30-150 qubit instances for 10 qubit increments.}
\label{fig:runtime_m}
\end{figure}

\begin{figure}[h]
\centering
%\setkeys{Gin}{width=\linewidth}
\begin{subfigure}{0.49\textwidth}
\includegraphics[width=\textwidth]{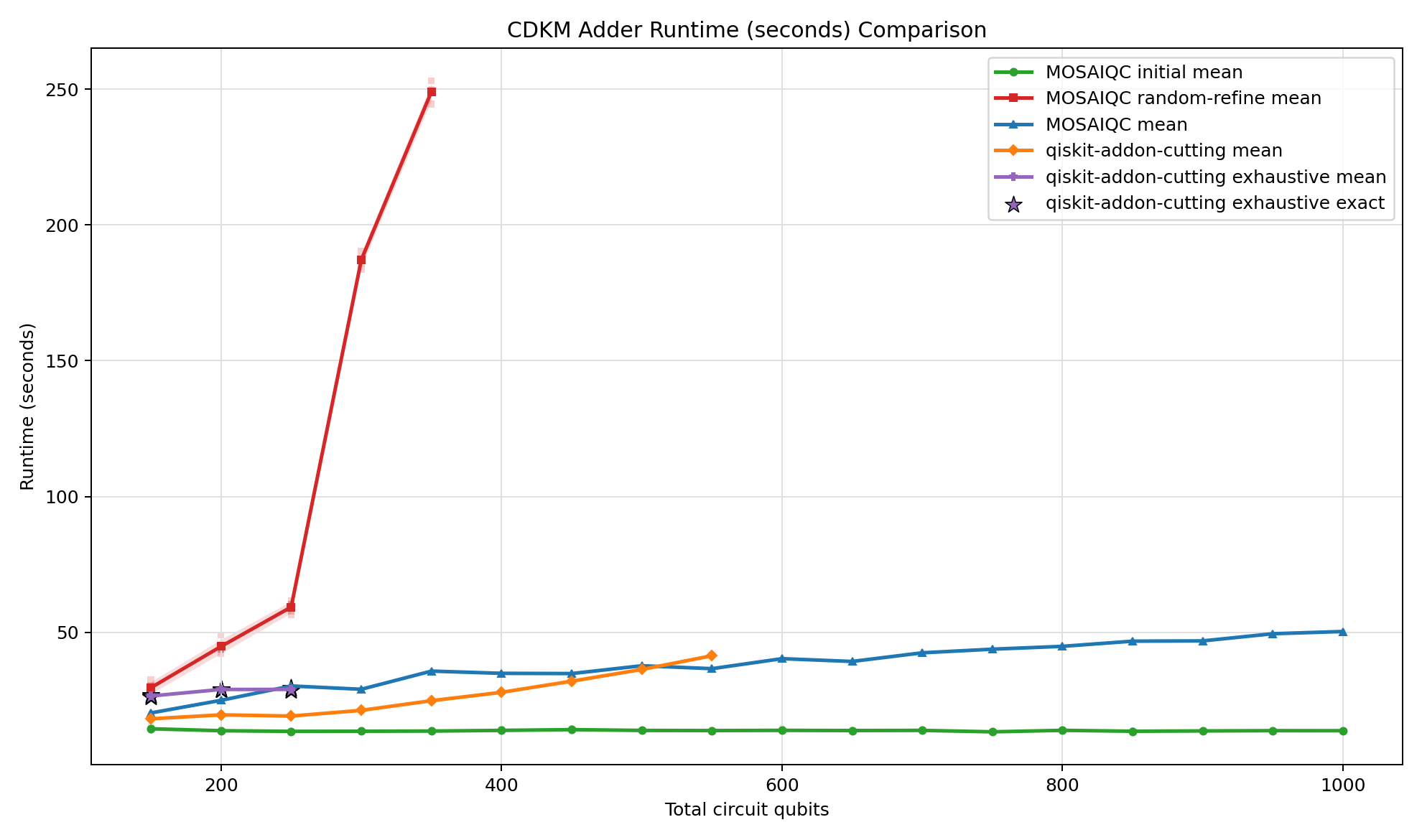}
\caption{Adder}
\label{fig:a_r_l}
\end{subfigure}
\begin{subfigure}{0.49\textwidth}
\includegraphics[width=\linewidth]{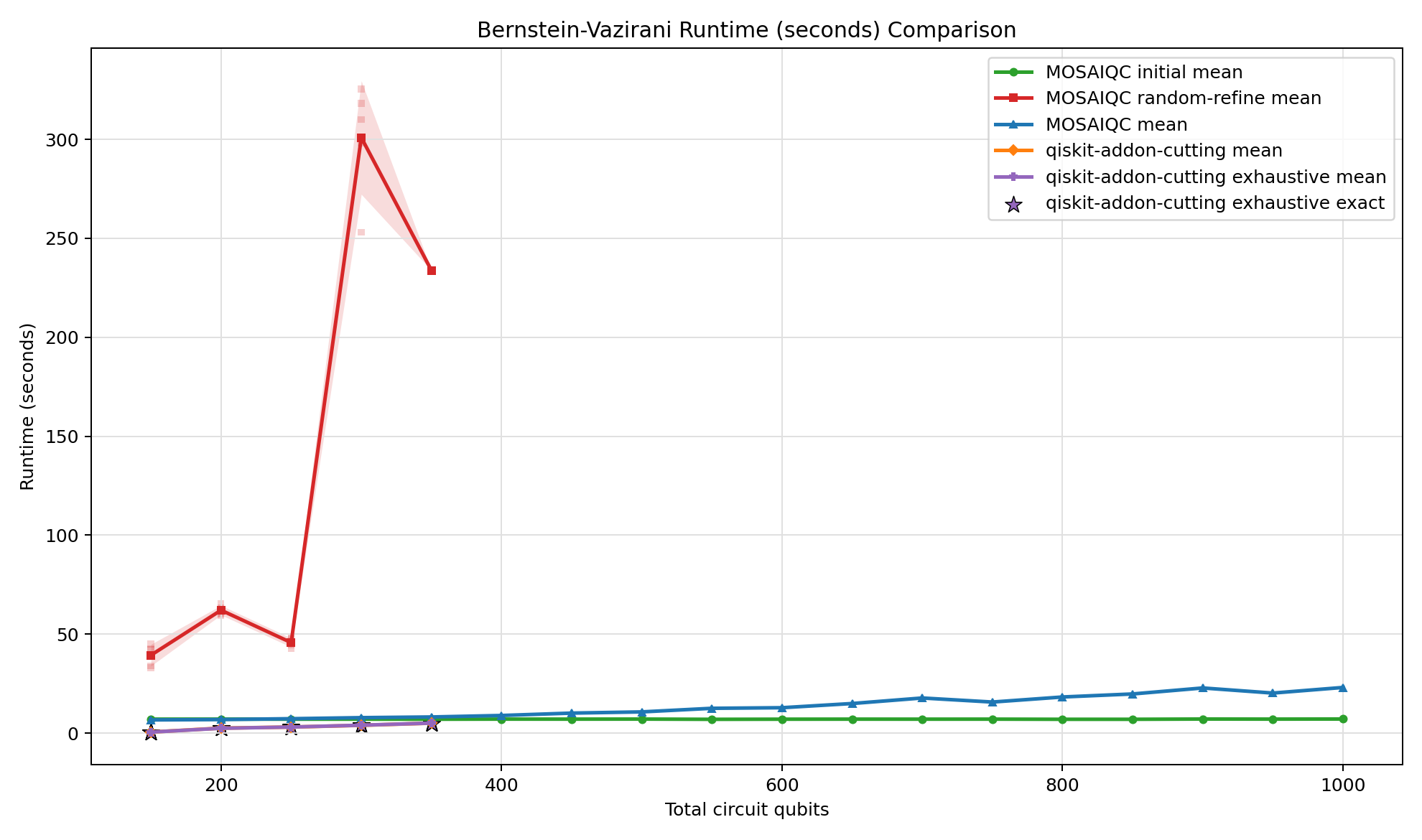}
\caption{BV}
\label{fig:b_r_l}
\end{subfigure}
\begin{subfigure}{0.49\textwidth}
\includegraphics[width=\linewidth]{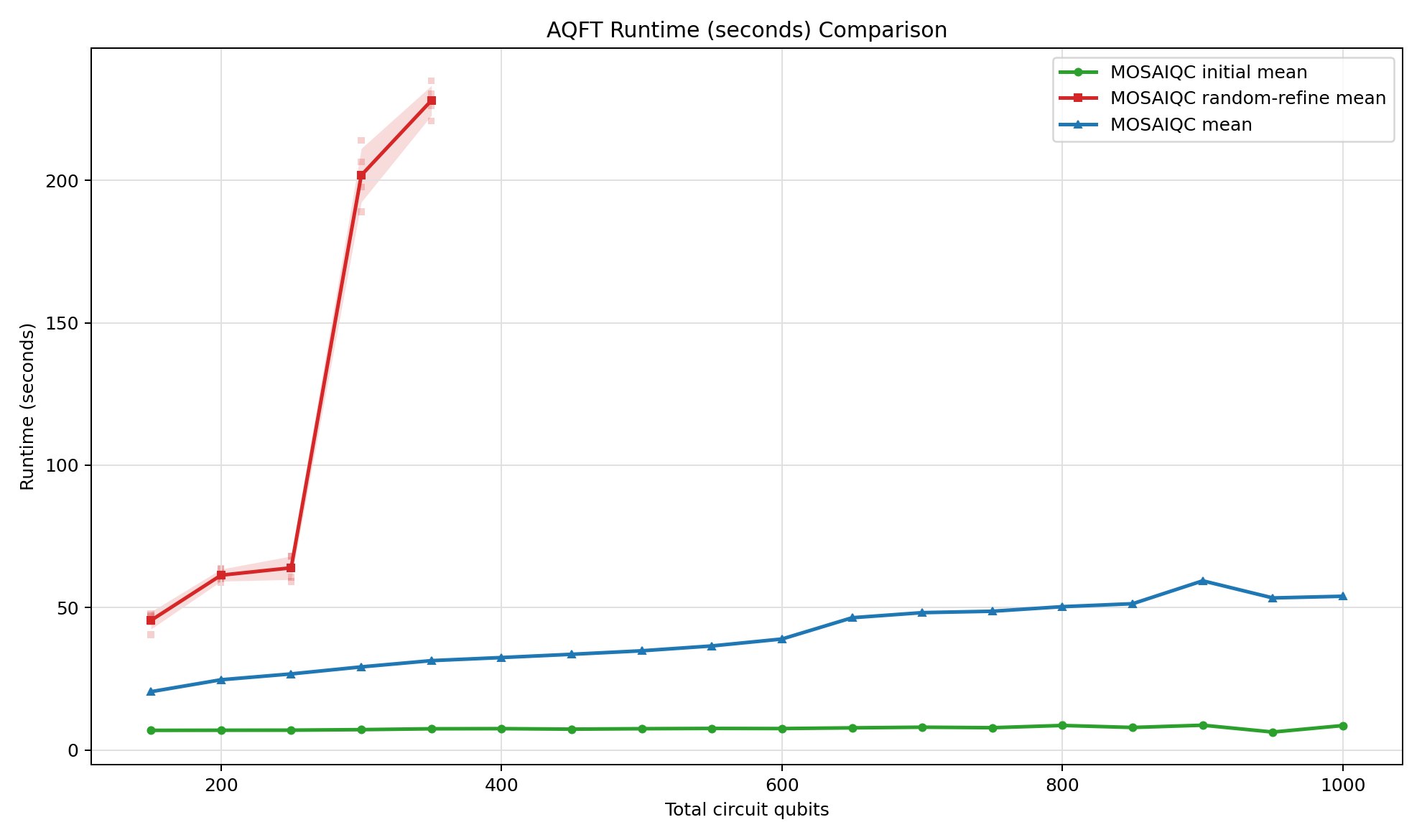}
\caption{AQFT}
\label{fig:c_r_l}
\end{subfigure}
\begin{subfigure}{0.49\textwidth}
\includegraphics[width=\linewidth]{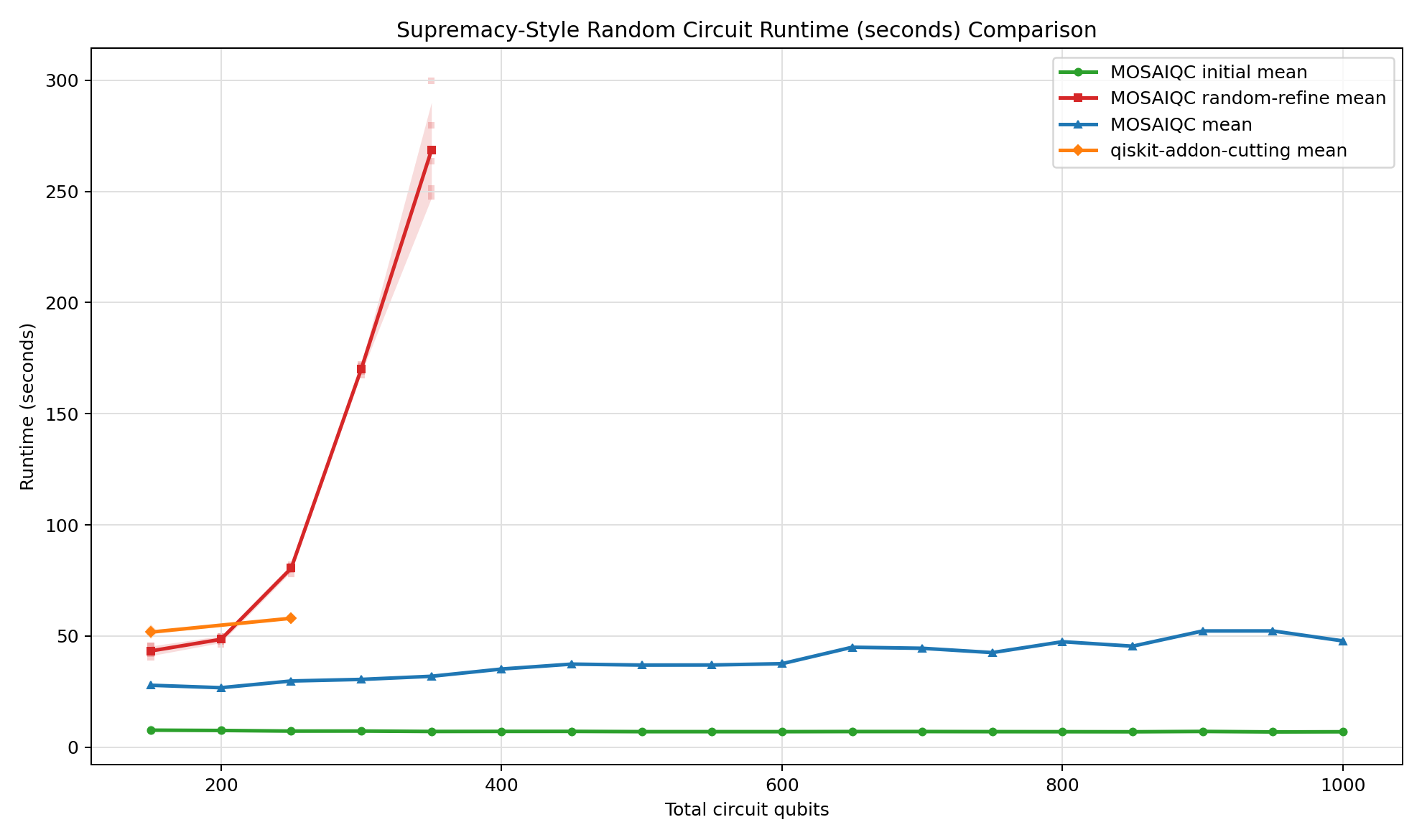}
\caption{Supremacy}
\label{fig:d_r_l}
\end{subfigure}
\begin{subfigure}{0.49\textwidth}
\includegraphics[width=\linewidth]{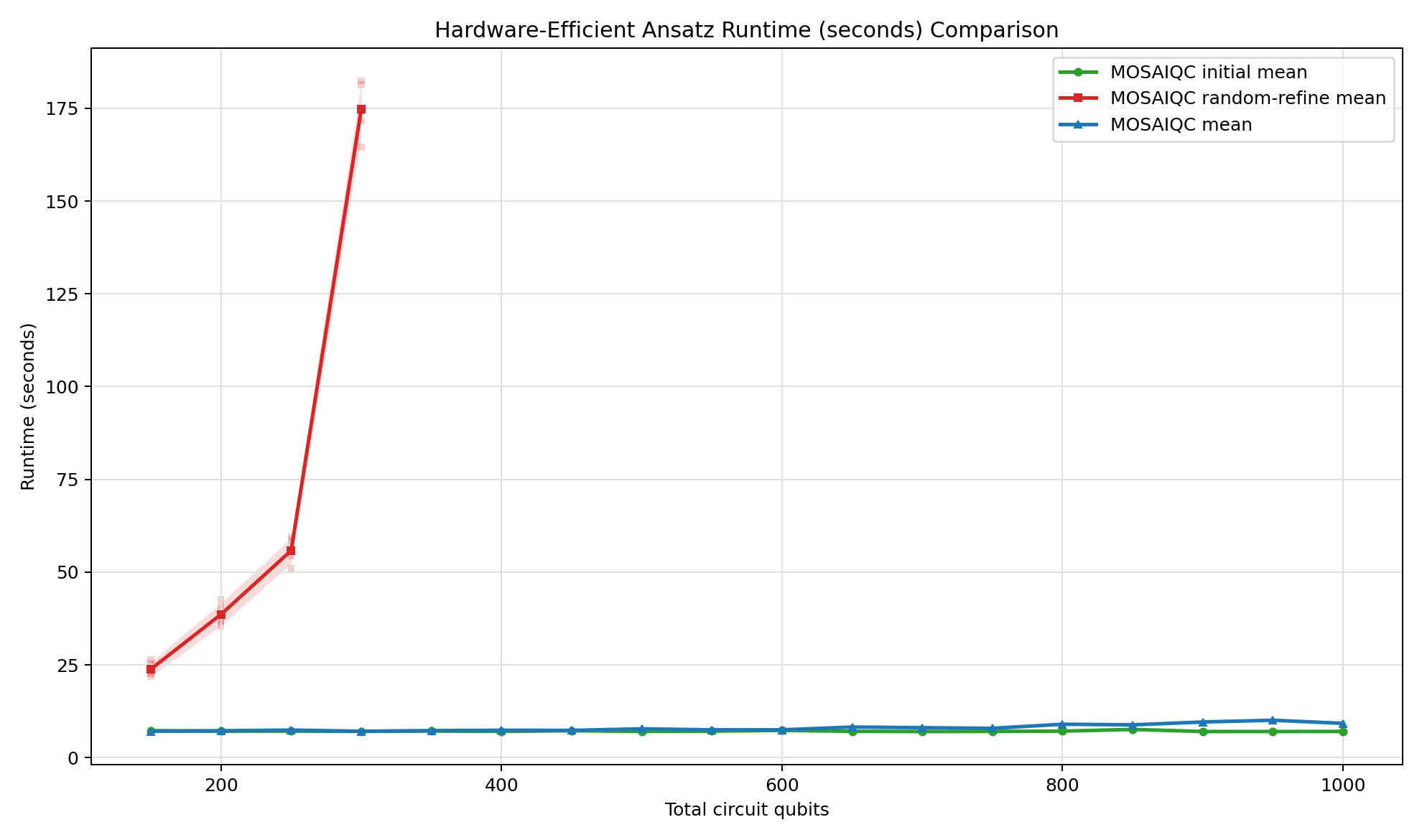}
\caption{HWEA}
\label{fig:e_r_l}
\end{subfigure}
\begin{subfigure}{0.49\textwidth}
\includegraphics[width=\linewidth]{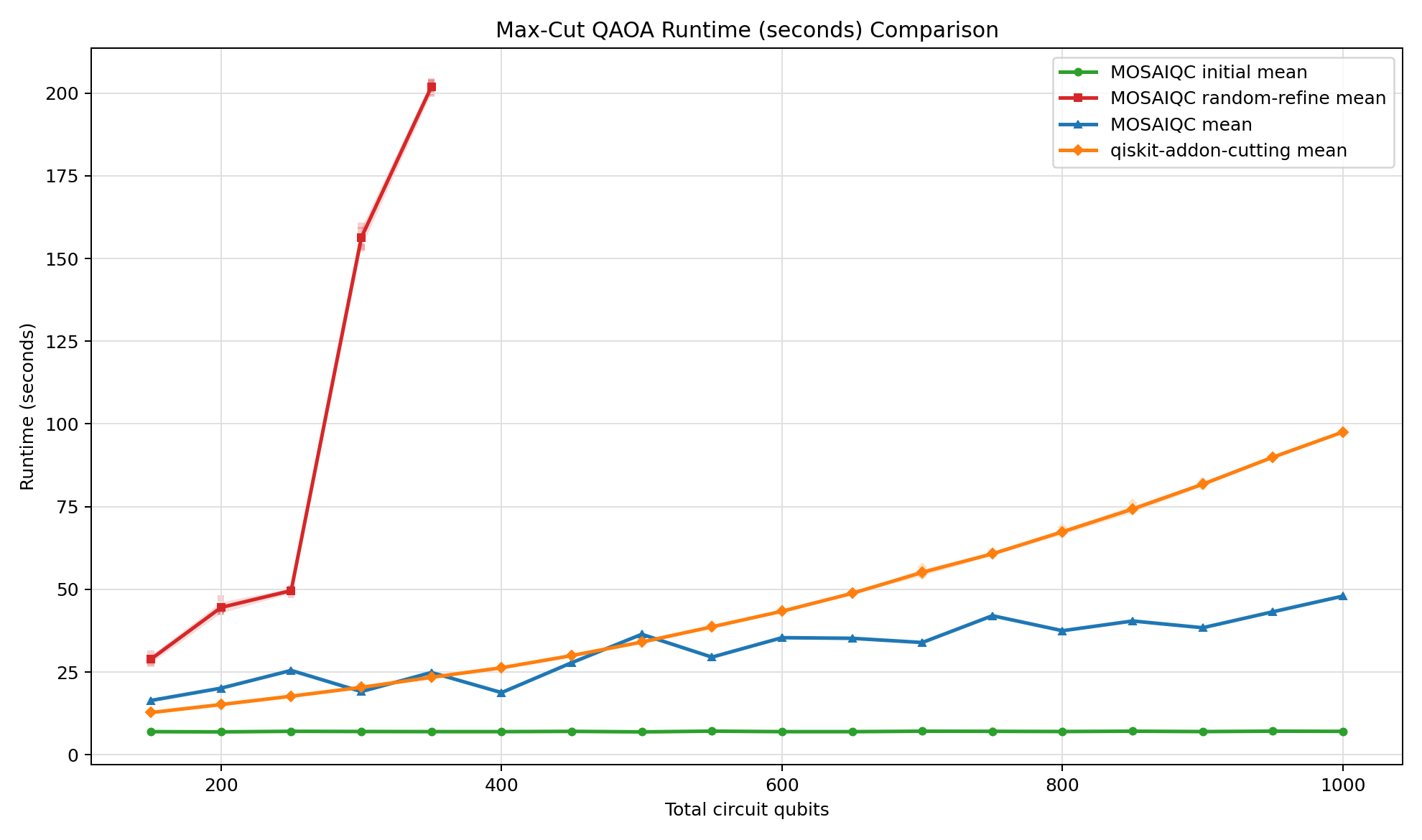}
\caption{QAOA MCP}
\label{fig:f_r_l}
\end{subfigure}

\begin{subfigure}{0.49\textwidth}
\includegraphics[width=\linewidth]{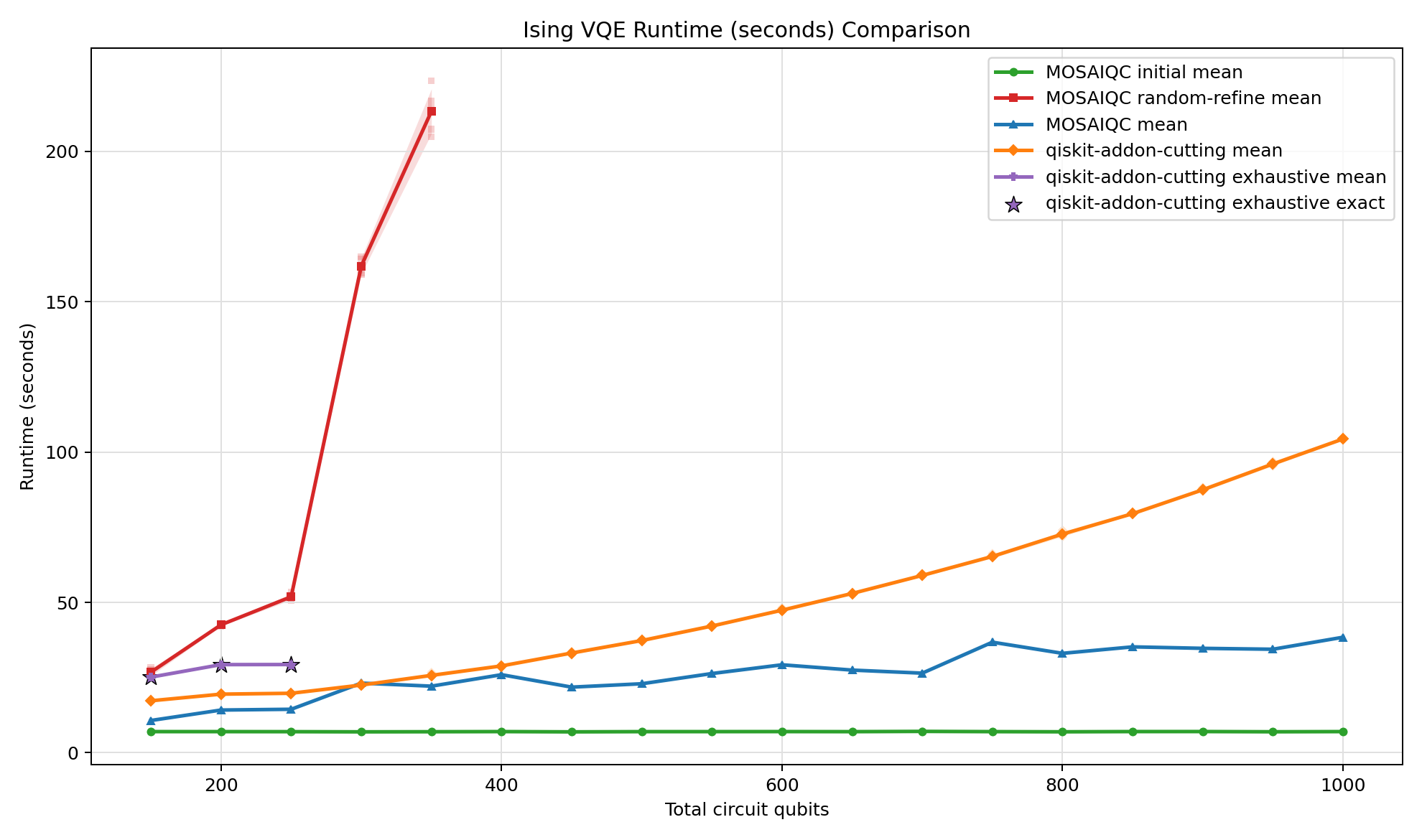}
\caption{Ising VQE HWAE}
\label{fig:g_r_l}
\end{subfigure}

\caption{Runtime results for 150-1000 qubit instances for 50 qubit increments.}
\label{fig:runtimes_l}
\end{figure}

\clearpage
\subsection{Detailed sampling overhead results}
\label{app:overhead}

\begin{figure}[h]
\centering

\begin{minipage}{0.88\linewidth}  
\centering
%\setkeys{Gin}{width=\linewidth}
\begin{subfigure}{0.49\textwidth}
\includegraphics[width=\textwidth]{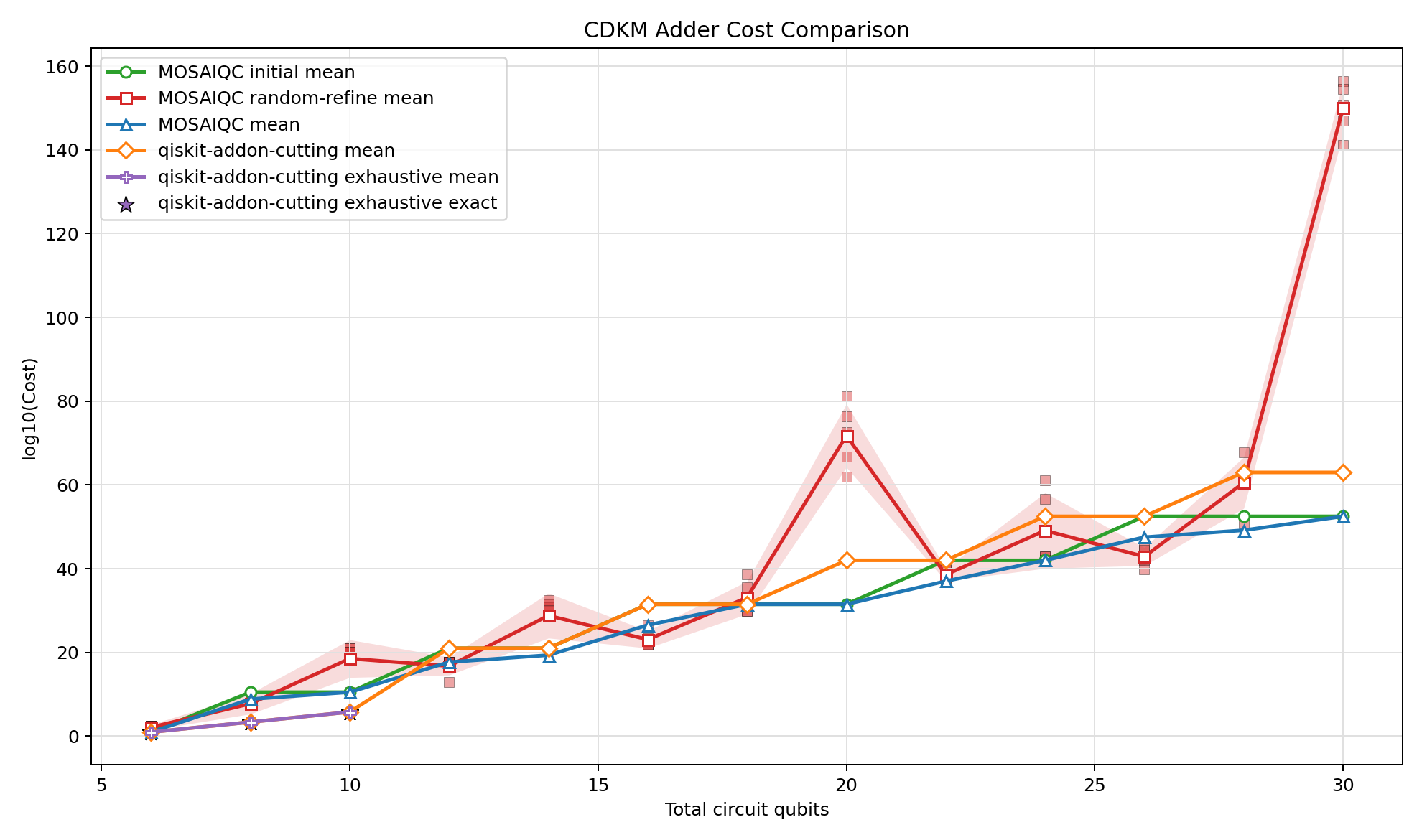}
\caption{Adder}
\label{fig:a_o_s}
\end{subfigure}
\begin{subfigure}{0.49\textwidth}
\includegraphics[width=\linewidth]{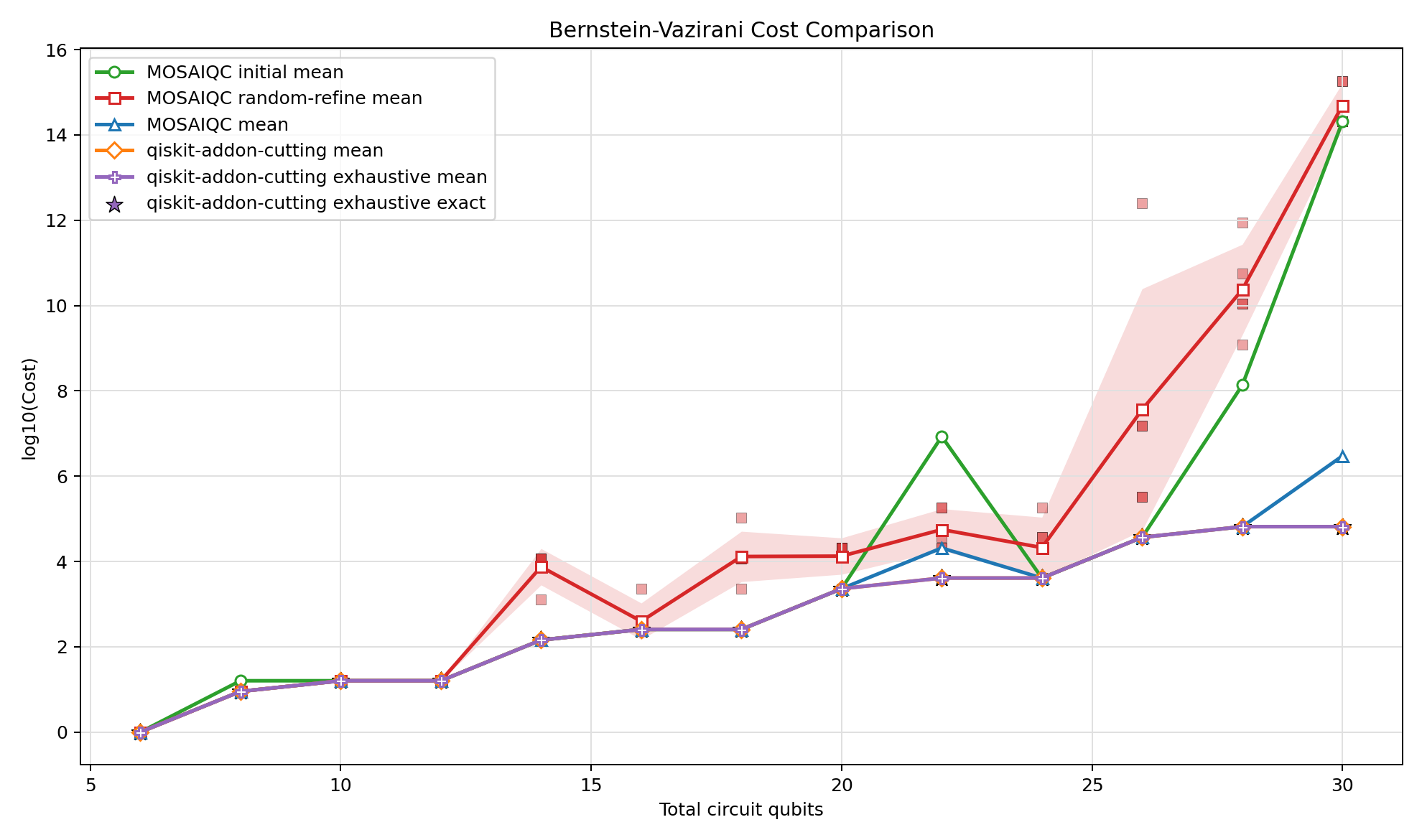}
\caption{BV}
\label{fig:b_o_s}
\end{subfigure}
\begin{subfigure}{0.49\textwidth}
\includegraphics[width=\linewidth]{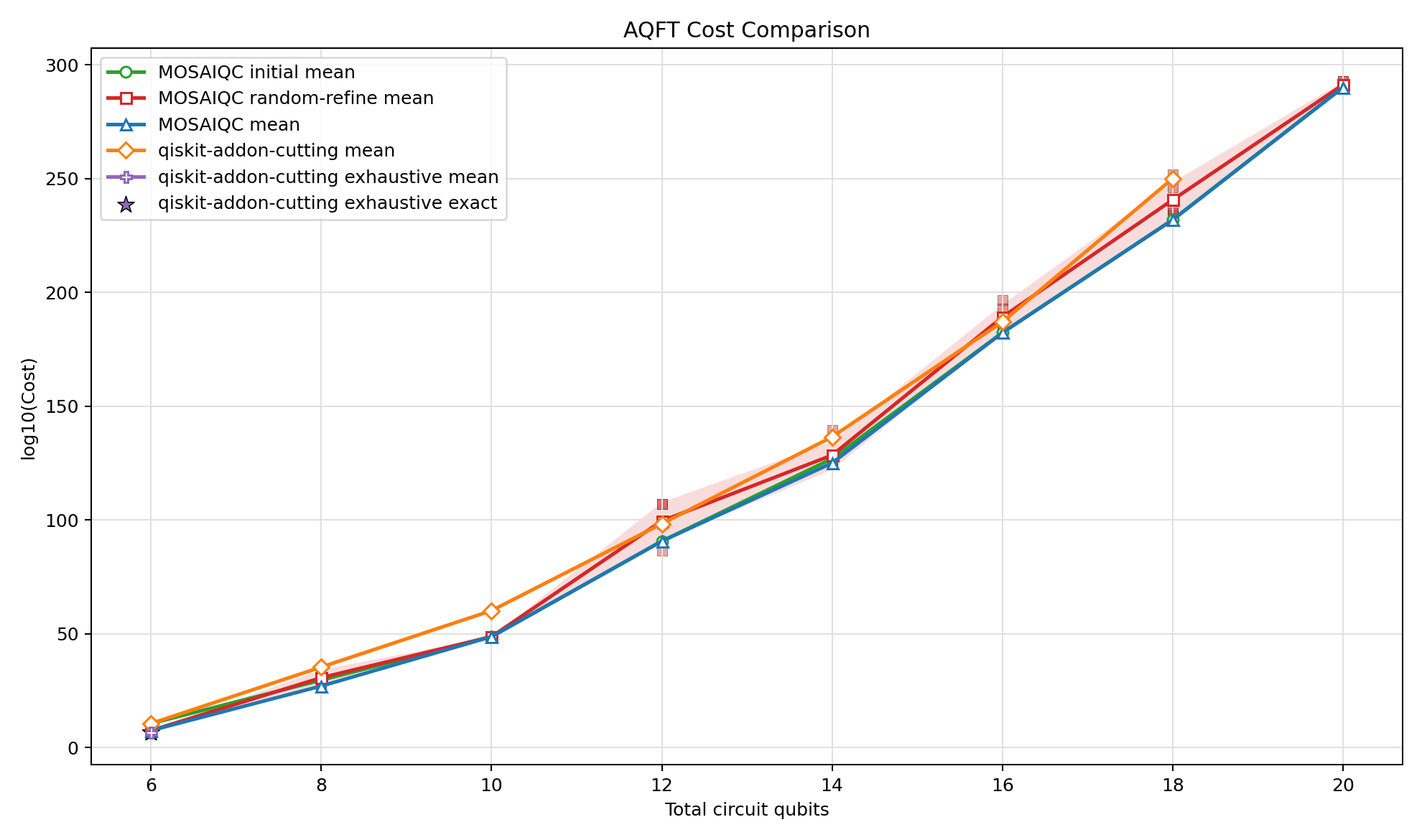}
\caption{AQFT}
\label{fig:c_o_s}
\end{subfigure}
\begin{subfigure}{0.49\textwidth}
\includegraphics[width=\linewidth]{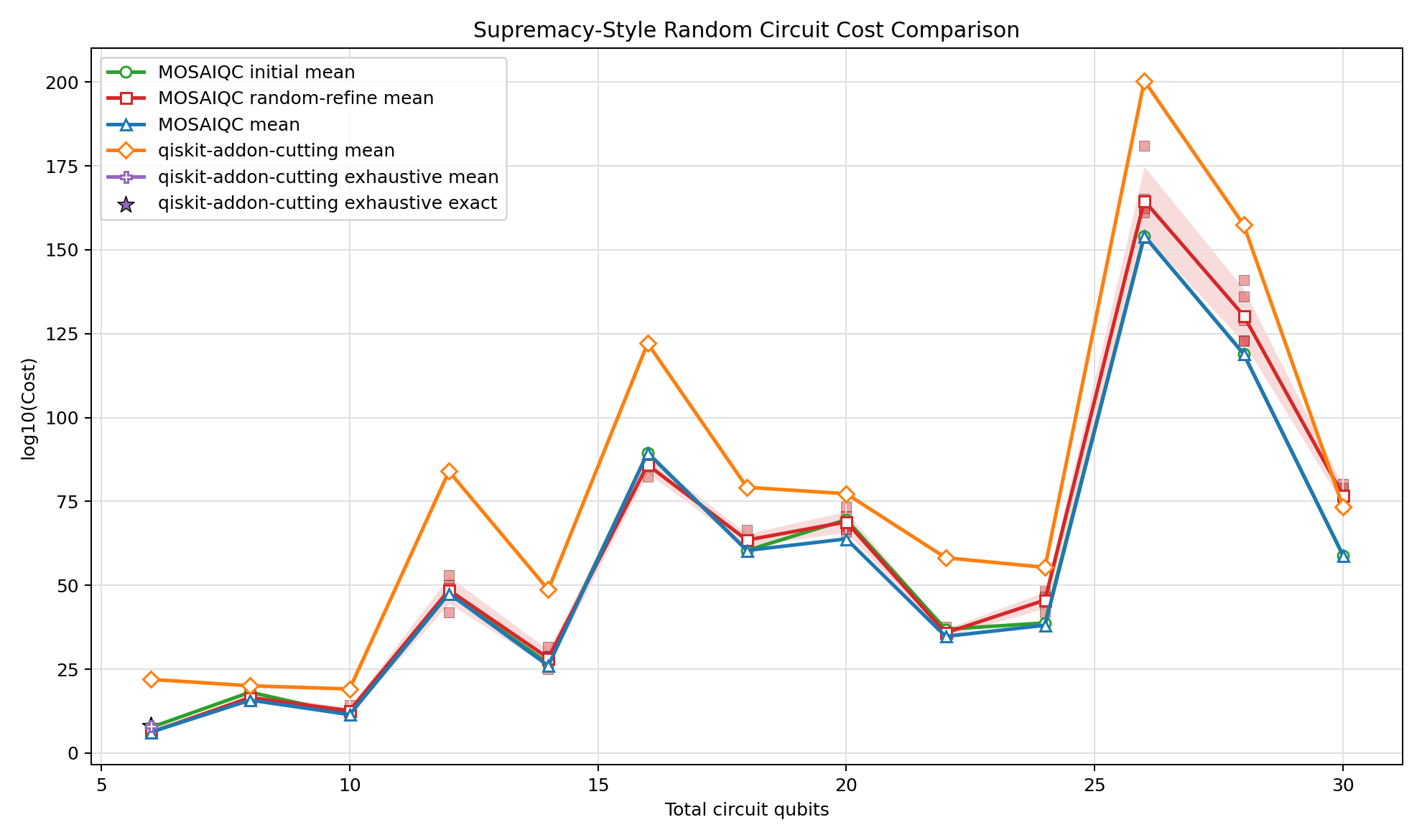}
\caption{Supremacy}
\label{fig:d_o_s}
\end{subfigure}
\begin{subfigure}{0.49\textwidth}
\includegraphics[width=\linewidth]{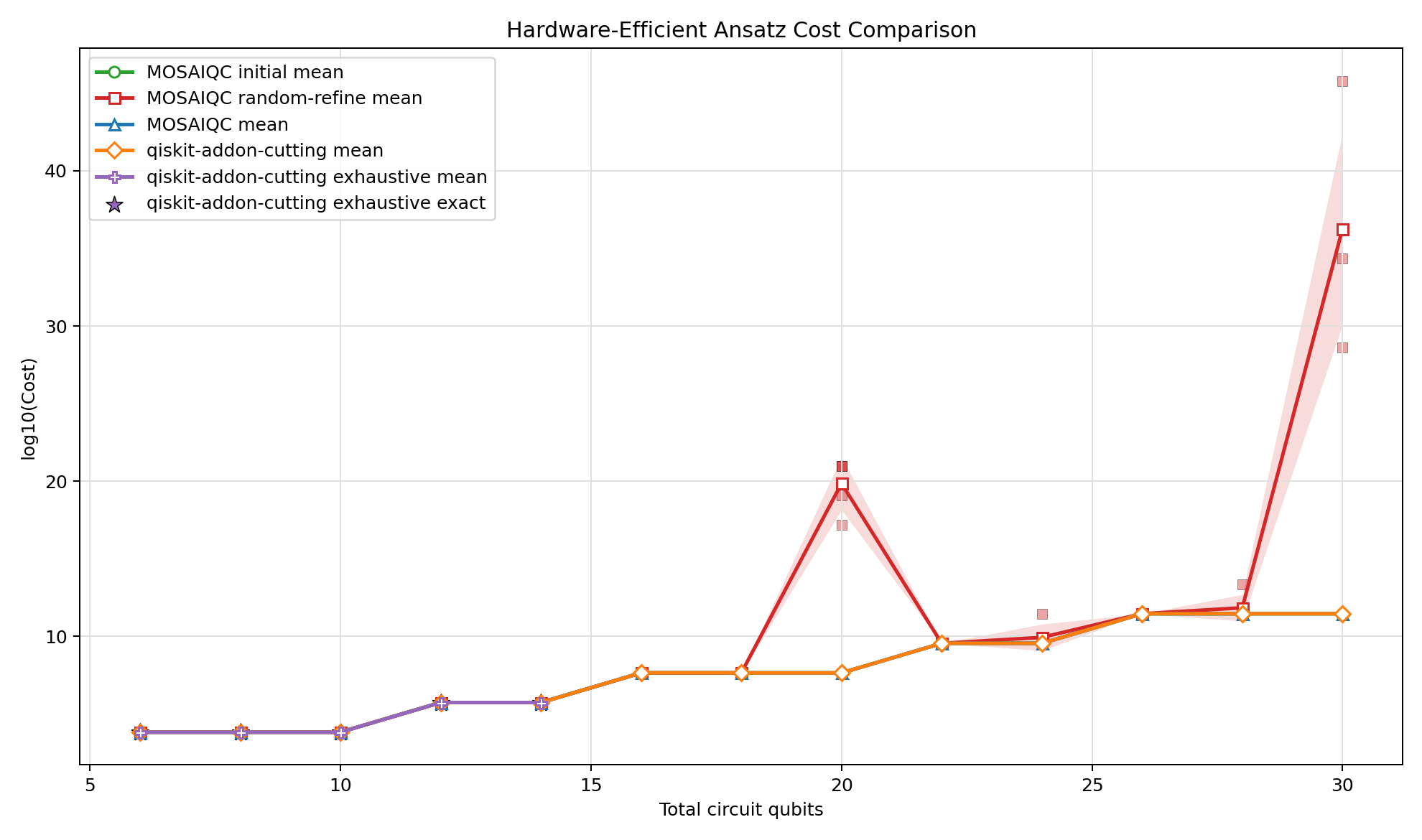}
\caption{HWEA}
\label{fig:e_o_s}
\end{subfigure}
\begin{subfigure}{0.49\textwidth}
\includegraphics[width=\linewidth]{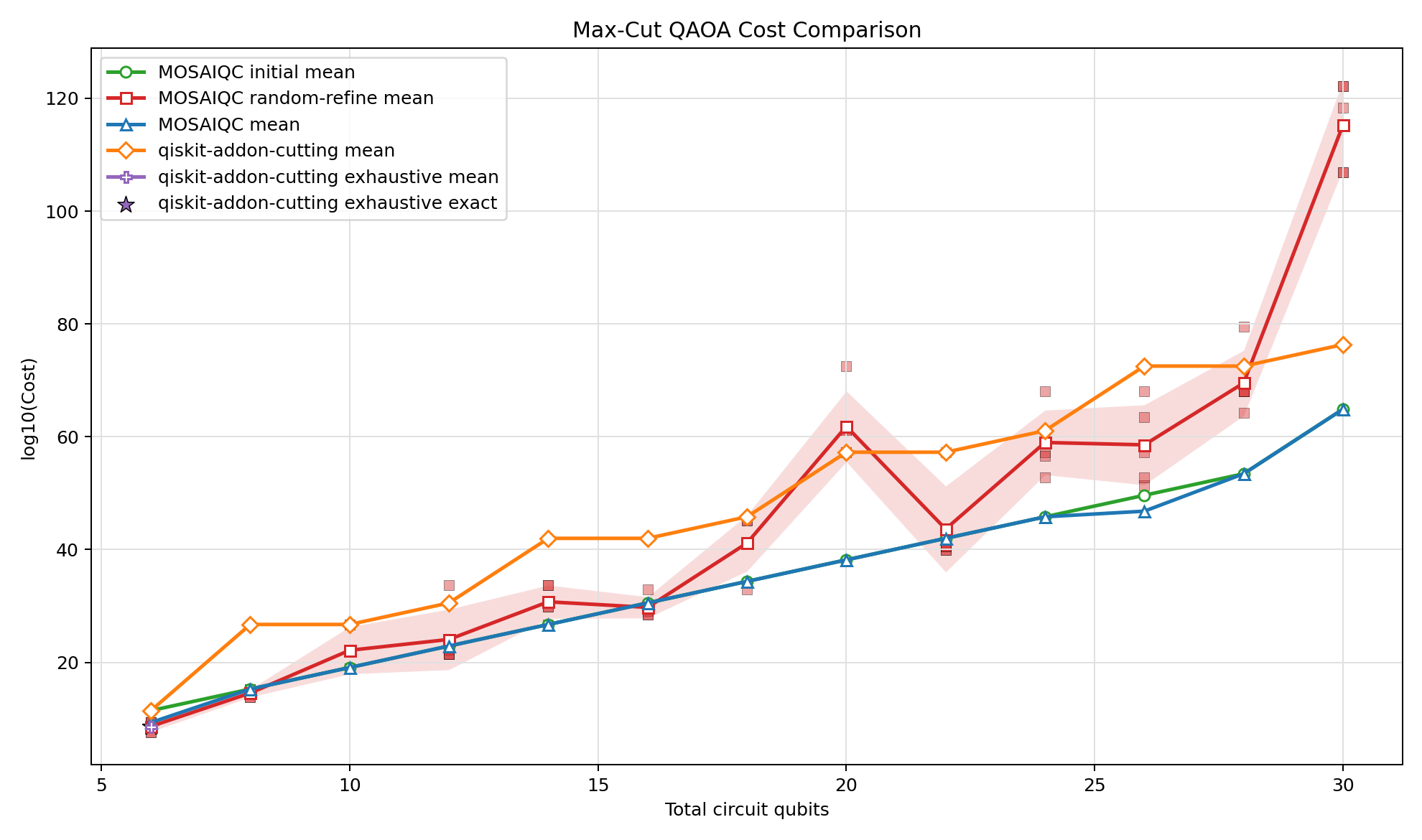}
\caption{QAOA MCP}
\label{fig:f_o_s}
\end{subfigure}
\\
\begin{subfigure}{0.49\textwidth}
\includegraphics[width=\linewidth]{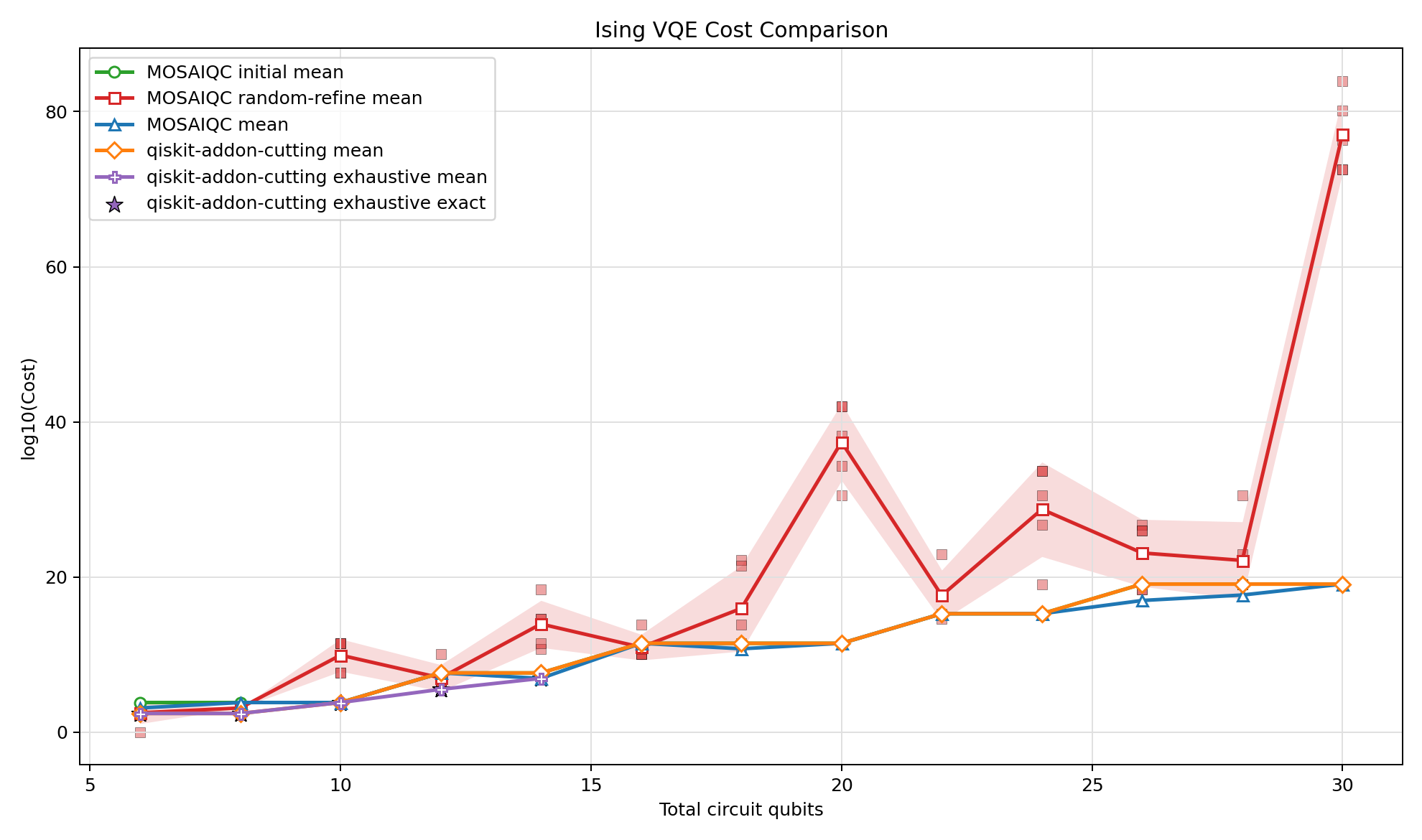}
\caption{Ising VQE HWAE}
\label{fig:g_o_s}
\end{subfigure}
\begin{subfigure}{0.49\textwidth}
\includegraphics[width=\linewidth]{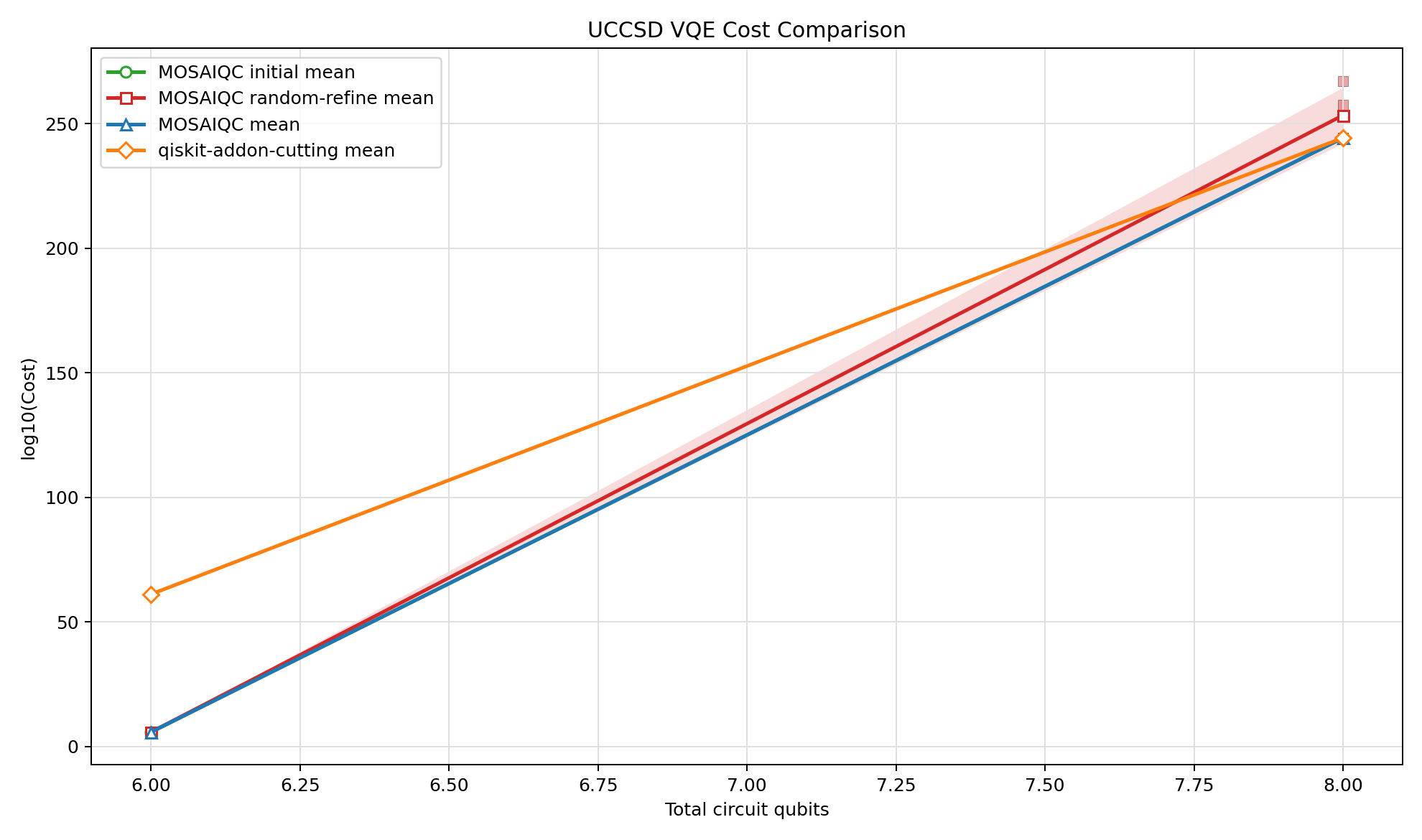}
\caption{Ising VQE UCCSD}
\label{fig:h_o_s}
\end{subfigure}

\end{minipage}
\caption{Sampling overhead results for 6-30 qubit instances for 2-qubit increments.}
\label{fig:sampling_s}
\end{figure}

\begin{figure}[h]
\centering
%\setkeys{Gin}{width=\linewidth}
\begin{subfigure}{0.49\textwidth}
\includegraphics[width=\textwidth]{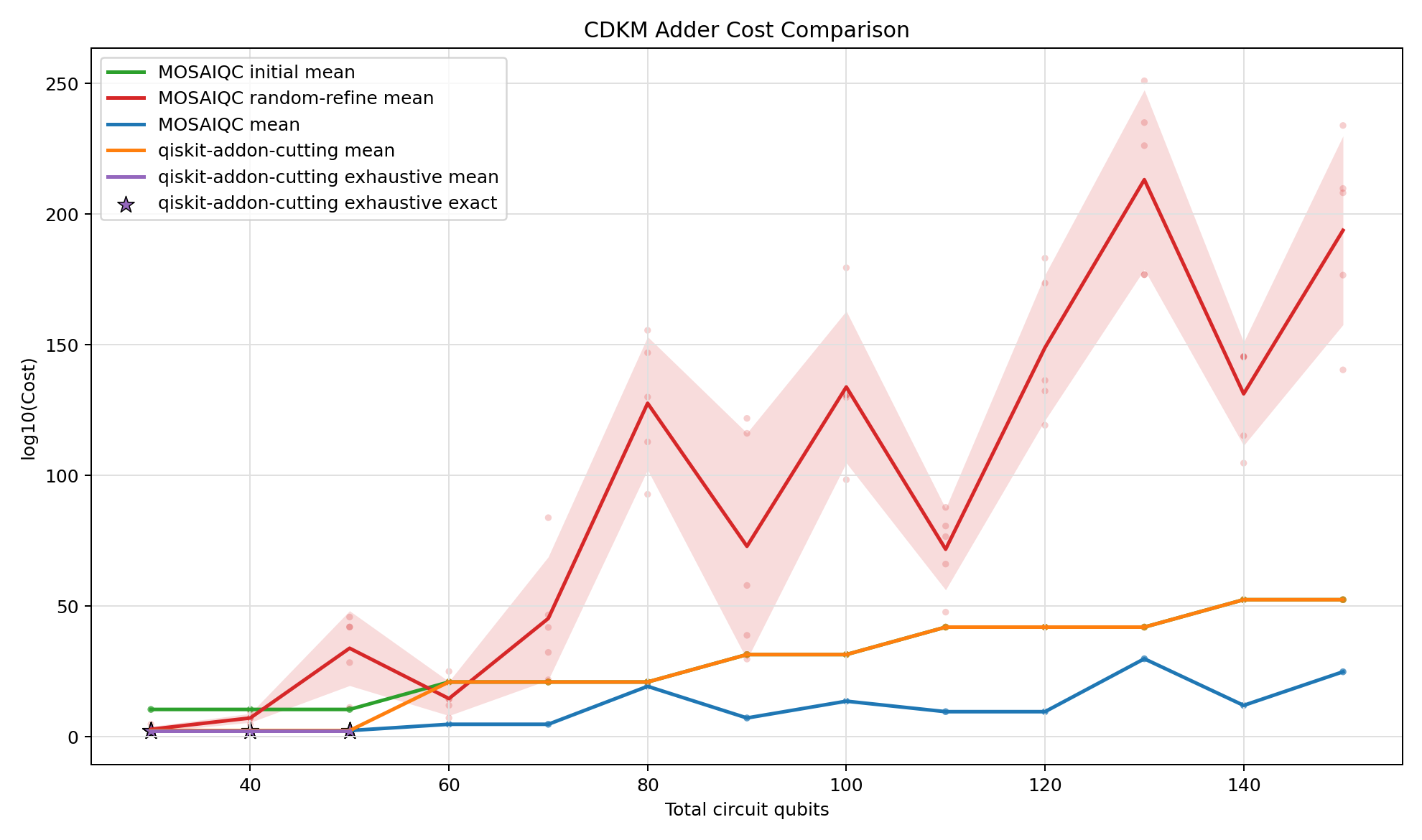}
\caption{Adder}
\label{fig:a_o_m}
\end{subfigure}
\begin{subfigure}{0.49\textwidth}
\includegraphics[width=\linewidth]{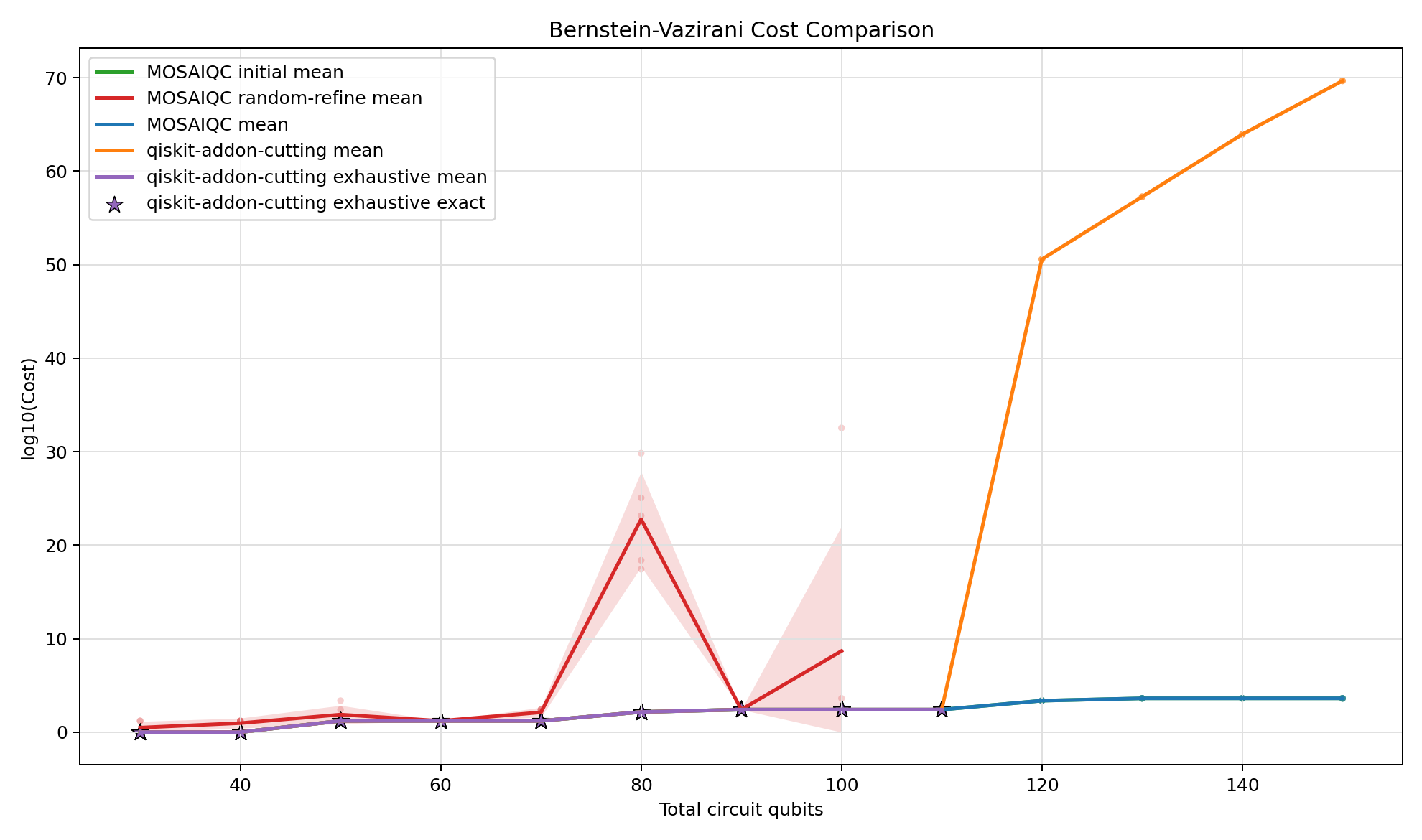}
\caption{BV}
\label{fig:b_o_m}
\end{subfigure}
\begin{subfigure}{0.49\textwidth}
\includegraphics[width=\linewidth]{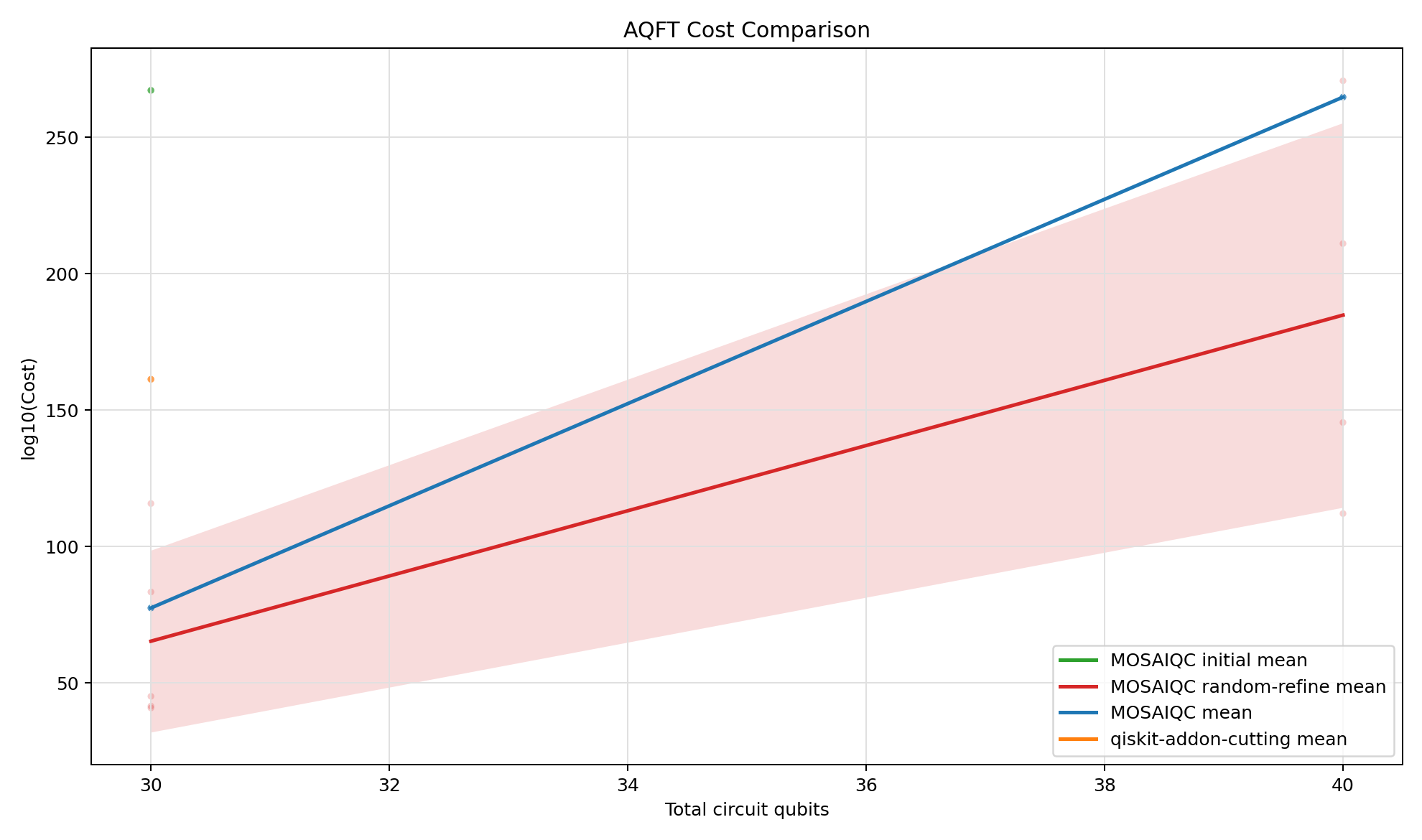}
\caption{AQFT}
\label{fig:c_o_m}
\end{subfigure}
\begin{subfigure}{0.49\textwidth}
\includegraphics[width=\linewidth]{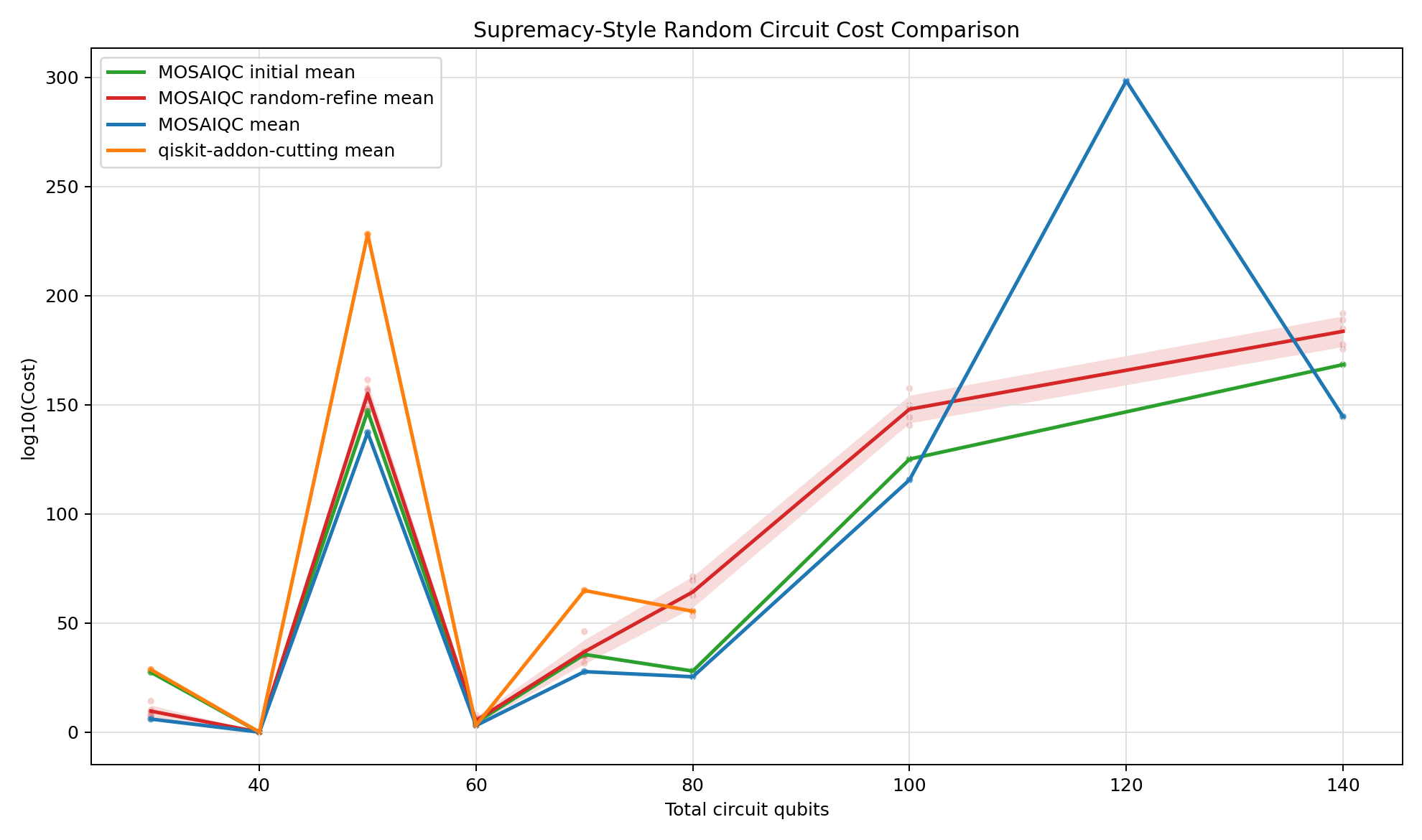}
\caption{Supremacy}
\label{fig:d_o_m}
\end{subfigure}
\begin{subfigure}{0.49\textwidth}
\includegraphics[width=\linewidth]{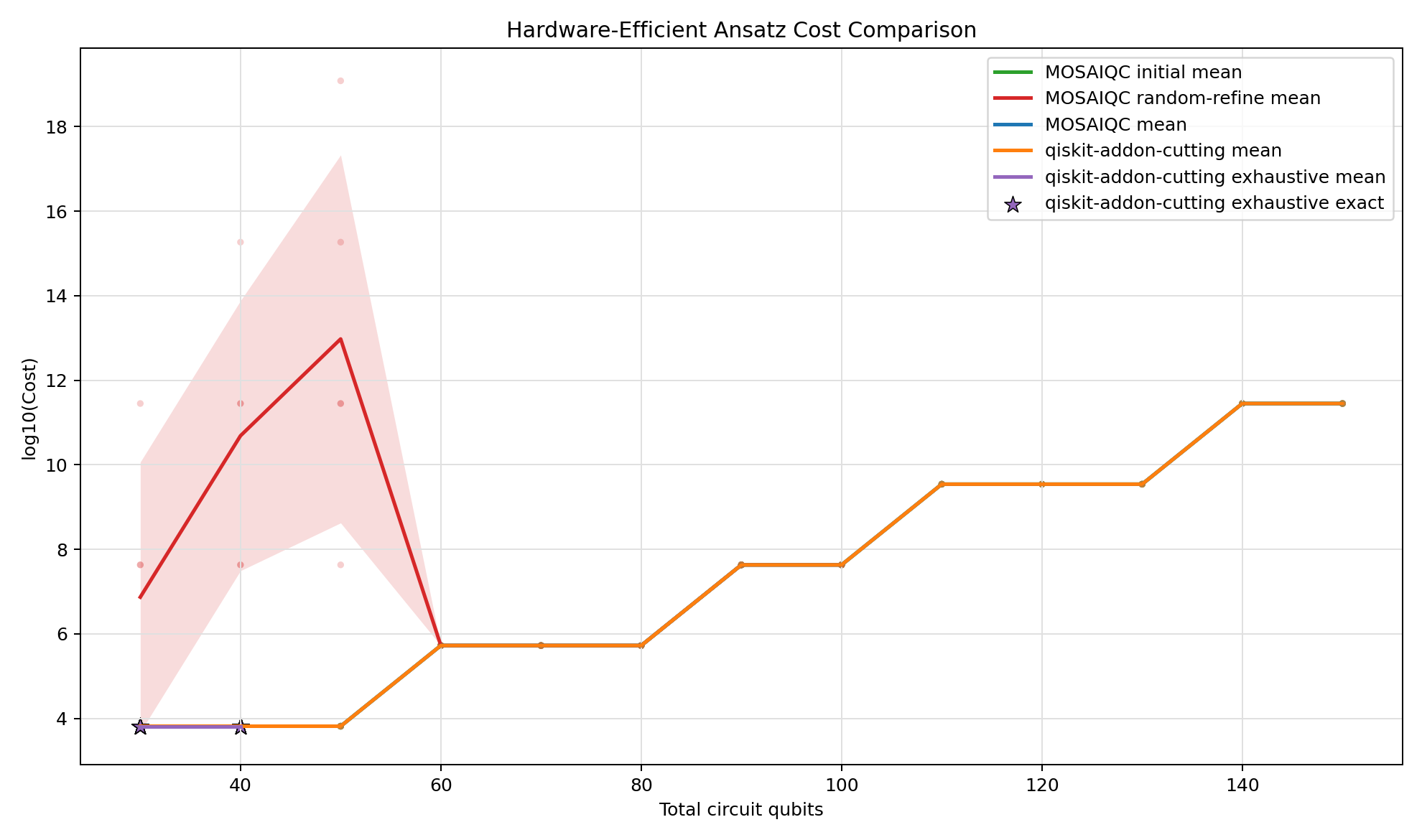}
\caption{HWEA}
\label{fig:e_o_m}
\end{subfigure}
\begin{subfigure}{0.49\textwidth}
\includegraphics[width=\linewidth]{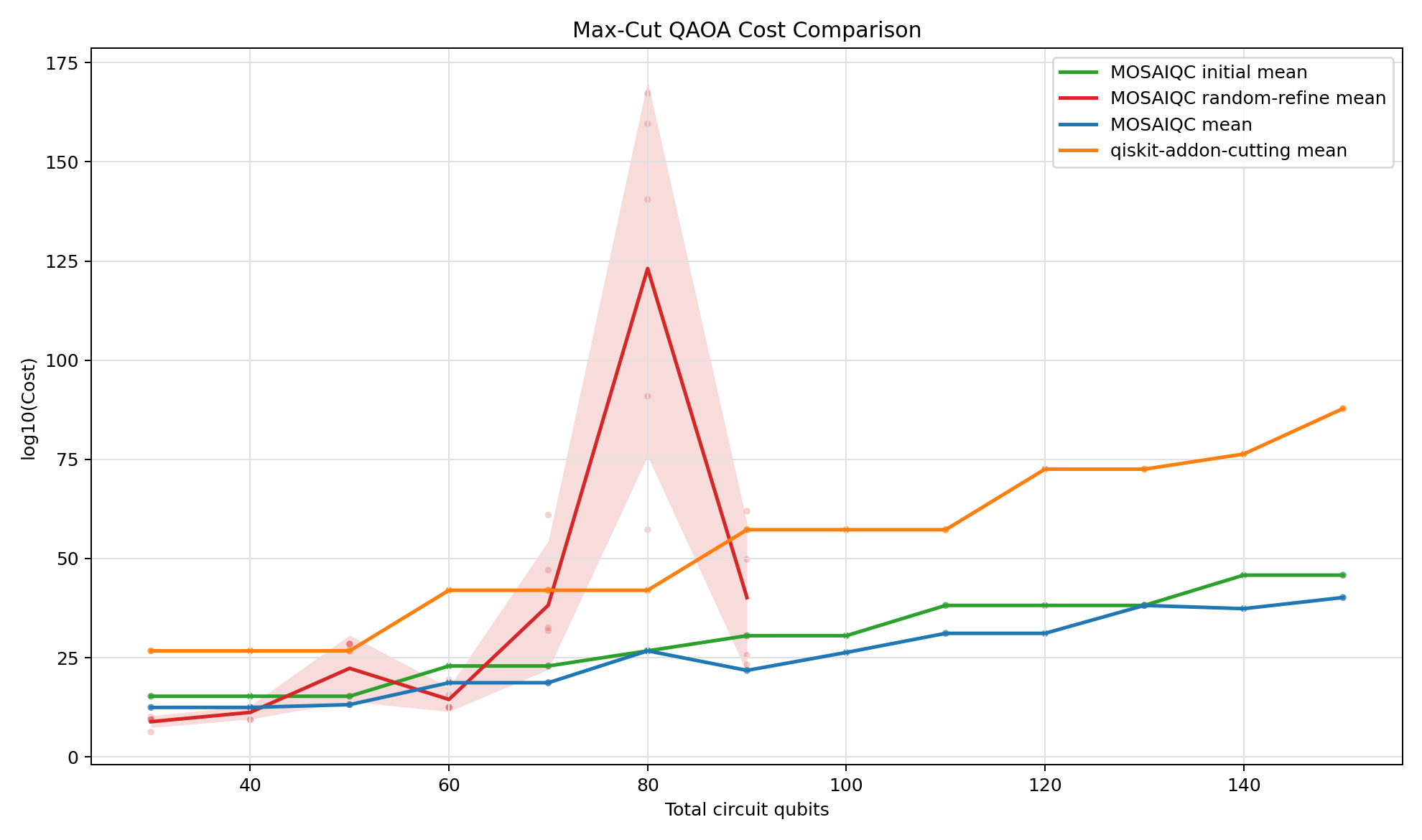}
\caption{QAOA MCP}
\label{fig:f_o_m}
\end{subfigure}
\\
\begin{subfigure}{0.49\textwidth}
\includegraphics[width=\linewidth]{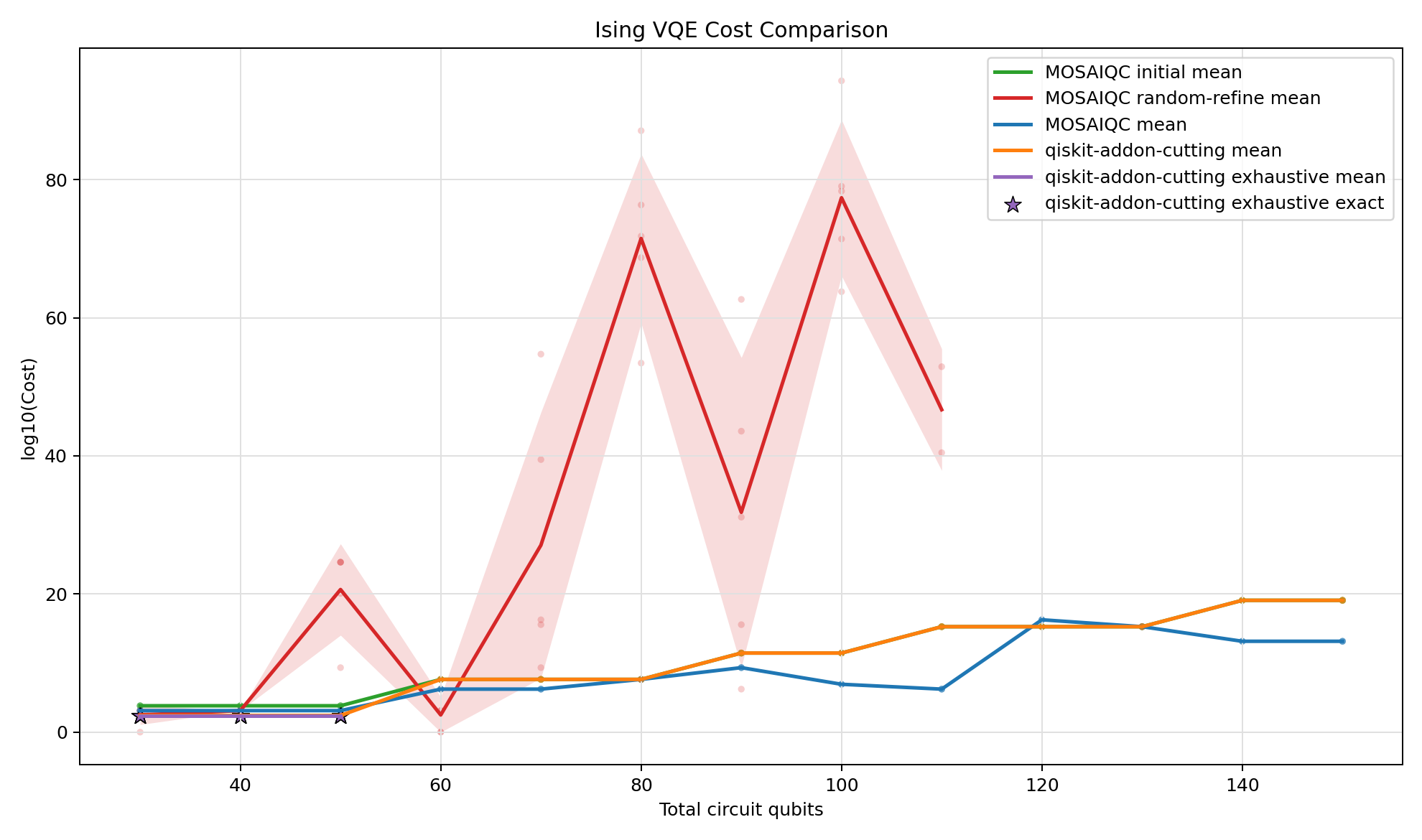}
\caption{Ising VQE HWAE}
\label{fig:g_o_m}
\end{subfigure}

\caption{Sampling overhead results for 30-150 qubit instances for 10 qubit increments.}
\label{fig:sampling_m}
\end{figure}

\begin{figure}[h]
\centering
%\setkeys{Gin}{width=\linewidth}
\begin{subfigure}{0.49\textwidth}
\includegraphics[width=\textwidth]{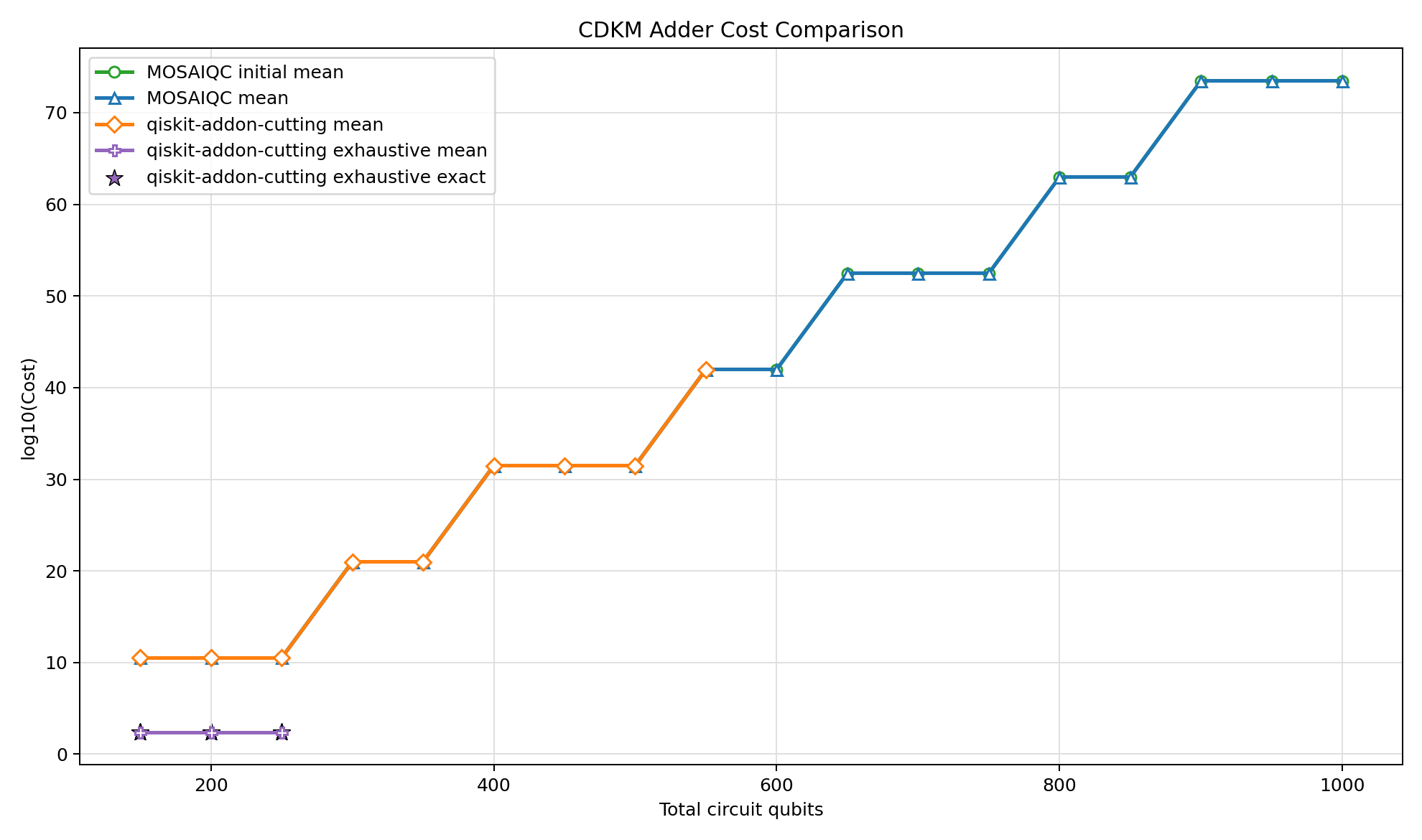}
\caption{Adder}
\label{fig:a_o_l}
\end{subfigure}
\begin{subfigure}{0.49\textwidth}
\includegraphics[width=\linewidth]{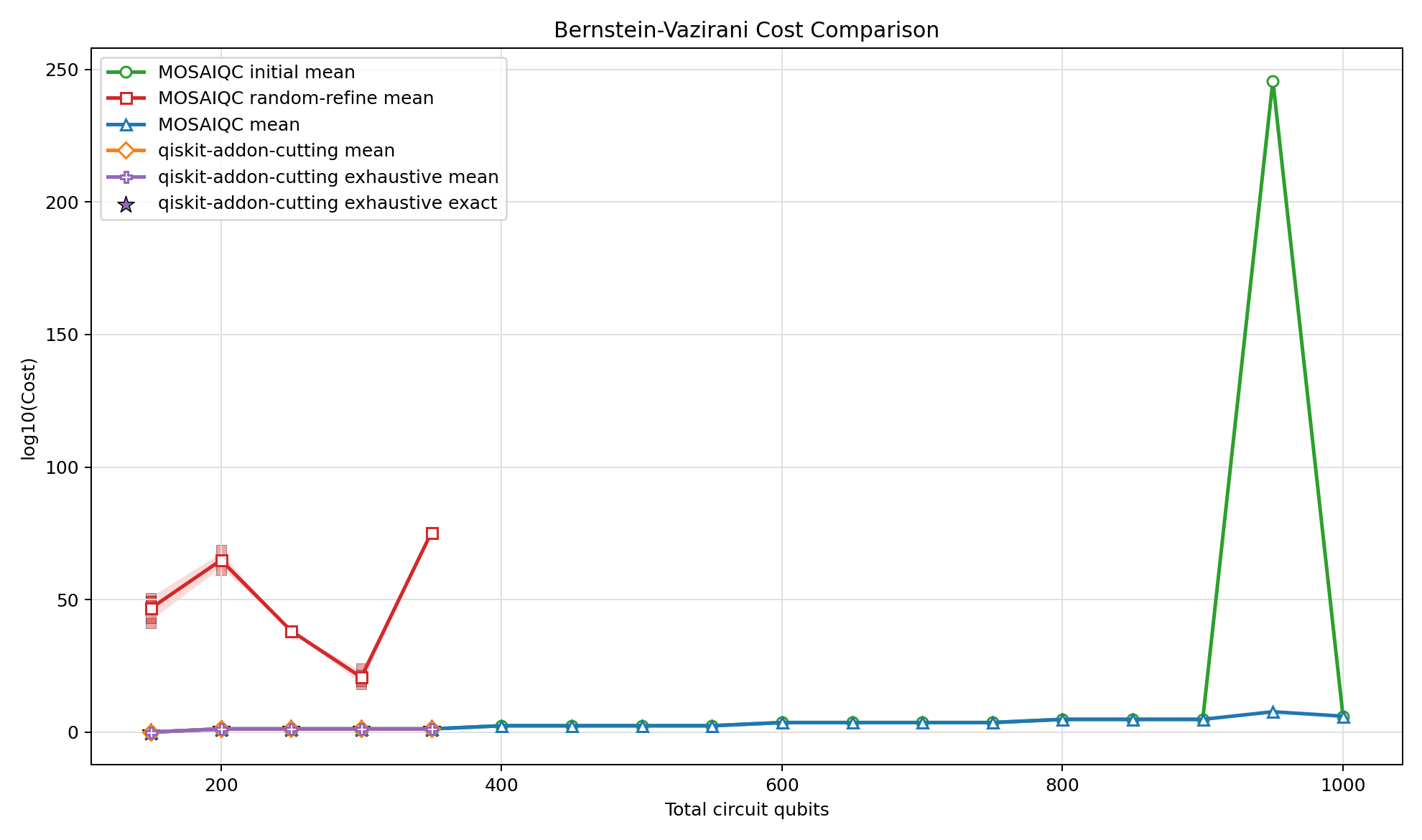}
\caption{BV}
\label{fig:b_o_l}
\end{subfigure}
\begin{subfigure}{0.49\textwidth}
\includegraphics[width=\linewidth]{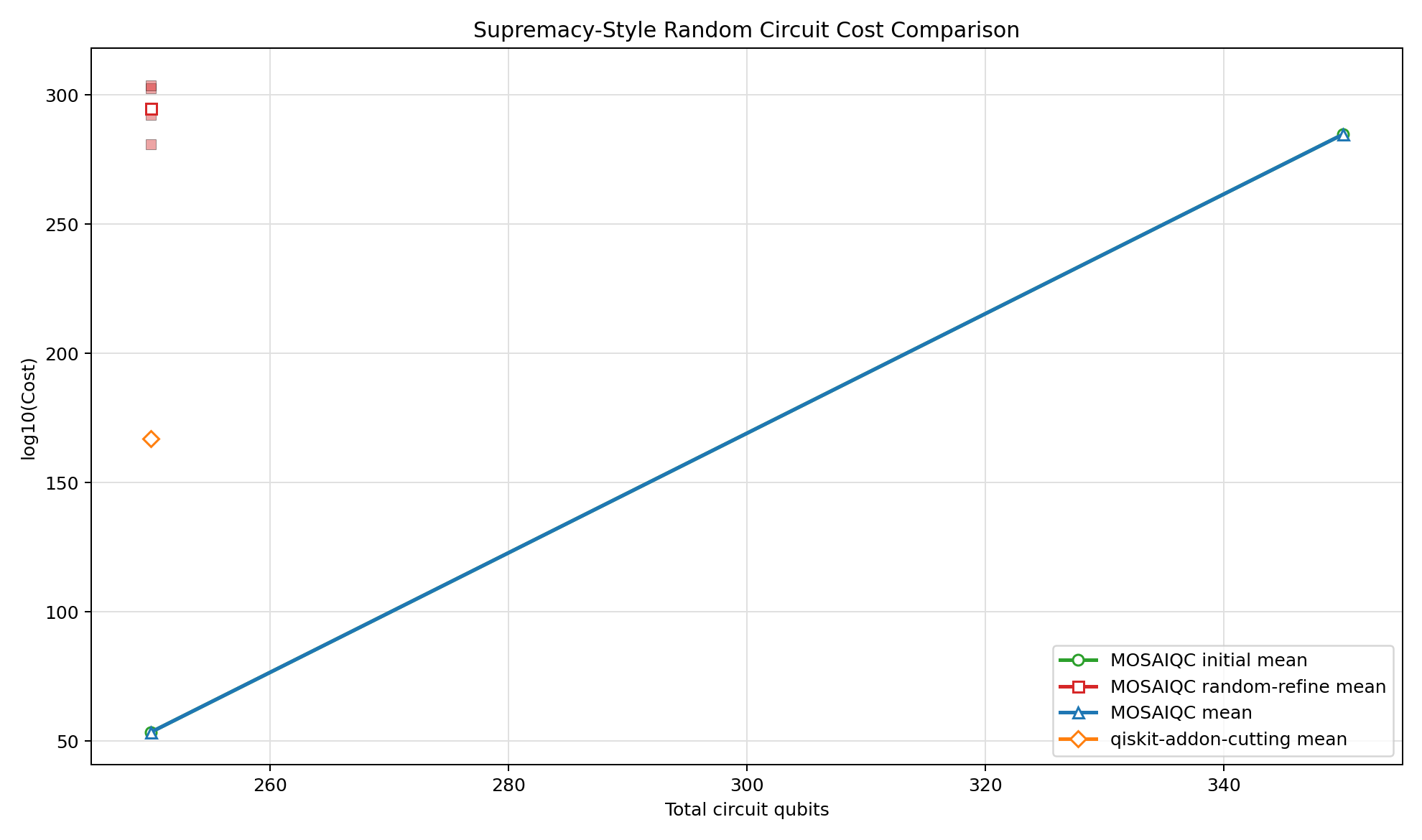}
\caption{Supremacy}
\label{fig:d_o_l}
\end{subfigure}
\begin{subfigure}{0.49\textwidth}
\includegraphics[width=\linewidth]{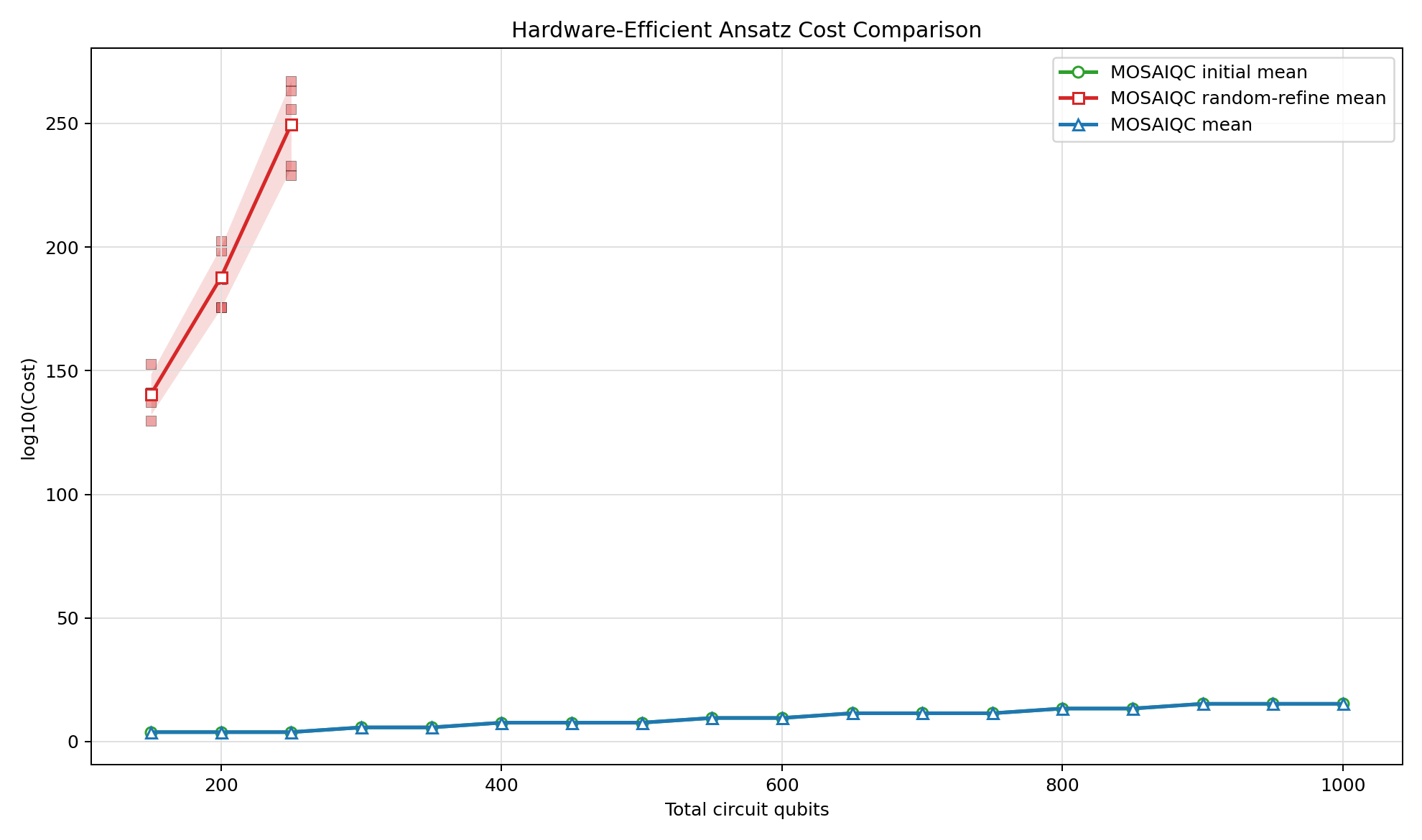}
\caption{HWEA}
\label{fig:e_o_l}
\end{subfigure}
\begin{subfigure}{0.49\textwidth}
\includegraphics[width=\linewidth]{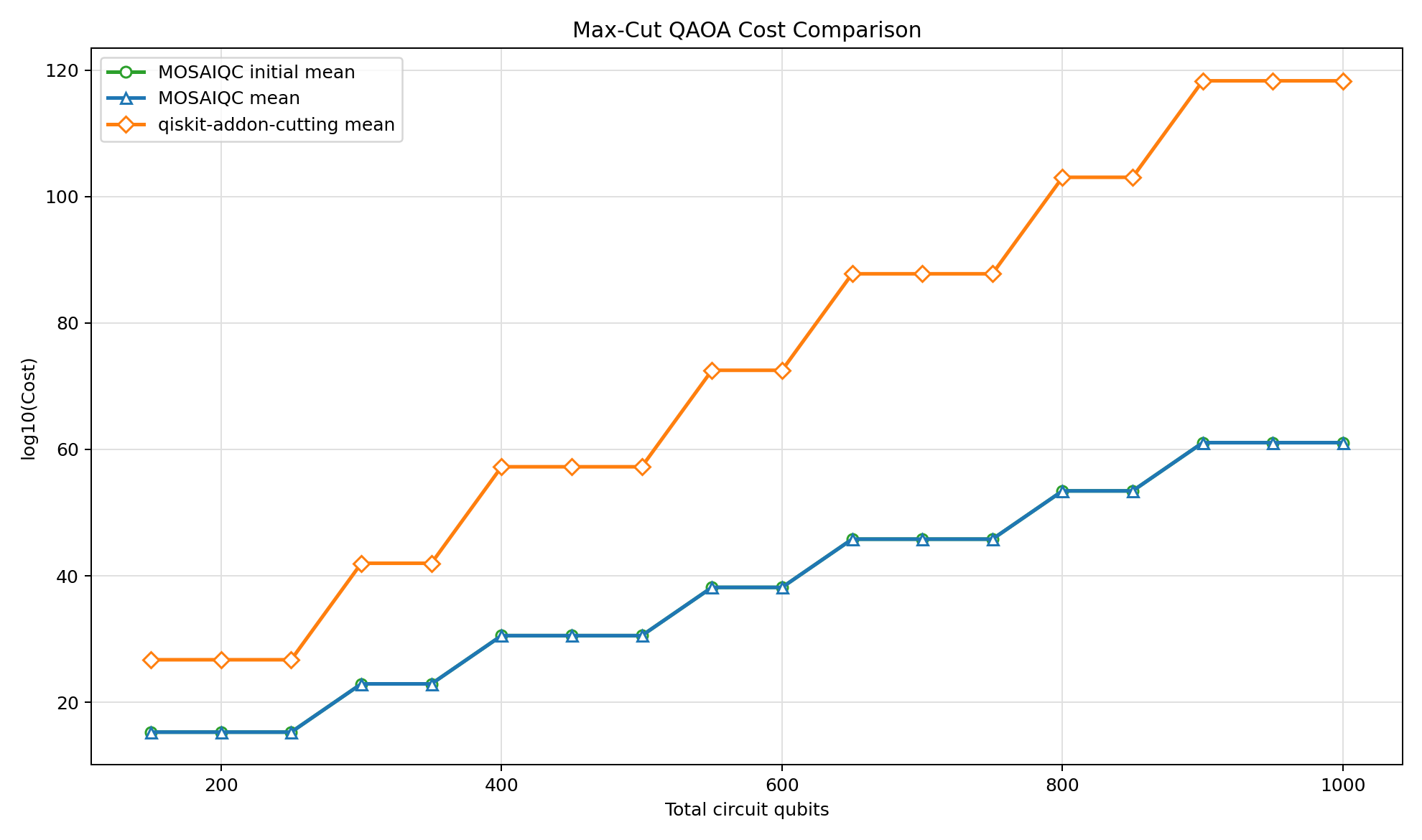}
\caption{QAOA MCP}
\label{fig:f_o_l}
\end{subfigure}
\begin{subfigure}{0.49\textwidth}
\includegraphics[width=\linewidth]{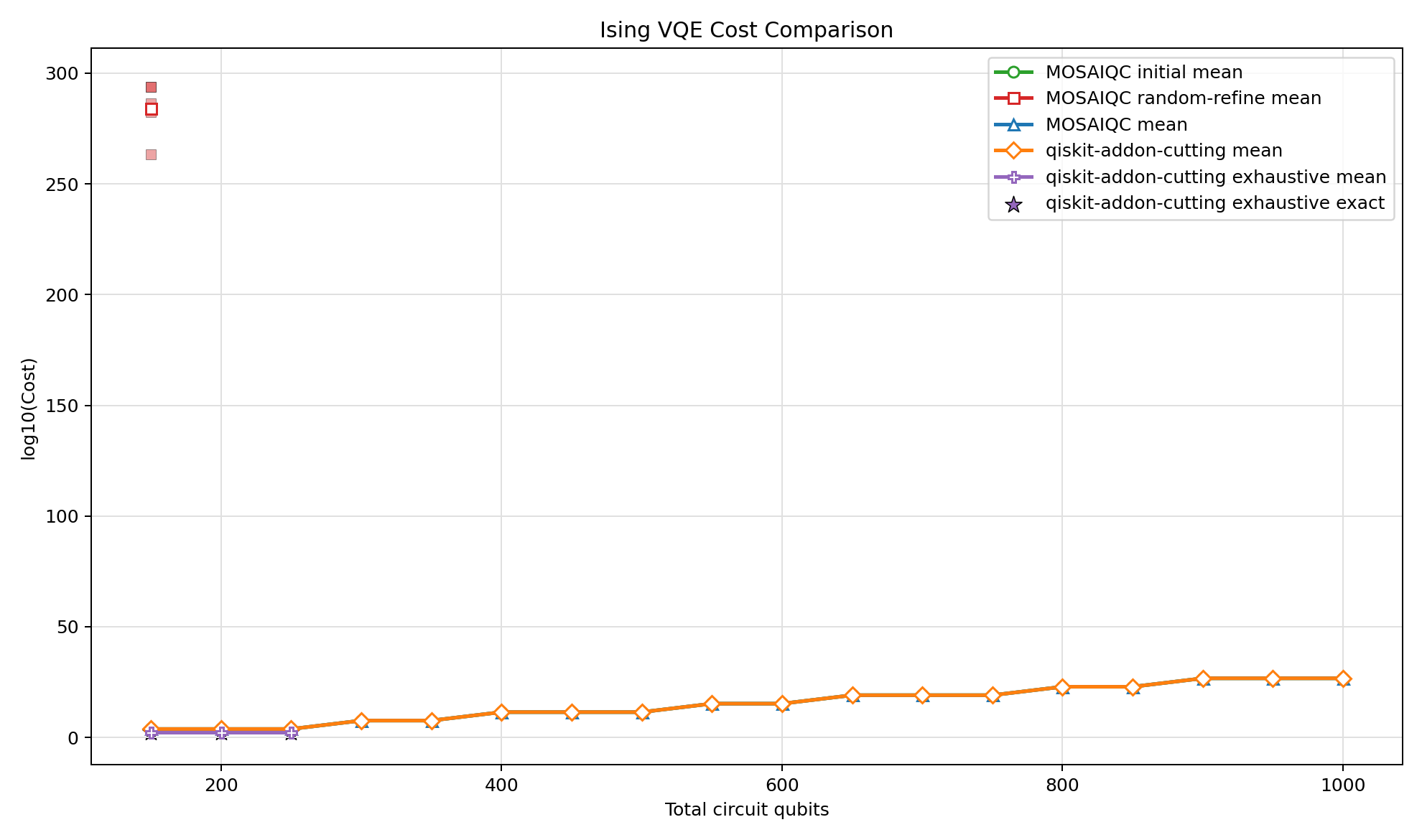}
\caption{Ising VQE HWAE}
\label{fig:g_o_l}
\end{subfigure}
\caption{Sampling overhead results for 150-1000 qubit instances for 50 qubit increments.}
\label{fig:sampling_l}
\end{figure}

%% file: Sections/WireCut_alg.tex
\begin{algorithm}[h]
\caption{Measurement overhead cost for wire cuts}
\begin{algorithmic}[1]
    \Require Array $A$ of two-qubit gates with associated costs and targets
    \Require Maximum number of allowed wire cuts, $\mathit{num\_allowed\_wire\_cuts}$
    \Require Wire-cut cost, $\mathit{wire\_cut\_cost}$
    \Ensure Sets of wire cuts and gate cuts

    \State $\mathit{wire\_cuts} \gets \emptyset$
    \State $\mathit{gate\_cuts} \gets \emptyset$

    \State $\mathit{placed\_wire\_cuts} \gets \Call{EvaluateWireCutPlacement}{A}$

    \While{$\mathit{placed\_wire\_cuts} > \mathit{num\_allowed\_wire\_cuts}$}
        \State $g \gets \Call{MinCostTwoQubitGate}{A}$
        \State $A \gets A \setminus \{g\}$
        \State $\mathit{gate\_cuts} \gets \mathit{gate\_cuts} \cup \{g\}$
        \State $\mathit{placed\_wire\_cuts} \gets \Call{EvaluateWireCutPlacement}{A}$
    \EndWhile

    \While{$A \neq \emptyset$}
        \State $g \gets \Call{MinCostTwoQubitGate}{A}$
        \If{$\Call{Cost}{g} < \mathit{wire\_cut\_cost}$}
            \State $A \gets A \setminus \{g\}$
            \State $\mathit{gate\_cuts} \gets \mathit{gate\_cuts} \cup \{g\}$
        \Else
            \State \textbf{break}
        \EndIf
    \EndWhile

    \For{each gate $g \in A$}
        \If{$\Call{PreviousTarget}{g} \neq \Call{CurrentTarget}{g}$}
            \State $\mathit{wire\_cuts} \gets \mathit{wire\_cuts} \cup \{g\}$
        \EndIf
    \EndFor

    \State \Return $\mathit{wire\_cuts}, \mathit{gate\_cuts}$
\end{algorithmic}
\end{algorithm}

%% file: bib.bib
@article{Basu2024FragQC,
  title   = {FragQC: An Efficient Quantum Error Reduction Technique using Quantum Circuit Fragmentation},
  author  = {Basu, Saikat and Das, Arnav and Saha, Amit and Chakrabarti, Amlan and Sur-Kolay, Susmita},
  journal = {Journal of Systems and Software},
  volume  = {214},
  pages   = {112085},
  year    = {2024},
  doi     = {10.1016/j.jss.2024.112085}
}

@article{Bhoumik2023DistributedScheduling,
  title   = {Distributed Scheduling of Quantum Circuits with Noise and Time Optimization},
  author  = {Bhoumik, Debasmita and Majumdar, Ritajit and Saha, Amit and Sur-Kolay, Susmita},
  journal = {arXiv preprint arXiv:2309.06005},
  year    = {2023},
  doi     = {10.48550/arXiv.2309.06005}
}

@article{Du2024DisMap,
  title   = {Hardware-aware Circuit Cutting and Distributed Qubit Mapping for Connected Quantum Systems},
  author  = {Du, Zefan and Li, Yanni and Mo, Zijian and Wei, Wenqi and Chen, Juntao and Buyya, Rajkumar and Mao, Ying},
  journal = {arXiv preprint arXiv:2412.18458},
  year    = {2024},
  doi     = {10.48550/arXiv.2412.18458}
}

@inproceedings{Marchisio2025CuttingAllYouNeed,
  title     = {Cutting is All You Need: Execution of Large-Scale Quantum Neural Networks on Limited-Qubit Devices},
  author    = {Marchisio, Alberto and Sychiuco, Emman and Kashif, Muhammad and Shafique, Muhammad},
  booktitle = {2025 IEEE International Conference on Quantum Artificial Intelligence (QAI)},
  pages     = {330--336},
  year      = {2025}
}

@inproceedings{Ren2024HardwareAwareGateCutting,
  title     = {A Hardware-Aware Gate Cutting Framework for Practical Quantum Circuit Knitting},
  author    = {Ren, Xiangyu and Zhang, Mengyu and Barbalace, Antonio},
  booktitle = {Proceedings of the 2024 IEEE/ACM International Conference on Computer-Aided Design (ICCAD)},
  year      = {2024},
  doi       = {10.1145/3676536.3676719}
}

@article{Sahu2025DevQCC,
  title   = {DevQCC: Device-Aware Quantum Circuit Cutting Framework with Applications in Quantum Machine Learning},
  author  = {Sahu, Himanshu and Gupta, Hari Prabhat and Puvvada, Vishnu Vardhan and others},
  journal = {Quantum Machine Intelligence},
  volume  = {7},
  pages   = {89},
  year    = {2025},
  doi     = {10.1007/s42484-025-00313-0}
}

@article{Yang2024ScalabilityCircuitCutting,
  title   = {Understanding the Scalability of Circuit Cutting Techniques for Practical Quantum Applications},
  author  = {Yang, Songqinghao and Murali, Prakash},
  journal = {arXiv preprint arXiv:2411.17756},
  year    = {2024},
  doi     = {10.48550/arXiv.2411.17756}
}

@article{Brandhofer2024OptimalPartitioning,
  title   = {Optimal Partitioning of Quantum Circuits using Gate Cuts and Wire Cuts},
  author  = {Brandhofer, Sebastian and Polian, Ilia and Krsulich, Kevin},
  journal = {IEEE Transactions on Quantum Engineering},
  volume  = {5},
  pages   = {1--10},
  year    = {2024},
  doi     = {10.1109/TQE.2023.3347106}
}

@inproceedings{Tang2021CutQC,
  title     = {CutQC: Using Small Quantum Computers for Large Quantum Circuit Evaluations},
  author    = {Tang, Wei and Tomesh, Teague and Suchara, Martin and Larson, Jeffrey and Martonosi, Margaret},
  booktitle = {Proceedings of the 26th ACM International Conference on Architectural Support for Programming Languages and Operating Systems (ASPLOS '21)},
  pages     = {473--486},
  year      = {2021},
  doi       = {10.1145/3445814.3446758}
}

@article{Tang2022ScaleQC,
  title   = {ScaleQC: A Scalable Framework for Hybrid Computation on Quantum and Classical Processors},
  author  = {Tang, Wei and Martonosi, Margaret},
  journal = {arXiv preprint arXiv:2207.00933},
  year    = {2022},
  doi     = {10.48550/arXiv.2207.00933}
}

@article{Cambiucci2025SpatialTemporalCutting,
  title   = {Spatial and Temporal Circuit Cutting with Hypergraphic Partitioning},
  author  = {Cambiucci, Waldemir and Silveira, Regina Melo and Ruggiero, Wilson Vicente},
  journal = {arXiv preprint arXiv:2504.09334},
  year    = {2025},
  doi     = {10.48550/arXiv.2504.09334}
}

@article{Dou2025LarQucut,
  title   = {LarQucut: A New Cutting and Mapping Approach for Large-sized Quantum Circuits in Distributed Quantum Computing (DQC) Environments},
  author  = {Dou, Xinglei and Liu, Lei and Wang, Zhuohao and Li, Pengyu},
  journal = {arXiv preprint arXiv:2502.21000},
  year    = {2025},
  doi     = {10.48550/arXiv.2502.21000}
}

@article{Kan2024FitCut,
  title   = {Scalable Circuit Cutting and Scheduling in a Resource-Constrained and Distributed Quantum System},
  author  = {Kan, Shuwen and Du, Zefan and Palma, Miguel and Stein, Samuel A. and Liu, Chenxu and Wei, Wenqi and Chen, Juntao and Li, Ang and Mao, Ying},
  journal = {arXiv preprint arXiv:2405.04514},
  year    = {2024},
  doi     = {10.48550/arXiv.2405.04514}
}

@inproceedings{Smith2023SuperSim,
  title     = {Clifford-based Circuit Cutting for Quantum Simulation},
  author    = {Smith, Kaitlin N. and Perlin, Michael A. and Gokhale, Pranav and Frederick, Paige and Owusu-Antwi, David and Rines, Richard and Omole, Victory and Chong, Frederic T.},
  booktitle = {Proceedings of the 2023 ACM/IEEE International Symposium on Computer Architecture (ISCA)},
  year      = {2023},
  doi       = {10.1145/3579371.3589352}
}

@article{Tejedor2025Qdislib,
  title   = {Distributed Quantum Circuit Cutting for Hybrid Quantum-Classical High-Performance Computing},
  author  = {Tejedor, Mar and Casas, Berta and Conejero, Javier and Cervera-Lierta, Alba and Badia, Rosa M.},
  journal = {arXiv preprint arXiv:2505.01184},
  year    = {2025},
  doi     = {10.48550/arXiv.2505.01184}
}

@article{Frank1956AnAF,
  title={An algorithm for quadratic programming},
  author={Marguerite Frank and Philip Wolfe},
  journal={Naval Research Logistics Quarterly},
  year={1956},
  volume={3},
  pages={95-110},
  url={https://api.semanticscholar.org/CorpusID:122654717}
}

@inproceedings{jaggi2013revisiting,
  title={Revisiting Frank-Wolfe: Projection-free sparse convex optimization},
  author={Jaggi, Martin},
  booktitle={International conference on machine learning},
  pages={427--435},
  year={2013},
  organization={PMLR}
}

@article{glover1989tabu,
  title={Tabu search—part I},
  author={Glover, Fred},
  journal={ORSA Journal on computing},
  volume={1},
  number={3},
  pages={190--206},
  year={1989},
  publisher={Informs}
}

@article{karypis1997metis,
  title={METIS: A software package for partitioning unstructured graphs, partitioning meshes, and computing fill-reducing orderings of sparse matrices},
  author={Karypis, George and Kumar, Vipin},
  year={1997}
}

@misc{pybind11,
  author       = {Wenzel Jakob and Jason Rhinelander and Dean Moldovan},
  title        = {pybind11 -- Seamless operability between C++11 and Python},
  year         = {2017},
  note         = {https://github.com/pybind/pybind11}
}

@misc{javadiabhari2024quantumcomputingqiskit,
      title={Quantum computing with Qiskit}, 
      author={Ali Javadi-Abhari and Matthew Treinish and Kevin Krsulich and Christopher J. Wood and Jake Lishman and Julien Gacon and Simon Martiel and Paul D. Nation and Lev S. Bishop and Andrew W. Cross and Blake R. Johnson and Jay M. Gambetta},
      year={2024},
      eprint={2405.08810},
      archivePrefix={arXiv},
      primaryClass={quant-ph},
      url={https://arxiv.org/abs/2405.08810}, 
}

@article{ayral2021quantum,
  title={Quantum divide and compute: exploring the effect of different noise sources},
  author={Ayral, Thomas and R{\'e}gent, Fran{\c{c}}ois-Marie Le and Saleem, Zain and Alexeev, Yuri and Suchara, Martin},
  journal={SN Computer Science},
  volume={2},
  number={3},
  pages={132},
  year={2021},
  publisher={Springer}
}

@misc{harada2025exponentialtopolynomialscalingmeasurementoverhead,
      title={Exponential-to-polynomial scaling of measurement overhead in circuit knitting via quantum tomography}, 
      author={Hiroyuki Harada and Kaito Wada and Naoki Yamamoto and Suguru Endo},
      year={2025},
      eprint={2512.19623},
      archivePrefix={arXiv},
      primaryClass={quant-ph},
      url={https://arxiv.org/abs/2512.19623}, 
}

@article{nakamura2025improved,
  title={Improved sampling bounds and scalable partitioning for quantum circuit cutting beyond bipartitions},
  author={Nakamura, Junya and Satoh, Takahiko and Sanji, Shinichiro},
  journal={Physical Review A},
  volume={112},
  number={4},
  pages={042422},
  year={2025},
  publisher={APS}
}

@article{lowe2023fast,
  title={Fast quantum circuit cutting with randomized measurements},
  author={Lowe, Angus and Medvidovi{\'c}, Matija and Hayes, Anthony and O'Riordan, Lee J and Bromley, Thomas R and Arrazola, Juan Miguel and Killoran, Nathan},
  journal={Quantum},
  volume={7},
  pages={934},
  year={2023},
  publisher={Verein zur F{\"o}rderung des Open Access Publizierens in den Quantenwissenschaften}
}

@misc{soloviev2025quantcut,
      title={Scaling Portfolio Diversification with Quantum Circuit Cutting Techniques}, 
      author={Vicente P. Soloviev and Antonio Márquez Romero and Josh Kirsopp and Michal Krompiec},
      year={2025},
      eprint={2506.08947},
      archivePrefix={arXiv},
      primaryClass={quant-ph},
      url={https://arxiv.org/abs/2506.08947}, 
}

@misc{periyasamy2025cutreglossregularizerenhancing,
      title={CutReg: A loss regularizer for enhancing the scalability of QML via adaptive circuit cutting}, 
      author={Maniraman Periyasamy and Christian Ufrecht and Daniel D. Scherer and Wolfgang Mauerer},
      year={2025},
      eprint={2506.14858},
      archivePrefix={arXiv},
      primaryClass={quant-ph},
      url={https://arxiv.org/abs/2506.14858}, 
}

@misc{mohseni2026buildquantumsupercomputerscaling,
      title={How to Build a Quantum Supercomputer: Scaling from Hundreds to Millions of Qubits}, 
      author={Masoud Mohseni and Artur Scherer and K. Grace Johnson and Oded Wertheim and Matthew Otten and Namit Anand and Navid Anjum Aadit and Yuri Alexeev and Gilad Ben-Shach and Kirk M. Bresniker and Kerem Y. Camsari and Barbara Chapman and Soumitra Chatterjee and Shuvro Chowdhury and Gebremedhin A. Dagnew and Tom Dvir and Aniello Esposito and Farah Fahim and Michael Ferguson and Marco Fiorentino and Archit Gajjar and Katerina Gratsea and Gaurav Gyawali and Christian Heiter and Ali H. Z. Kavaki and Abdullah Khalid and Xiangzhou Kong and Bohdan Kulchytskyy and Elica Kyoseva and Ruoyu Li and P. Aaron Lott and Igor L. Markov and Robert F. McDermott and Lucas Morais and Giacomo Pedretti and Pooja Rao and Eleanor Rieffel and Allyson Silva and John Sorebo and Panagiotis Spentzouris and Ziv Steiner and Boyan Torosov and Davide Venturelli and Robert J. Visser and Zak Webb and Xin Zhan and Yonatan Cohen and Pooya Ronagh and Alan Ho and Raymond G. Beausoleil and John M. Martinis},
      year={2026},
      eprint={2411.10406},
      archivePrefix={arXiv},
      primaryClass={quant-ph},
      url={https://arxiv.org/abs/2411.10406}, 
}

@inproceedings{tomesh2023divide,
  title={Divide and conquer for combinatorial optimization and distributed quantum computation},
  author={Tomesh, Teague and Saleem, Zain H and Perlin, Michael A and Gokhale, Pranav and Suchara, Martin and Martonosi, Margaret},
  booktitle={2023 IEEE International Conference on Quantum Computing and Engineering (QCE)},
  volume={1},
  pages={1--12},
  year={2023},
  organization={IEEE}
}

@misc{johnson2026distributedquantumcomputingadaptive,
      title={Distributed Quantum Computing via Adaptive Circuit Knitting}, 
      author={K. Grace Johnson and Aniello Esposito and Gaurav Gyawali and Xin Zhan and Rohit Ganti and Namit Anand and Raymond G. Beausoleil and Masoud Mohseni},
      year={2026},
      eprint={2603.12411},
      archivePrefix={arXiv},
      primaryClass={quant-ph},
      url={https://arxiv.org/abs/2603.12411}, 
}

@article{mitarai2021constructing,
  title={Constructing a virtual two-qubit gate by sampling single-qubit operations},
  author={Mitarai, Kosuke and Fujii, Keisuke},
  journal={New Journal of Physics},
  volume={23},
  number={2},
  pages={023021},
  year={2021},
  publisher={IOP Publishing}
}

@article{peng2020simulating,
  title={Simulating large quantum circuits on a small quantum computer},
  author={Peng, Tianyi and Harrow, Aram W and Ozols, Maris and Wu, Xiaodi},
  journal={Physical review letters},
  volume={125},
  number={15},
  pages={150504},
  year={2020},
  publisher={APS}
}

@inproceedings{bechtold2023patterns,
  title={Patterns for Quantum Circuit Cutting},
  author={Bechtold, Marvin and Barzen, Johanna and Beisel, Martin and Leymann, Frank and Weder, Benjamin},
  booktitle={Proceedings of the 30th Conference on Pattern Languages of Programs},
  pages={1--12},
  year={2023}
}

@inproceedings{kan2025circuit,
  title={Circuit Folding: Scalable and Graph-Based Circuit Cutting via Modular Structure Exploitation},
  author={Kan, Shuwen and Li, Yanni and Wang, Hao and Mouradian, Sara and Mao, Ying},
  booktitle={2025 IEEE/ACM International Conference On Computer Aided Design (ICCAD)},
  pages={1--9},
  year={2025},
  organization={IEEE}
}

@misc{chen2025enhancedquantumcircuitcutting,
      title={Enhanced Quantum Circuit Cutting Framework for Sampling Overhead Reduction}, 
      author={Po-Hung Chen and Dah-Wei Chiou and Bo-Hung Chen and Jie-Hong Roland Jiang},
      year={2025},
      eprint={2412.17704},
      archivePrefix={arXiv},
      primaryClass={quant-ph},
      url={https://arxiv.org/abs/2412.17704}, 
}

@article{Piveteau_2025,
   title={Circuit cutting with classical side information},
   volume={7},
   ISSN={2643-1564},
   url={http://dx.doi.org/10.1103/38mx-36k6},
   DOI={10.1103/38mx-36k6},
   number={3},
   journal={Physical Review Research},
   publisher={American Physical Society (APS)},
   author={Piveteau, Christophe and Schmitt, Lukas and Sutter, David},
   year={2025},
   month=July }

@misc{qiskit-addon-cutting,
  author = {
    Agata M. Bra\'{n}czyk
    and Almudena {Carrera Vazquez}
    and Daniel J. Egger
    and Bryce Fuller
    and Julien Gacon
    and James R. Garrison
    and Jennifer R. Glick
    and Caleb Johnson
    and Saasha Joshi
    and Edwin Pednault
    and C. D. Pemmaraju
    and Pedro Rivero
    and Ibrahim Shehzad
    and Stefan Woerner
  },
  title = {{Qiskit addon: circuit cutting}},
  howpublished = {\url{https://github.com/Qiskit/qiskit-addon-cutting}},
  year = {2024},
  doi = {10.5281/zenodo.7987997}
}

@article{CALEFFI2024110672,
title = {Distributed quantum computing: A survey},
journal = {Computer Networks},
volume = {254},
pages = {110672},
year = {2024},
issn = {1389-1286},
doi = {https://doi.org/10.1016/j.comnet.2024.110672},
url = {https://www.sciencedirect.com/science/article/pii/S1389128624005048},
author = {Marcello Caleffi and Michele Amoretti and Davide Ferrari and Jessica Illiano and Antonio Manzalini and Angela Sara Cacciapuoti},
}

@article{escofet2023hungarian,
  title={Hungarian qubit assignment for optimized mapping of quantum circuits on multi-core architectures},
  author={Escofet, Pau and Ovide, Anabel and Almudever, Carmen G and Alarc{\'o}n, Eduard and Abadal, Sergi},
  journal={IEEE Computer Architecture Letters},
  volume={22},
  number={2},
  pages={161--164},
  year={2023},
  publisher={IEEE}
}

@article{fujii2022deep,
  title={Deep variational quantum eigensolver: A divide-and-conquer method for solving a larger problem with smaller size quantum computers},
  author={Fujii, Keisuke and Mizuta, Kaoru and Ueda, Hiroshi and Mitarai, Kosuke and Mizukami, Wataru and Nakagawa, Yuya O},
  journal={PRX Quantum},
  volume={3},
  number={1},
  pages={010346},
  year={2022},
  publisher={APS}
}

@misc{idan2026quantumcircuitcuttingcomplexity,
      title={Quantum Circuit Cutting: Complexity and Optimization}, 
      author={Yuval Idan and Eitan Zahavi and Elad Mentovich and Eliahu Cohen and Shmuel Zaks},
      year={2026},
      eprint={2604.23700},
      archivePrefix={arXiv},
      primaryClass={quant-ph},
      url={https://arxiv.org/abs/2604.23700}, 
}

@inproceedings{ito2023algorithmic,
  title={Algorithmic theory of qubit routing},
  author={Ito, Takehiro and Kakimura, Naonori and Kamiyama, Naoyuki and Kobayashi, Yusuke and Okamoto, Yoshio},
  booktitle={Algorithms and Data Structures Symposium},
  pages={533--546},
  year={2023},
  organization={Springer}
}

@article{zhu2022complexity,
  title={The complexity of quantum circuit mapping with fixed parameters},
  author={Zhu, Pengcheng and Zheng, Shenggen and Wei, Lihua and Cheng, Xueyun and Guan, Zhijin and Feng, Shiguang},
  journal={Quantum Information Processing},
  volume={21},
  number={10},
  pages={361},
  year={2022},
  publisher={Springer}
}

@article{arute2019quantum,
  title={Quantum supremacy using a programmable superconducting processor},
  author={Arute, Frank and Arya, Kunal and Babbush, Ryan and Bacon, Dave and Bardin, Joseph C and Barends, Rami and Biswas, Rupak and Boixo, Sergio and Brandao, Fernando GSL and Buell, David A and others},
  journal={nature},
  volume={574},
  number={7779},
  pages={505--510},
  year={2019},
  publisher={Nature Publishing Group UK London}
}
